\documentclass[journal]{IEEEtran}
\ifCLASSINFOpdf
\else
\fi
\ifCLASSOPTIONcompsoc
 \usepackage[caption=false,font=normalsize,labelfont=sf,textfont=sf]{subfig}
\else
 \usepackage[caption=false,font=footnotesize]{subfig}
\fi

\def\one{{\boldsymbol{1}}}
\def\zero{{\boldsymbol{0}}}
\def\bell{{\boldsymbol{\ell}}}

\def\bb0{{\mathbb{0}}}

\def\ba{{\boldsymbol{a}}}
\def\bb{{\boldsymbol{b}}}

\def\bd{{\boldsymbol{d}}}
\def\bee{{\boldsymbol{e}}}

\def\bk{{\boldsymbol{k}}}

\def\bmm{{\boldsymbol{m}}}
\def\bn{{\boldsymbol{n}}}

\def\bv{{\boldsymbol{v}}}
\def\bw{{\boldsymbol{w}}}

\def\bz{{\boldsymbol{z}}}
\def\b0{{\boldsymbol{0}}}

\def\bC{{\boldsymbol{C}}}

\def\bE{{\boldsymbol{E}}}

\def\bH{{\boldsymbol{H}}}
\def\bI{{\boldsymbol{I}}}
\def\bJ{{\boldsymbol{J}}}
\def\bK{{\boldsymbol{K}}}

\def\bN{{\boldsymbol{N}}}

\def\bQ{{\boldsymbol{Q}}}

\def\bS{{\boldsymbol{S}}}
\def\bT{{\boldsymbol{T}}}

\def\x{{\mathrm{x}}}
\def\y{{\mathrm{y}}}

\def\b{{\mathrm{b}}}

\def\r0{{\mathbf{0}}}

\def\bbC{{\mathbb{C}}}

\def\bbE{{\mathbb{E}}}

\def\bbN{{\mathbb{N}}}

\def\bbR{{\mathbb{R}}}

\def\bbZ{{\mathbb{Z}}}

\def\cD{\mathcal{D}}

\def\cM{\mathcal{M}}
\def\cN{\mathcal{N}}
\def\cO{\mathcal{O}}

\def\cT{\mathcal{T}}

\def\sfD{\mathsf{D}}

\def\btau{\bm \tau}

\def\bdelta{\bm \delta}

\def\bmu{\bm \mu}

\def\bsf0{{\bm{\mathsf{0}}}}

\def\N0{{N_{\mathrm{0}}}}

\def\bsf{{\boldsymbol{s}_\mathrm{f}}}

\newcommand{\be}{\begin{equation}}
\newcommand{\ee}{\end{equation}}
\newcommand{\bal}{\begin{align}}
\newcommand{\eal}{\end{align}}
\def\tr {{\rm tr}}

\def\SNR    {{\mathsf{SNR}}}

\def\diag   {\mbox{\rm diag}}

\usepackage{graphicx}
\usepackage{amsmath}
\usepackage{amsthm}
\usepackage{mathtools}
\usepackage{epsfig}
\usepackage{float}
\usepackage{lipsum}
\usepackage{stfloats}
\usepackage{array}
\usepackage{amssymb}
\usepackage{cite}
\usepackage{url}
\usepackage{algorithm}
\usepackage{algpseudocode}
\usepackage{multirow}
\usepackage{fancyh}
\usepackage{flushend}
\usepackage{upgreek}
\usepackage{dsfont}
\usepackage{gensymb}
\usepackage{booktabs}
\usepackage{bm}
\usepackage[shortlabels]{enumitem}
\usepackage{threeparttable}
\usepackage{makecell}
\usepackage{multicol, blindtext}
\usepackage{mathtools, cuted}
\usepackage[table]{xcolor}

\usepackage{adjustbox}
\usepackage{color}
\usepackage{cancel}
\usepackage{tikz}
\usetikzlibrary{
intersections, arrows.meta,
automata,er,calc,
backgrounds,
mindmap,folding,
patterns,
decorations.markings,
fit,
shapes,matrix,
positioning,
shapes.geometric,
arrows,through
}
\usepackage{pifont}% http://ctan.org/pkg/pifont

\theoremstyle{remark}

\theoremstyle{remark}

\allowdisplaybreaks
\renewcommand{\Im}{\operatorname{Im}}
\normalsize
\graphicspath{ {./images/} }

\newcommand\gray{\cellcolor{gray!20}}
\newcommand\ggray{\cellcolor{gray!60}}
\newcommand\gggray{\cellcolor{gray!100}}

\begin{document}
\title{Multidimensional Polynomial Phase Estimation
\thanks{
Heedong Do and Namyoon Lee are with Korea University, 02841 Seoul (e-mail: \{doheedong, namyoon\}@korea.ac.kr). Their work is supported by the National Research Foundation of Korea (NRF) grant (No. 2020R1C1C1013381) funded by the Korea
government (MSIT).
A. Lozano is with Univ. Pompeu Fabra, 08018 Barcelona (e-mail: angel.lozano@upf.edu). His work is supported by the Maria de Maeztu Units of Excellence Programme (CEX2021-001195-M) and by ICREA.}
}
\author{\IEEEauthorblockN{Heedong~Do},
 {\it Member,~IEEE},
\and
\IEEEauthorblockN{Namyoon~Lee},
{\it Senior Member,~IEEE},
\and\\
\IEEEauthorblockN{Angel~Lozano},
{\it Fellow,~IEEE}
}
\maketitle

\maketitle

\begin{abstract}
An estimation method is presented for polynomial phase signals, i.e., those adopting the form of a complex exponential whose phase is polynomial in its indices.
Transcending the scope of existing techniques, the proposed estimator can handle an arbitrary number of dimensions and an arbitrary set of polynomial degrees along each dimension;
the only requirement is that the number of observations per dimension exceeds the highest degree thereon.
%only the mild restriction that the highest such degree not exceed the number of signal observations on that dimension.
Embodied by a highly compact sequential algorithm,
this estimator exhibits a strictly linear computational complexity in the number of observations, and is efficient at high signal-to-noise ratios (SNRs). To reinforce the performance at low and medium SNRs,
where any phase estimator is bound to be hampered by the inherent ambiguity caused by phase wrappings, suitable functionalities are incorporated and shown to be highly effective.
%For low and medium SNRs, where the performance of any phase estimator is hampered by the inherent ambiguity caused by phase wrappings, the proposed estimator incorporates suitable (functionalities)
%Altogether, the presented estimator is essentially universal, of linear complexity, and optimal at high SNR in the mean-square sense.
%and superior to existing solutions for those dimensionalities and polynomial orders
%enabling the reconstruction of any polynomial phase signal from its noisy observation with the sole requisite of having a number of observations that matches the polynomial order on each dimension.
%\angel{Yes, the scaling with polynomial degrees is pending refinement. I was trying to see if I could bound it above in some useful way. I'd say an upper bound is $\cO (\Pi_{d=0}^{\sfD-1} M_d \sum_{d=0}^{\sfD-1} M^2_d ) $, which in turn is bounded by $M^{\sfD+2}_{\max}$, but none of the two need be particularly tight.} \heedong{One ``interpretable'' (albeit not analytically attractive) way to think of $\sum_{\bmm\in\cM}|\bmm|$ is that realizing
%\begin{align*}
%    \sum_{\bmm\in\cM}|\bmm| = \bigg|\sum_{\bmm\in\cM} \bmm\,\,\bigg| = |\cM| \bigg|\underbrace{\frac{1}{|\cM|}\sum_{\bmm\in\cM} \bmm}_{\mathclap{\text{center of mass of }\cM}}\,\,\bigg|. 
%\end{align*}
%Having said that, I'm a bit reluctant to even state the complexity scaling with respect to $\cM$ given that the utmost interest is where the degrees are $\leq 3$.
%}
\end{abstract}



% \fi
\IEEEpeerreviewmaketitle

\section{Introduction}

A complex exponential function whose phase is polynomial in its indices can model a variety of \emph{polynomial phase signals}, including images of planar surfaces \cite{francos2001parametric}, magnetic resonance outputs \cite{liang1996model}, interferometric synthetic aperture radar signals \cite{katkovnik2008phase}, burst-mode transmissions \cite{luise1995carrier}, or pulse signals experiencing Doppler effects \cite{lank1973semicoherent}. It can further model communication channels of interest, for instance the expansion of a spherical radio wavefront over an array of antennas \cite{do2023parabolic}; the ensuing channel adopts a multidimensional polynomial phase form (two-dimensional with linear transmit and receive arrays \cite{bohagen2007design, sarris2007design, do2020reconfigurable} and four-dimensional with rectangular arrays \cite{bohagen2007optimal, song2015spatial,  do2021rotatable}, with those dimension counts augmenting respectively to three and five if multiple frequency bands are present).

In any of the above contexts, a crucial problem is often the estimation of the polynomial coefficients from a noisy observation of the signal.
Such \emph{polynomial phase estimation} problem has been studied extensively (see excellent reviews in \cite{mckilliam2010lattice, djurovic2018review, madsen2021finite}), yet the focus has been largely limited to its one-dimensional incarnation. In particular, one-dimensional linear polynomial signals, amounting to complex sinusoids, have attracted the most interest; the problem then reduces to an estimation of the frequency, and readers are referred to \cite{kootsookos1993review, morelli1998feedforward, fowler2002phase} for excellent overviews on the subject.
\nocite{rife1974single, tretter1985estimating, lang1989frequency, kay1989fast, lovell1992statistical, quinn1994threshold, kim1996improved, quinn1997estimation, macleod1998fast, abeysekera1998performance, fowler1999extending, clarkson1999frequency, volker2002frequency, brown2002iterative, aboutanios2005iterative, jacobsen2007fast, quinn2008maximizing, mckilliam2010frequency, candan2011method, serbes2018fast, vincent2023improved}
\nocite{abatzoglou1986fast, djuric1990parameter, peleg1991linear}
\nocite{o2004fast}
% \nocite{solak2022fast}
\nocite{peleg1991cramer, kitchen1994method, peleg1995discrete, porat1996asymptotic, ikram1998estimating, farquharson2005computationally, ju2007generalized, o2010refining, djurovic2012hybrid, mckilliam2014cramer, mckilliam2014polynomial, madsen2019finite}
\nocite{kay1990efficient, so2006approximate}
\nocite{friedlander1996estimation, friedlander1996model}
\nocite{francos1998two}
\nocite{solak2022fast}
A myriad of frequency estimation methods have been set forth, to wit: finding the peak of a periodogram \cite{rife1974single}, interpolating a sampled periodogram \cite{jacobsen2007fast}, harnessing properties of finite differences \cite{kay1989fast}, or finding the nearest lattice point \cite{mckilliam2010frequency}.
Some of these methods have been extended to polynomials of arbitrary order,
but always within the confines of either one or two dimensions;
a comprehensive list of references on these extensions, along with their precise scope, is furnished in Table~\ref{table:priorArt}.
To the best of our knowledge, only the recent work in \cite{solak2022fast} can handle an arbitrary dimensionality, but the polynomial is then circumscribed to being an affine function.

\begin{table}
\setlength{\tabcolsep}{4pt}
\addtolength{\leftskip} {-2cm}
\addtolength{\rightskip} {-2cm}
\centering
\begin{threeparttable}
\caption{Prior Art on Polynomial Phase Estimation
}
\label{table:priorArt}
\begin{tabular}{l l l } 
\toprule
$\!\!$Dimension & Set of polynomial degrees & References %& Multi-component
\\
\midrule
% $1$ & $\{0\}$ & \cite{mckilliam2012direction} & - \\
$1$ & $\{0,1\}$ & \cite{rife1974single, tretter1985estimating, lang1989frequency, kay1989fast, lovell1992statistical, quinn1994threshold, kim1996improved, quinn1997estimation, macleod1998fast, abeysekera1998performance, fowler1999extending, clarkson1999frequency, volker2002frequency, brown2002iterative, aboutanios2005iterative, jacobsen2007fast, quinn2008maximizing, mckilliam2010frequency, candan2011method, serbes2018fast, vincent2023improved}\tnote{*}
%& \cite{rife1976multiple, tufts1982estimation, hua1990matrix, nishiyama1997nonlinear, kay2000mean, malioutov2005sparse, bhaskar2013atomic}\tnote{$\dagger$}
\\
$1$ & $\{0,1,2\}$ & \cite{abatzoglou1986fast, djuric1990parameter, peleg1991linear}
%& \cite{xia2000discrete, saha2002maximum}
\\
$1$ & $\{0,1,2,3\}$ & \cite{o2004fast} 
%& -
\\
$1$ & $\{0,1,\ldots,M\}$ & \cite{peleg1991cramer, kitchen1994method, peleg1995discrete, porat1996asymptotic, ikram1998estimating, farquharson2005computationally, ju2007generalized, o2010refining, djurovic2012hybrid, mckilliam2014cramer, mckilliam2014polynomial, madsen2019finite}
%& \cite{peleg1996multicomponent, barbarossa1998product, pham2006analysis}
\\
$2$ & $\{(m_0,m_1):m_0+m_1\leq 1\}$ & \cite{kay1990efficient, so2006approximate}
%& \cite{hua1992estimating, sacchini1993two, rouquette2001estimation, chi2014compressive}\tnote{$\dagger$}
\\
$2$ & $\{(m_0,m_1):m_0+m_1\leq M\}$ & \cite{friedlander1996estimation, friedlander1996model}
%& \cite{barbarossa2014parameter, simeunovic2015parameter}
\\
$2$ & $\{0,1,\ldots,M_0\}\times\{0,1,\ldots,M_1\}$ & \cite{francos1998two}
%& -
\\
Any $\sfD$ & $\{(m_0,\ldots,m_{\sfD-1}):m_0+\cdots+m_{\sfD-1}\leq 1\}$ & \cite{solak2022fast}
%& \cite{liu2006eigenvector, chen2013spectral, sahnoun2017multidimensional} 
\\
Any $\sfD$ & Any set satisfying \eqref{moreSampleThanDegree} & \textbf{This paper} %& -
\\
\bottomrule
\end{tabular}
\begin{tablenotes} \footnotesize
\item[*] Further references dealing only with real-valued signals are \cite{quinn1994estimating, handel2000properties}.
%\item[$\dagger$] Direction-of-arrival estimation methods \cite{schmidt1986multiple, roy1989esprit} with spatial smoothing \cite{shan1985spatial} can also be used for this purpose.
\end{tablenotes}
\end{threeparttable}
\end{table}

%However, for the sake of generality, the complex-valued formulation is more relevant.

%\heedong{Almost correct but there are some exceptions such as \cite{peleg1995discrete}, which can be generalized to 2D \cite{friedlander1996model}.}

%\heedong{The number of references is now over a hundred! It is perhaps too comprehensive and looks like more of a review article. Would it be okay?}
%\angel{It's a possibility, yes. But the paper contains a lot of original work, which a survey paper does not. We'll see about what to do with it...}

This paper considers multidimensional polynomial phase estimation in broad generality. Arbitrary dimensionalities and polynomial orders are encompassed, subject only to the very mild condition that the number of noisy signal observations per dimension exceeds the number of polynomial terms thereon.
In such broad generality, a major challenge of the multidimensional problem is the exploding number of observations.
%What distinguishes the multidimensional case from its one-dimensional counterpart is the signal size.
For instance, a four-dimensional signal containing $32\times32\times32\times32$ observations has about a million entries, and algorithms with superlinear complexity in the number of observations would be extremely taxing computationally.

Among the aforementioned one-dimensional methods, those inspired by properties of finite differences are compelling owing precisely to their linear complexity in the number of observations, hence they are the approach of choice here. In addition to their favorable complexity scaling, it is found that repeated application of finite differencing serves to naturally identify the fundamental ambiguity of the estimation problem.
Altogether, the contributions of this work are:
\begin{itemize}
    \item The fundamental ambiguity of the multidimensional polynomial phase estimation is identified, generalizing the one-dimensional result in \cite{mckilliam2009identifiability}.
    \item Armed with a toolkit that, besides finite differences \cite{kay1989fast}, includes sequential estimation \cite{friedlander1996model} and a modification to account for the circular nature \cite{madsen2019finite}, a computationally affordable method is devised for multidimensional polynomial phase estimation. Precisely, the complexity is strictly linear in the number of observations.
    \item The resulting estimator is shown to be efficient at high SNR, meaning that the covariance of the estimate attains the Cramer-Rao bound (CRB).
    \item At low and medium SNRs, where a shortfall from the CRB does exist, a progressive refinement of the estimates via repeated application of finite differences with multiple lags is shown to shrink this gap considerably.
    %It is shown how, by applying finite differences with multiple lags to sequentially refine the estimates, the shortfall from the CRB at low and medium SNRs can be shrunk considerably.
\end{itemize}

The manuscript is organized as follows.
Sec.~\ref{SecI} describes the structure of a polynomial phase signal and various orders on the set of its polynomial degrees. Subsequently, Sec.~\ref{sec:choiceOfPolynomialBasis} deals with the choice of a basis for the vector space of polynomials given a set of degrees, and with change-of-basis transformations,
finally settling on a binomial representation as the most convenient. Then, Sec.~\ref{sec:differenceOperator} introduces shift and difference operators and it unveils the inherent ambiguity created by phase wrappings---only the fractional parts of the polynomial coefficients can be determined.
Sec.~\ref{sec:polynomialPhaseEstimation} is the heart of the paper, in the sense that the basic version of the proposed phase estimation algorithm is expounded therein and its complexity is assessed. After that, Secs.~\ref{sec:crb} and \ref{sec:performanceEvaluation} deal with the CRB, respectively deriving it for the problem at hand and showing that it is attained by the proposed estimator at high SNR, when ambiguity is not an issue. Sec.~\ref{sec:accountingForTheCircularNature} is devoted to refining the estimator to improve its performance outside the high-SNR regime, when ambiguity does become an issue. Ramping down the paper, Secs.~\ref{sec:LiftingTechnicalCondition} and \ref{sec:introducingLagParameters} further refine the estimator to render it more general and
to further enhance its performance at low-to-medium SNRs, and Sec.~\ref{sec:conclusion} concludes the paper.

\subsection{Notation}
\label{sec:notation}
The set of nonnegative integers is denoted by $\bbN_0$ while the first $N$ nonnegative integers are compactly denoted by
\begin{align}
    [N]\equiv \begin{cases}
    \{0,1,\ldots,N-1\} & \;\; \text{integer }N>0\\
    \emptyset & \;\; \text{integer }N\leq 0
    \end{cases}.
\end{align}
A second bracket notation is used throughout the paper, namely the Iverson bracket \cite{knuth1992two}
\begin{align}
    [\text{condition}] \equiv \begin{cases}
    1 &\text{the condition is true}\\
    0 &\text{otherwise}
    \end{cases} .
\end{align} 
Yet a third bracket notation, $[\cdot]_n$, indicates the $n$th entry of a vector and, by extension, $[\cdot]_{n,n'}$ indicates the $(n,n')$th entry of a matrix.
There should be no confusion among these various bracket notations, as their arguments are of a different nature.

The generalized binomial coefficient is denoted, for $n\in\bbZ$, by \cite{graham1994concrete}
\begin{align}
    {n \choose k} \equiv \begin{cases}
    \frac{n(n-1)\cdots(n-k+1)}{k!} & \;\; \text{integer }k\geq0\\
    0 & \;\; \text{integer }k<0 
    \end{cases} .
    \label{EOS}
\end{align}
%for $n\in\bbZ$. % with $k!$ the factorial of $k$.

Multidimensional counterparts to the above and to other one-dimensional notations are used extensively in the sequel. Specifically, for
\begin{align}
    &\bN=(N_0,\ldots,N_{\sfD-1})\in\bbZ^\sfD \qquad \bn=(n_0,\ldots,n_{\sfD-1})\in\bbZ^\sfD\nonumber\\
    &\bmm=(m_0,\ldots,m_{\sfD-1})\in\bbZ^\sfD \qquad \bk=(k_0,\ldots,k_{\sfD-1})\in\bbN_0^\sfD, \nonumber
\end{align}
% $\bN=(N_0,\ldots,N_{\sfD-1})\in\bbZ^\sfD$, $\bn=(n_0,\ldots,n_{\sfD-1})\in\bbZ^\sfD$, $\bmm=(m_0,\ldots,m_{\sfD-1})\in\bbZ^\sfD$, and $\bk=(k_0,\ldots,k_{\sfD-1})\in\bbN_0^\sfD$,
we define $[\bN]\equiv [N_0]\times \cdots \times [N_{\sfD-1}]$ with $\times$ the Cartesian product of sets, and also
\begin{align}
    \bn^\bk &\equiv n_0^{k_0}\cdots n_{\sfD-1}^{k_{\sfD-1}} 
    &{\bn \choose \bmm}&\equiv {n_0 \choose m_0}\ldots{n_{\sfD-1} \choose m_{\sfD-1}}\nonumber\\
    \bk!&\equiv k_0!\cdots k_{\sfD-1}! 
    &|\bn|&\equiv \sum_{d=0}^{\sfD-1} |n_d|.\nonumber
\end{align}

% Readers are referred \ref{:notations} for the notations used throughout the paper.

% \begin{table}
% % \setlength{\tabcolsep}{2pt}
% \renewcommand{\arraystretch}{0}
% \centering
% \begin{threeparttable}
% \caption{Notations} %\vspace*{-2mm}
% \begin{tabular}{ll} 
% \toprule
% Notation & Arguments\\
% \midrule
% % \addlinespace
% % $|\balpha|\equiv \alpha_1+\ldots+\alpha_n$ & $\balpha \in \bbZ_0^n$\\
% \addlinespace
% $\balpha !\equiv \alpha_1!\ldots\alpha_n!$ & $\balpha \in \bbN_0^n$\\
% \addlinespace
% ${\bx \choose \balpha} \equiv {x_1 \choose \alpha_1}\ldots{x_n \choose \balpha_n}$ & $\bx \in \bbR^n, \balpha \in \bbZ^n$\\
% \addlinespace
% $\bx^\balpha \equiv x_1^{\alpha_1}\ldots x_n^{\alpha_n}$ & $\bx \in \bbR^n, \balpha \in \bbN_0^n$\\
% \addlinespace
% $[\balpha]\equiv [\alpha_1]\times \cdots \times [\alpha_n]$  & $\balpha \in \bbN_0^n$\\
% % \addlinespace
% % $\cT^{\balpha} \equiv \cT_1^{\alpha_1}\ldots\cT_n^{\alpha_n} $  & $\balpha \in \bbN_0^n$, $\cT_k$ is operator\tnote{*}\\
% \bottomrule
% \end{tabular}
% % \begin{tablenotes} \footnotesize 
% % \item[*] Superscripts on the right-hand side denote composition of operators. 
% % \end{tablenotes}
% \label{table:notations}
% \end{threeparttable}
% %\vspace*{-4mm}
% \end{table}

\section{Polynomial Phase Signal}
\label{SecI}

Let us consider, in broad generality, the $\sfD$-dimensional polynomial phase signal $\big[\bn\in[\bN]\big]e^{j2\pi x(\bn)}$, with the Iverson bracket reflecting that the signal is observed over a finite window, and with
\begin{align}
    x(\bn)&= \sum_{\bmm} 
 a_{\bmm}\frac{\bn^{\bmm}}{\bmm !}.  \label{polynomialMonomial}
\end{align}
% with $\bn\equiv(n_0,\dots,n_{\sfD-1}) \in \bbZ^{\sfD}$.
% while the outer bracket in $\big[\bn\in[\bN]\big]$ denotes the Iverson bracket \cite{knuth1992two} and $[\bN] \equiv [N_0]\times\cdots \times [N_{\sfD-1}]$ with $\bN = (N_0,\ldots,N_{\sfD-1})$.
% In turn, $\bmm\equiv(m_0,\ldots,m_{\sfD-1})$ and
The coefficient $a_\bmm$ vanishes outside a prescribed set $\cM\subset \bbN_0^\sfD$, hence the summation in \eqref{polynomialMonomial} is implicitly over $\cM$.
This is the set of polynomial degrees.

The above reduces to the familiar one-dimensional formulation for $\sfD=1$ and $\cM=[M+1]$, whereby \eqref{polynomialMonomial} becomes
%\footnote{Throughout the paper, for $\sfD=1$, we shall use plain fonts for brevity as long as no confusion arises.}
\begin{align}
    x(n) = a_0 + a_1 n+\cdots+\frac{a_{M}}{M!}n^{M}
\end{align}
while, for $\sfD=2$ and $\cM = \{0,\ldots,M_0 \}\times\{0,\ldots,M_1\}$,
\begin{equation}
    x(n_0,n_1) = \sum_{m_0 = 0}^{M_0} \sum_{m_1 = 0}^{M_1} a_{m_0,m_1} \frac{n_0^{m_0}}{m_0!} \frac{n_1^{m_1}}{m_1!} .
\end{equation}
In turn, the case of an affine polynomial of arbitrary dimension $\sfD$ corresponds to
\begin{align}
    \cM = \left \{ (m_0,\ldots,m_{\sfD-1})\in\bbN_0^{\sfD}:m_0+\cdots+m_{\sfD-1}\leq 1 \right\}, \nonumber
\end{align}
for which, letting $\bee_d\in\bbR^{\sfD}$ be the $d$th standard unit vector,
\begin{align}
    x(\bn) = a_{\zero} + a_{\bee_0}n_0 + a_{\bee_1}n_1 + \cdots + a_{\bee_{\sfD-1}}n_{\sfD-1} .
\end{align}
% \angel{Is there a reason why you wanted to introduce these unit vectors here? There aren't really needed here, we could just give the above as
% \begin{align}
%     x(\bn) = a_{\zero} + a_{0}n_0 + a_{1}n_1 + \cdots + a_{\sfD-1}n_{\sfD-1} .
% \end{align}
% }

Returning to the general form in \eqref{polynomialMonomial},
the task at hand is to estimate the polynomial coefficients, $a_{\bmm}$, %for $m \in \cM$,
from the noisy observation
\begin{align}
    y(\bn) = \big[\bn\in[\bN]\big] \big(e^{j2\pi x(\bn)}  + w_{\bbC}(\bn)\big), \label{signalModel}
\end{align}
where $w_{\bbC}(\bn) \overset{\mathrm{iid}}{\sim} \cN_{\bbC}(0,\frac{1}{\SNR})$.

\subsection{Orders on the Set of Polynomial Degrees}
\label{sec:order}

For later use, let us define a partial order \cite[Ch. 9.6]{rosen2012discrete} on $\bbN_0^{\sfD}$, denoted by $\leq$, as
\begin{align}
 % \bmm' \leq \bmm \Leftrightarrow m_d' \leq m_d \text{ for all } d. \label{partialOrder} 
  \bmm' \leq \bmm \Leftrightarrow [\bmm'] \subset [\bmm]. \label{partialOrder} 
\end{align}
% \angel{We can also use $\bbN_0$ instead of $\bbZ_{\geq 0}$. Regardless of which one you prefer, the definition should be given the first time we use it. It's fairly obvious, but also the typical thing that annoying reviewers pick on :-)}
% We can think of the partial order on $\cM$ inherited from $\bbN_0^{\sfD}$. 
Inherited from $\bbN_0^{\sfD}$, one can naturally construct a partial order on $\cM$, which is the partial order used hereafter.

There exists a total order $\preccurlyeq$ (not necessarily unique) that is compatible with the partial order. This holds for any finite partially ordered set 
\cite[Ch. 9.6]{rosen2012discrete}, and it is instructive to explain how it can be obtained. As $\cM$ is finite, it is always possible to find a minimal element (not necessarily unique) with respect to the partial order \eqref{partialOrder}  and label it $\bmm_1$. Then, consider the set $\cM\setminus \{\bmm_1\}$, pick one of its minimal elements, and label it $\bmm_2$. Repeating the procedure, one obtains a total order
\begin{align}
    \bmm_1 \prec \bmm_2 \prec \ldots \prec \bmm_{|\cM|}
\end{align}
that, by construction, is compatible with the original partial order.
% The last thing to be checked is that the total order $\preccurlyeq$ is compatible with the partial order $\leq$. It is enough to show that
% % \begin{align}
% %     \bmm_\ell \succ \bmm_{\ell'} \Rightarrow \bmm_\ell \nleq \bmm_{\ell'},
% % \end{align}
% % which is the contrapositive of
% \begin{align}
%     \bmm_\ell \leq \bmm_{\ell'} \Rightarrow \bmm_\ell \preccurlyeq \bmm_{\ell'}.
% \end{align}
% We proceed by contraposition. Let $\bmm_\ell \succ \bmm_{\ell'}$, or equivalently $\ell > \ell'$. By construction, $\bmm_{\ell'}$ is a minimal element of $\{\bmm_{\ell'}, \ldots, \bmm_{\ell}, \ldots, \bmm_{|\cM|}\}$. We therefore have $\bmm_\ell \nleq \bmm_{\ell'}$.
A specific instance of this procedure is employed in \cite{friedlander1996model, francos1998two} for $\sfD = 2$, with an element having minimum absolute value being chosen at each step.
% Let $\bmm$ is a element with minimum 1-norm. For any $\bmm'\leq \bmm$, we have $m_d'\leq m_d$ for all $d$ and $|\bmm'|\geq |\bmm|$, which implies $\bmm=\bmm'$. Therefore, $\bmm$ is a minimal element.

\begin{figure}
\centering
\subfloat[Condition \eqref{containingAllLowerDegrees} holds.]{
\begin{tikzpicture}
\path (-1.2,0) rectangle (2.8,2.5);
\draw[step=0.5cm,gray,thin] (0,0) grid (2.1,2.1);
\node[align=center] at (1,-0.6) {$m_0$};
\node[align=center] at (-0.7,1) {$m_1$};
\node[align=center] at (0,-0.3) {\footnotesize $0$};
\node[align=center] at (0.5,-0.3) {\footnotesize $1$};
\node[align=center] at (1,-0.3) {\footnotesize $2$};
\node[align=center] at (1.5,-0.3) {\footnotesize $3$};
\node[align=center] at (2,-0.3) {\footnotesize $4$};
\node[align=center] at (-0.3,0) {\footnotesize $0$};
\node[align=center] at (-0.3,0.5) {\footnotesize $1$};
\node[align=center] at (-0.3,1) {\footnotesize $2$};
\node[align=center] at (-0.3,1.5) {\footnotesize $3$};
\node[align=center] at (-0.3,2) {\footnotesize $4$};
\fill[gray, opacity = 0.7] (0,0) rectangle (1.5,1);
\fill (0,0) circle (0.07cm);
\fill (0.5,0) circle (0.07cm);
\fill (1,0) circle (0.07cm);
\fill (1.5,0) circle (0.07cm);
\fill (2,0) circle (0.07cm);
\fill (0,0.5) circle (0.07cm);
\fill (0.5,0.5) circle (0.07cm);
\fill (1,0.5) circle (0.07cm);
\fill (1.5,0.5) circle (0.07cm);
\fill (2,0.5) circle (0.07cm);
\fill (0,1) circle (0.07cm);
\fill (0.5,1) circle (0.07cm);
\fill (1,1) circle (0.07cm);
\fill (1.5,1) circle (0.07cm);
\fill (0,1.5) circle (0.07cm);
\fill (0.5,1.5) circle (0.07cm);
\fill (0,2) circle (0.07cm);
\end{tikzpicture}
}
\subfloat[Condition \eqref{containingAllLowerDegrees} does not hold.]{
\begin{tikzpicture}
\path (-1.2,0) rectangle (2.8,2.5);
\draw[step=0.5cm,gray,thin] (0,0) grid (2.1,2.1);
\node[align=center] at (1,-0.6) {$m_0$};
\node[align=center] at (-0.7,1) {$m_1$};
\node[align=center] at (0,-0.3) {\footnotesize $0$};
\node[align=center] at (0.5,-0.3) {\footnotesize $1$};
\node[align=center] at (1,-0.3) {\footnotesize $2$};
\node[align=center] at (1.5,-0.3) {\footnotesize $3$};
\node[align=center] at (2,-0.3) {\footnotesize $4$};
\node[align=center] at (-0.3,0) {\footnotesize $0$};
\node[align=center] at (-0.3,0.5) {\footnotesize $1$};
\node[align=center] at (-0.3,1) {\footnotesize $2$};
\node[align=center] at (-0.3,1.5) {\footnotesize $3$};
\node[align=center] at (-0.3,2) {\footnotesize $4$};
\fill[gray, opacity = 0.7] (0,0) rectangle (1.5,1);
\fill (0,0) circle (0.07cm);
\fill (0.5,0) circle (0.07cm);
\fill (1,0) circle (0.07cm);
\fill (1.5,0) circle (0.07cm);
\fill (2,0) circle (0.07cm);
\fill (0,0.5) circle (0.07cm);
\fill (0.5,0.5) circle (0.07cm);
\fill (1,0.5) circle (0.07cm);
\fill (1.5,0.5) circle (0.07cm);
\fill (2,0.5) circle (0.07cm);
\fill (0,1) circle (0.07cm);
\fill (0.5,1) circle (0.07cm);
% \fill (1,1) circle (0.07cm);
\fill (1.5,1) circle (0.07cm);
\fill (0,1.5) circle (0.07cm);
\fill (0.5,1.5) circle (0.07cm);
\fill (0,2) circle (0.07cm);
\end{tikzpicture}
}
\caption{Depictions of $\cM$ for $\sfD=2$, with the dots representing its constituent elements.
For $\sfD=2$, condition \eqref{containingAllLowerDegrees} amounts to every rectangle whose upper-right corner is in $\cM$ being entirely within $\cM$. As an example, the shaded rectangle represents the set $[\bmm+\one]$ for $\bmm = (3,2)$.}
\label{fig:firstProvision}
\end{figure}
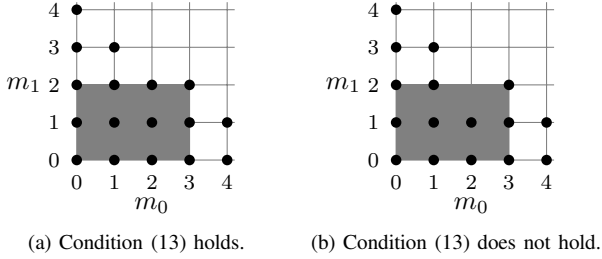

\subsection{Conditions on the Set of Polynomial Degrees}
\label{sec:assumptions}

Only a very mild condition is imposed on $\cM$, namely that, to prevent ambiguities related to the finiteness of the observation window, the monomials
\begin{align}
    \bigg\{\big[\bn\in[\bN]\big]\frac{\bn^{\bmm}}{\bmm !}: \bmm \in \cM\bigg\} \label{monomialBasis}
    % \bigg\{ p: [\bN]\rightarrow \bbR : p(\bn) = \frac{\bn^{\bmm}}{\bmm !}, \bmm \in \cM\bigg\} \label{monomialBasis}
\end{align}
are linearly independent. As shown in App.~\ref{app:secondConditionProof}, it follows that, for every $\bmm\in\cM$,
\begin{align}
\bN \geq \bmm+\one  \label{moreSampleThanDegree}    
\end{align}
or, equivalently, $\cM\subset[\bN]$. This, in turn, will be later shown %in App.~\ref{FCB} %Sec.~\ref{sec:integerValuedPolynomials}
to suffice for the linear independence of \eqref{monomialBasis}. The imposed condition thus translates to a certain number of observations being needed on each dimension, with such number depending on the highest polynomial degree on that dimension. 
%Conversely, the highest polynomial degrees that can be tackled along the $\sfD$ dimensions are determined by the respective numbers of observations.

An additional condition, which will be shown to be unnecessary in Sec. \ref{sec:LiftingTechnicalCondition}, is assumed in the interim to facilitate the free conversion of polynomial basis. This technical condition is that
every $\bmm$ within $\cM$ satisfies (see Fig. \ref{fig:firstProvision})
\begin{align}
    [\bmm+\one] \subset \cM. \label{containingAllLowerDegrees}
\end{align}
with $\one$ henceforth denoting the all-ones vector of appropriate length. 
%To the best of our knowledge, all prior work on polynomial phase estimation abides by (\ref{containingAllLowerDegrees}). 
%This simply imposes a certain smoothness on the boundary of $\cM$, preventing excessive jaggedness. 
We hasten to reiterate that this condition will be lifted later in the paper.

\section{Choice of Polynomial Basis}
\label{sec:choiceOfPolynomialBasis}

Given \eqref{containingAllLowerDegrees} and the linear independence of the monomials in \eqref{monomialBasis}, these monomials are a basis for the vector space of polynomials whose degrees are in $\cM$.
For the sake of analysis, however, other bases are sometimes more convenient.
For instance, in some prior art \cite{peleg1991cramer, farquharson2005computationally, mckilliam2014polynomial} (see also \cite{ristic1998comments}), a shifted observation window is considered, i.e., the signal model in \eqref{signalModel} is modified into
\begin{align}
    y(\bn) = \big[\bn-\bdelta \in[\bN]\big] \big(e^{j2\pi x(\bn)}  + w_{\bbC}(\bn)\big)
\end{align}
for some constant $\bdelta \in\bbZ^{\sfD}$. Shifting the indices, we obtain
\begin{align}
    &y(\bn+ \bdelta) 
    = \big[\bn\in[\bN]\big]\\
    &\qquad\cdot\bigg( \! \exp\!\bigg[j2\pi\bigg(\sum_{\bmm} a_{\bmm}\frac{(\bn+\bdelta)^\bmm}{\bmm!} \bigg)\bigg] + w_{\bbC}(\bn+\bdelta)\bigg), \nonumber
\end{align}
which amounts to using another polynomial basis. This section addresses the choice of a basis in full generality.

%\heedong{In some papers, such as \cite{ristic1998comments} and references therein, people choose shifted monomials for a basis. It can serve as another good motivation of this section.}

Let us consider a set of polynomials
\begin{align}
    \big\{\big[\bn\in[\bN]\big] \, p_{\bmm}(\bn):\bmm\in\cM\big\}, \label{setOfPolynomials}    
\end{align}
where
\begin{align}
    p_{\bmm}(\bn) = \sum_{\bell} t_{\bell,\bmm} \frac{\bn^\bell}{\bell!}\label{anotherBasis}
\end{align}
with the summation being implicitly over $\cM$. 
If \eqref{setOfPolynomials} is a basis of the space spanned by the monomials in \eqref{monomialBasis}, the matrix\footnote{With a total order defined on it, $\cM$ can be identified with $\big[|\cM|\big]$. Hereafter, and with a slight abuse of notation, this identification is embraced. In particular, $\bmm\in\cM$ is allowed as an index for a vector or a matrix.} $\bT\in\bbR^{|\cM|\times|\cM|}$, with $(\bell,\bmm)$th entry given by $t_{\bell,\bmm}$, is invertible \cite[Thm. 2.22 (a)]{friedberg2019linear}. Conversely, if $\bT$ is invertible, \eqref{setOfPolynomials} is a basis \cite[Exerc. 2.5.13]{friedberg2019linear}.
A change-of-basis formula emerges readily; letting
\begin{align}
    x(\bn) = \sum_{\bmm} a_{\bmm}\frac{\bn^{\bmm}}{\bmm !}= \sum_\bmm d_\bmm p_\bmm(\bn) \label{representations}
\end{align}
and plugging \eqref{anotherBasis} in, it is found that \cite[Thm. 2.22 (b)]{friedberg2019linear}
\begin{align}
    a_{\bell} = \sum_{\bmm} t_{\bell,\bmm}d_{\bmm}.
\end{align}
Vectorization converts the above into
\begin{align}
    \ba = \bT\bd,
\end{align}
which is intuitive given the significance of the quantities involved, namely \cite[Thm. 2.14]{friedberg2019linear}
\begin{align}
    \ba &= \text{coordinate vector of \eqref{representations} relative to } \big\{\! \tfrac{\bn^\bmm}{\bmm!} \! \big\} \nonumber\\
    \bd &= \text{coordinate vector of \eqref{representations} relative to }\{p_\bmm\} \nonumber\\
    % \bC &: \text{matrix representation of the identity map}\nonumber\\
    % &\quad\! \text{between the span of }\{p_\bmm\}\text{ and that of }\big\{\! \tfrac{\bn^\bmm}{\bmm!} \! \big\}\nonumber.
    \bT &= \text{change-of-basis matrix that converts}\nonumber\\
    &\quad\, \{p_\bmm\}\text{-coordinates into }\big\{\! \tfrac{\bn^\bmm}{\bmm!} \! \big\}\text{-coordinates}.\nonumber
\end{align}
%For future use, we introduce the notations $[\cdot]_\bell$ and $[\cdot]_{\bell,\bmm}$ for the $\bell$th component of a vector and the $(\bell,\bmm)$th entry of the matrix.

%\heedong{I think we're run out of ``good-looking'' alphabets. The notations $a_{\bmm}$ and $c_{\bell,\bmm}$ are overwritten (recall \eqref{representations}), and it may confuse readers... Any suggestion?}
%\angel{Right. We should keep $b_\bmm$ for the bases $p_\bmm$ and find new coefficients for the binomial bases, perhaps $d_\bmm$.} \heedong{How about using $\{a_\bmm'\}$ for $\{p_\bmm\}$ and $\{b_\bmm\}$ for ${\bn \choose \bmm}$? (``b'' for ``b''inomial) It's still not very satisfactory... I'll give it some thought.} \angel{Ok, let's see what it looks like this, and then perhaps make final adjustments.}

%\angel{Also, how is this an "identity map"?}
%\heedong{Because both polynomials in \eqref{representations} are identical. So we can think of it as a matrix representation of
%\begin{align*}
%    {\sf id}:\vspan\{p_\bmm\} &\to \vspan\{\tfrac{\bn^\bmm}{\bmm!}\}.\\
%    \sum_\bmm b_\bmm p_\bmm(\bn) &\mapsto \sum_\bmm a_{\bmm}\tfrac{\bn^{\bmm}}{\bmm !}
%\end{align*}
%}

\subsection{Convenient Bases}

We hereafter consider %polynomials satisfying
\begin{align}
    p_\bmm(\bn) = \frac{\bn^\bmm}{\bmm!} + (\text{terms with degree}\in\cM \text{ and}\prec\bmm), \label{goodBasis}
\end{align}
which corresponds to $\bT$ being upper unitriangular, i.e., upper triangular with all-one diagonal entries. 
The usefulness of this form for $\bT$ is to become clear in Sec. \ref{sec:changeOfBasisRevisited}.

A special case of \eqref{goodBasis}
% , which essentially amounts to a change of basis of polynomial space, 
turns out to be analytically more convenient, 
namely the binomial representation
\begin{align}
    p_{\bmm}(\bn) = {\bn\choose\bmm}, \label{binomialBasis}
\end{align}
which can be seen to indeed abide by \eqref{goodBasis} by recalling from Sec.~\ref{sec:notation} that
\begin{align}
    {\bn\choose\bmm} &= {n_0\choose m_0}\cdots{n_{\sfD-1}\choose m_{\sfD-1}}\\
    &=\frac{n_0(n_0-1)\ldots(n_0-m_0+1)}{m_0!}\nonumber\\
    &\quad \cdots\frac{n_{\sfD-1}(n_{\sfD-1}-1)\ldots(n_{\sfD-1}-m_{\sfD-1}+1)}{m_{\sfD-1}!}\\
    &= \bigg(\frac{n_0^{m_0}}{m_0!}+(\text{terms with degree}<m_0)\bigg)\\
    &\quad \cdots \bigg(\frac{n_{\sfD-1}^{m_{\sfD-1}}}{m_{\sfD-1}!}+(\text{terms with degree}<m_{\sfD-1})\bigg)\nonumber\\
    &=\frac{\bn^\bmm}{\bmm!} + (\text{terms with degree}<\bmm), \label{strongerThanGoodBasis}
\end{align}
and by invoking \eqref{containingAllLowerDegrees} and the compatibility of the total order.
In fact, \eqref{binomialBasis} is somewhat stronger than \eqref{goodBasis} in how high the lower-degree terms may be.

In the sequel, we adhere to the binomial representation with coefficients $\{b_\bmm\}$, hence the problem at hand becomes estimating $\{b_\bmm\}$ in
\begin{align}
    x(\bn) =\sum_{\bmm} b_{\bmm} {\bn \choose \bmm}. \label{binomialRepresentation}
\end{align}
from the noisy observation $y(\bn)$. This is equivalent to the original problem of estimating $\{a_{\bmm}\}$ in (\ref{polynomialMonomial}), and the corresponding matrix $\bT$ relates the two solutions.

\section{Difference Operator and Ambiguity}
\label{sec:differenceOperator}

\subsection{Difference Operators and Their Composition}

Let us define some operators that play a central role in the formulation. For 
a general (not necessarily polynomial) signal $s$ defined over $\bbZ^\sfD$,
the shift operator $E_d$ for the $d$th dimension outputs
\begin{align}
    (E_d s)(\bn) \equiv s(\bn+\bee_d).
\end{align}
The difference operator $\Delta_d$ can subsequently be defined as
\begin{align}
    \Delta_d \equiv E_d-1
\end{align}
where $1$ is the identity operator.
For notational compactness, for $\bk\in\bbN_0^\sfD$ let 
\begin{align}
    \Delta^{\bk} \equiv \Delta_0^{k_0}\cdots\Delta_{\sfD-1}^{k_{\sfD-1}} \qquad\quad
    E^{\bk} \equiv E_0^{k_0}\cdots E_{\sfD-1}^{k_{\sfD-1}}
\end{align}
where the superscripts indicate operator compositions, i.e., how many times the operator is applied on each dimension. 

By means of the binomial theorem along with the commutativity of operators, it can be verified that
\begin{align}
    (\Delta^{\bk}s)(\bn)
    & = \big((E-1)^{\bk}s\big)(\bn)\\
    & = \bigg( \! \bigg(\sum_{\bell} (-1)^{|\bk+\bell|}{\bk \choose \bell}E^{\bell}\bigg)s\bigg)(\bn) \label{LY} \\
    & = \sum_{\bell} (-1)^{|\bk+\bell|}{\bk \choose \bell} s(\bn+\bell). \label{binomialExpansion}
\end{align}
% As the signals we are considering are constrained over $[\bN]$, it is instrumental to apply the identical operator to $\tilde{s}(\bn) \equiv \big[\bn \in [\bN] \big] s(\bn)$. As per \eqref{binomialExpansion}, the resultant signal equals
% \begin{align}
%     \sum_{\bell} (-1)^{|\bmm+\bell|}{\bmm \choose \bell} \big[\bn+\bell \in [\bN] \big] s(\bn+\bell),
% \end{align}
% which coincides with $(\Delta^{\bmm}s)(\bn)$ over $[\bN-\bmm']$.
% where we have implicitly assumed that $\bmm' < \bN$ as $\Delta^{\bmm'} s$ is defined over $[\bN-\bmm']$.
% where we have implicitly assumed that $N_d > m_d'$ for all $d$ as $\Delta^{\bmm'} s$ is defined over $[N_0-m_0']\times\ldots\times[N_{\sfD-1}-m_{\sfD-1}']$.

For polynomial signals in the binomial representation, the difference operator behaves very conveniently.
The distributive law gives
\begin{align}
    \Delta^{\bk}{\bn \choose \bmm} &= \bigg(\!\Delta_0^{k_0}{n_0 \choose m_0}\!\bigg)\cdots \bigg(\!\Delta_{\sfD-1}^{k_{\sfD-1}}{n_{\sfD-1} \choose m_{\sfD-1}}\!\bigg)\\
    &={\bn \choose \bmm-\bk}, \label{pascalsRule}
\end{align}
where the last step follows from Pascal's rule \cite[Table 174]{graham1994concrete}
\begin{align}
    \Delta_d {n_d \choose m_d} = {n_d+1 \choose m_d} - {n_d \choose m_d}={n_d \choose m_d-1}.
\end{align}
Applied to the signal of interest in \eqref{binomialRepresentation},
we have that
\begin{align}
    (\Delta^{\bk} x)(\bn)
    =\sum_{\bmm} b_{\bmm} {\bn \choose \bmm-\bk}. \label{polynomialDifference}
\end{align}
% for $\bn\in[\bN-\bmm]$. The set $[\bN-\bmm]$ is not empty from the assumption \eqref{moreSampleThanDegree}.

\subsection{Multidimensional Inversion Formula}

Let us consider $(\Delta^{\bk} x)(\zero)$, whose calculation can unfold in two ways.
From \eqref{polynomialDifference},
it readily follows that
\begin{align}
    (\Delta^{\bk} x)(\zero)
    &= \sum_{\bmm} b_{\bmm} {\zero \choose \bmm-\bk}\\
    &= \sum_{\bmm} b_{\bmm} [\bmm=\bk]\\
    &= b_{\bk}. \label{iversonIntroduction}
\end{align}
Alternatively, applying \eqref{binomialExpansion},
\begin{align}
    (\Delta^{\bk} x)(\zero) = \sum_{\bell} (-1)^{|\bk+\bell|}{\bk \choose \bell} x(\bell)
\end{align}
and therefore
\begin{align}
    b_{\bk} = \sum_{\bell} (-1)^{|\bk+\bell|}{\bk \choose \bell} x(\bell) . \label{binomialTransform}
\end{align}
This result, which is a multidimensional extension of the inversion formula in \cite[Eqn. 5.48]{graham1994concrete}, 
% Recalling the observation made at the end of Sec. \ref{sec:differenceOperator},
% \begin{align}
%     (\Delta^{\bmm'}\tilde{x})(\bn) = (\Delta^{\bmm'}x)(\bn)
% \end{align}
% over $[\bN-\bmm']$, where $\tilde{x}(\bn) = \big[\bn\in[\bN]\big]x(\bn)$. The set
% $[\bN-\bmm']$ is nonempty for $\bmm'\in \cM$ by virtue of the assumption \eqref{moreSampleThanDegree}. This gives
% \begin{align}
%     b_{\bmm'} = \sum_{\bell} (-1)^{|\bmm'+\bell|}{\bmm' \choose \bell} \tilde{x}(\bell) . \label{binomialTransformBox}
% \end{align}
is applied in App. \ref{FCB} to show that \eqref{moreSampleThanDegree} suffices for the linear independence of \eqref{monomialBasis},
or equivalently the linear independence of
\begin{align}
    \bigg\{\big[\bn\in[\bN]\big]{\bn \choose \bmm}: \bmm \in \cM\bigg\}. 
\end{align}
% \angel{I moved the proof to an appendix because, although short, it was a distraction here. But I think the mini-appendix needs some massaging, take a look at it}

\subsection{Integer-Valued Polynomial and Ambiguity}
\label{sec:integerValuedPolynomials}

% \angel{Readers will have forgotten about $a_{\bmm}$ by now, in this section you've been working with $b_{\bmm}$}
Even in the absence of noise, $b_{\bmm}$ cannot be uniquely determined due to the phase wrapping. Take the simplest possible example, for $\cM=\{0\}\subset \bbN_0^{1}$, such that $e^{j 2\pi x(\bn)} = e^{j 2\pi b_0}$. Since $e^{j 2\pi b_0} = e^{j 2\pi (b_0+1)}$, only the fractional part of $b_0$ can be estimated.
Such ambiguity exists more generally.
Let the polynomials 
\begin{align}
    x(\bn) = \sum_{\bmm} b_{\bmm} {\bn \choose \bmm} \qquad x'(\bn) = \sum_{\bmm} b_{\bmm}' {\bn \choose \bmm}.
\end{align}
be ambiguous, i.e., such that
\begin{align}
   e^{j 2\pi x(\bn)} = e^{j 2\pi x'(\bn)}
\end{align}
for all $\bn \in [\bN]$; put differently, $(x-x')(\bn)$ is integer-valued over $[\bN]$.
As a difference of integers is itself integer, it follows that $(\Delta^{\bmm}(x-x'))(\bn)$ is integer-valued over $[\bN-\bmm] \supset [\one]$ (recall the condition in \eqref{moreSampleThanDegree}).
Thus, invoking \eqref{iversonIntroduction},
\begin{align}
    b_{\bmm}-b'_{\bmm} =  (\Delta^{\bmm}(x-x'))(\zero)
\end{align}
should be integer-valued as well.
% Note that $\Delta^{\bmm} (x-x')$ is defined over
% $[\bN-\bmm]$, which is nonempty for $\bmm\in \cM$ by virtue of \eqref{moreSampleThanDegree}.
%This implies that only the fractional parts of the coefficients can be estimated.
Conversely, if $b_\bmm - b_\bmm'$ is integer for all $\bmm\in \cM$, then
\begin{align}
    (x-x')(\bn) = \sum_{\bmm} (b_{\bmm}-b_{\bmm}') {\bn \choose \bmm}
\end{align}
is integer for all $\bn\in[\bN]$ and the polynomials are ambiguous.

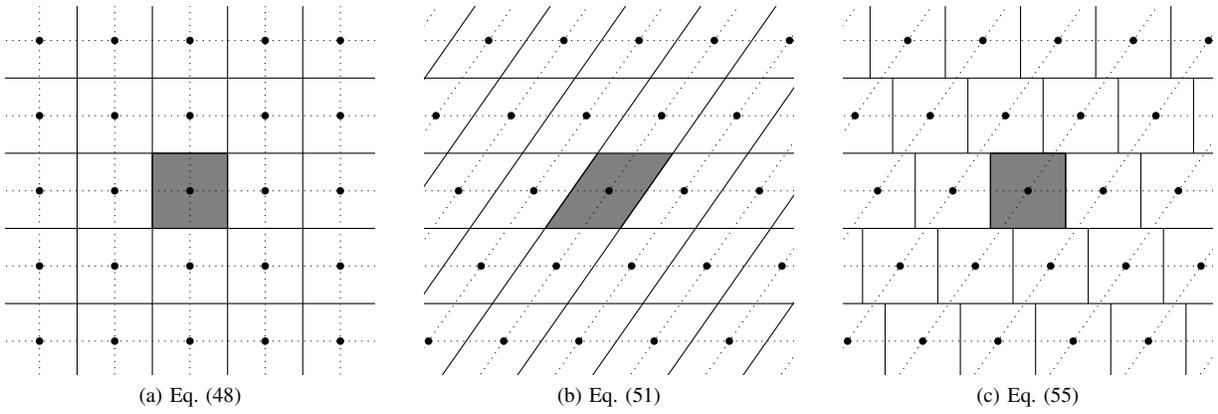
\begin{figure*}
\centering
\subfloat[Eq. \eqref{cubic}]
{
    \begin{tikzpicture}
    \begin{scope}
    \clip (-2.45,-2.45) rectangle (2.45,2.45); 
    \filldraw[fill=gray, draw=black] (-0.5,-0.5) rectangle (0.5,0.5);
    \foreach \x in {-7,-6,...,7}{                           
        \foreach \y in {-7,-6,...,7}{                      
        \fill (\x,\y) circle (0.05cm);
        }
    }
    \foreach \x in {-7,-6,...,7}{                           
        \draw[dotted] (\x,-7) -- (\x,7);
    }
    \foreach \y in {-7,-6,...,7}{                           
        \draw[dotted] (-7,\y) -- (7,\y);
    }
    \foreach \x in {-7,-6,...,7}{                           
        \foreach \y in {-7,-6,...,7}{                      
        \draw (\x-0.5,\y-0.5) -- (\x-0.5,\y+0.5);
        }
    }
    \foreach \y in {-7,-6,...,7}{                           
        \draw (-7,\y-0.5) -- (7,\y-0.5);
    }
    \end{scope}
    \end{tikzpicture}
}\hspace{3mm}
\subfloat[Eq. \eqref{slantedParallelotope}]
{
    \begin{tikzpicture}
    \begin{scope}
    \clip (-2.45,-2.45) rectangle (2.45,2.45); 
    \filldraw[fill=gray, draw=black] (0.5+0.7*0.5,0.5)--(0.5-0.7*0.5,-0.5)--(-0.5-0.7*0.5,-0.5)--(-0.5+0.7*0.5, 0.5);
    \foreach \x in {-7,-6,...,7}{                           
        \foreach \y in {-7,-6,...,7}{                      
        \fill (\x+0.7*\y,\y) circle (0.05cm);
        }
    }
    \foreach \x in {-7,-6,...,7}{                           
        \draw[dotted] (\x-0.7*7,-7) -- (\x+0.7*7,7);
    }
    \foreach \y in {-7,-6,...,7}{                           
        \draw[dotted] (-7,\y) -- (7,\y);
    }
    \foreach \x in {-7,-6,...,7}{                           
        \draw (\x-0.7*7-0.5,-7) -- (\x+0.7*7-0.5,7);
    }
    \foreach \y in {-7,-6,...,7}{                           
        \draw (-7,\y-0.5) -- (7,\y-0.5);
    }
    \end{scope}
    \end{tikzpicture}
}\hspace{3mm}
\subfloat[Eq. \eqref{slantedCubic}]
{
    \begin{tikzpicture}
    \begin{scope}
    \clip (-2.45,-2.45) rectangle (2.45,2.45); 
    \filldraw[fill=gray, draw=black] (-0.5,-0.5) rectangle (0.5,0.5);
    \foreach \x in {-7,-6,...,7}{                           
        \foreach \y in {-7,-6,...,7}{                      
        \fill (\x+0.7*\y,\y) circle (0.05cm);
        }
    }
    \foreach \x in {-7,-6,...,7}{                           
        \draw[dotted] (\x-0.7*7,-7) -- (\x+0.7*7,7);
    }
    \foreach \y in {-7,-6,...,7}{                           
        \draw[dotted] (-7,\y) -- (7,\y);
    }
    \foreach \x in {-7,-6,...,7}{                           
        \foreach \y in {-7,-6,...,7}{                      
        \draw (\x+0.7*\y-0.5,\y-0.5) -- (\x+0.7*\y-0.5,\y+0.5);
        }
    }
    \foreach \y in {-7,-6,...,7}{                           
        \draw (-7,\y-0.5) -- (7,\y-0.5);
    }
    \end{scope}
    \end{tikzpicture}
}
\caption{Two-dimensional lattices and cells for $\bT = \begin{bmatrix}
    1 & 0.7\\ 0 & 1\end{bmatrix}$.}
\label{fig:tessellation}
\end{figure*}

To uniquely determine the fractional parts of the coefficients, and given that
\begin{align}
    \big\{\bz  + \left[ \textstyle -\frac{1}{2},\frac{1}{2}\right)^{|\cM|}: \bz \in \bbZ^{|\cM|} \big\} \label{cubic}
    % \big\{\bb  + [0,1)^{|\cM|}: \bb \in \bbZ^{|\cM|} \big\}
\end{align}
tessellates $\bbR^{|\cM|}$, an additional constraint such as
\begin{align}
    b_\bmm \in \left[ \textstyle -\frac{1}{2},\frac{1}{2}\right) \label{bUnique}
\end{align}
for all $\bmm$, or equivalently $\bb \in \big[-\frac{1}{2},\frac{1}{2}\big)^{|\cM|}$ when vectorized, is needed.
This is the one considered hereafter and illustrated in Fig. \ref{fig:tessellation}a, although the restriction to other unit intervals would work just as well.
And, as an alternative to this cubic cell, a Voronoi cell would also work just as well, but
the computation of the closest lattice point is expensive \cite{agrell2002closest};
in the sequel, we therefore adhere to the convenient cubic cell.

% \heedong{Note to self: Add some text related to branch consideration here...}
% Put another way, the function
% \begin{align}
%     \bb \mapsto \big[\bn\in[\bN]\big]\exp\!\bigg(j2\pi \sum_{\bmm} b_{\bmm} {\bn \choose \bmm}\bigg) \nonumber
% \end{align}
% is not injective, and we must restrict its domain to construct the inverse.

%We hereafter assume the condition \eqref{bUnique}.
The foregoing characterization of the ambiguity, i.e., that $b_\bmm - b_\bmm'$ is integer for all $\bmm$, generalizes \cite[Thm. 1]{mckilliam2009identifiability} in two ways: the signal is multidimensional, and the observation window is finite.

% \begin{figure}
% \centering
% \begin{tikzpicture}
% \begin{scope}
% \clip (-2.5,-2.5) rectangle (2.5,2.5); 
% \filldraw[fill=gray, draw=black] (-0.5,-0.5) rectangle (0.5,0.5);
% \foreach \x in {-7,-6,...,7}{                           
%     \foreach \y in {-7,-6,...,7}{                      
%     \fill (\x+0.7*\y,\y) circle (0.05cm);
%     }
% }
% \foreach \x in {-7,-6,...,7}{                           
%     \draw[dotted] (\x-0.7*7,-7) -- (\x+0.7*7,7);
% }
% \foreach \y in {-7,-6,...,7}{                           
%     \draw[dotted] (-7,\y) -- (7,\y);
% }
% \foreach \x in {-7,-6,...,7}{                           
%     \foreach \y in {-7,-6,...,7}{                      
%     \draw (\x+0.7*\y-0.5,\y-0.5) -- (\x+0.7*\y-0.5,\y+0.5);
%     }
% }
% \foreach \y in {-7,-6,...,7}{                           
%     \draw (-7,\y-0.5) -- (7,\y-0.5);
% }
% \end{scope}
% \end{tikzpicture}
% \caption{Illustration of \eqref{rectangle} for $\bT' = \begin{bmatrix}
%     1 & 0.7\\ 0 & 1\end{bmatrix}$.}
% \label{fig:tessellation}
% \end{figure}

% \begin{algorithm}
% \caption{Lattice Point Finding}\label{algo:computeLatticePoint}
% \KwData{$\{a_\bmm\}$}
% \KwResult{$\{z_\bmm\}$}
% \While{$\cM = \emptyset$}{
%     Pick the largest element $\bell\in \cM$\\
%     $z_\bell \gets \lfloor a_\bell-\sum_{\bmm\succ \bell} c_{\ell,\bmm}z_\bmm \rceil$\\
%     $\cM \gets \cM \setminus \{\bell\}$
% }
% \end{algorithm}

\subsection{Change of Basis Revisited}
\label{sec:changeOfBasisRevisited}

%\heedong{Please take a look into the paragraph below and Fig. \ref{fig:tessellation}. It's still unsatisfactory in that the motivation of cubic cell (over parallelepiped one) is not explained. The motivation in my mind is application-specific, so I find it hard to explain in a general estimation context. 
%Let us consider a ULA-spawned MIMO channel. Then the polynomial can be written as
%\begin{align*}
%    x(\bn) = a_{0,0} + a_{1,0}n_0 + a_{0,1}n_1.
%\end{align*}
%From the physical constraints, we have
%\begin{align*}
%    |a_{0,1}|, |a_{1,0}| \leq \frac{\text{inter-antenna spacing}}{\text{wavelength}},
%\end{align*}
%which corresponds to the cubic cell. Note that the condition is somewhat different for UPA-single antenna case:
%\begin{align*}
%    a_{0,1}^2 + a_{1,0}^2 \leq \frac{\text{inter-antenna spacing}}{\text{wavelength}}.
%\end{align*}
%(But anyway the cubic cell would work as long as $\frac{\text{inter-antenna spacing}}{\text{wavelength}} < \frac{1}{2}$.)
%}

The previous subsection has evidenced that,
in the binomial representation,
%$\bb$ can be uniquely determined up to the cubic lattice $\bbZ^{|\cM|}$ and thus
the inherent ambiguity can be resolved by confining $\bb$ to a cubic cell. 
When mapping solutions back to the original monomial basis, %recalling the change-of-basis formula,
the lattice undergoes a linear transformation and becomes slanted as illustrated in Fig. \ref{fig:tessellation}b; however, as shown next, the cubic cell can still be employed in lieu of the parallelepiped one in the figure.
%\angel{What happens if \eqref{goodBasis} is not satisfied? (given that you rely on the unitriangularity).} 
%\heedong{As long as $\bT'$ relating two representations is invertible, we can extend the result. Let $\bT' = \bQ\bR$ be the QR decomposition and assume that $\bT'$ is invertible, i.e., the diagonal entries of $\bR$ is nonzero. For brevity, let us write $\bR = \bD\tilde{\bR}$ with a diagonal matrix $\bD$ to make $\tilde{\bR}$ unitriangular. Then,
%\begin{align*}
%    &\big\{\tilde{\bR}\bz  + \left[ \textstyle -\frac{1}{2},\frac{1}{2}\right)^{|\cM|}: \bz \in \bbZ^{|\cM|} \big\} \text{ tessellates } \bbR^{|\cM|}\\
%    &\Rightarrow \big\{\bT\bz  + \bQ\bD\left[ \textstyle -\frac{1}{2},\frac{1}{2}\right)^{|\cM|}: \bz \in \bbZ^{|\cM|} \big\} \text{ tessellates } \bbR^{|\cM|},\\
%\end{align*}
%where we make use of the unitriangularity of $\tilde{\bR}$ in the first step. The cell is now a (rotated) rectangular cuboid.}

\begin{algorithm}[t]
\caption{Computation of the lattice point $\bz$ satisfying $\ba - \bT\bz\in [- \frac{1}{2},\frac{1}{2})^{|\cM|}$ for an upper unitriangular matrix $\bT\in\bbR^{|\cM|\times |\cM|}$}\label{algo:computeLatticePoint}
\begin{algorithmic}
\Procedure{Compute-Lattice-Point}{$\ba,\bT$}
    \For{$\bell\in\cM$ (in descending order)}
        \State $[\bz]_\bell \gets \big\lfloor [\ba]_\bell-\sum_{\bmm \succ \bell} [\bT]_{\bell,\bmm}[\bz]_{\bmm} \big\rceil$
    \EndFor
\State \Return $\bz$
\EndProcedure
\end{algorithmic}
\end{algorithm}

Let us consider, once more,
\begin{align}
    x(\bn) = \sum_{\bmm} a_{\bmm}\frac{\bn^{\bmm}}{\bmm !} = \sum_\bmm b_\bmm{\bn \choose \bmm}. \label{newBasis}
\end{align}
From $\ba=\bT\bb$, the integer ambiguity translates to a lattice ambiguity. Precisely, $\ba$ and $\ba - \bT\bz$ are ambiguous for any $\bz\in\bbZ^{|\cM|}$ and
\begin{align}
    \big\{\bT\bz  + \bT\left[ \textstyle -\frac{1}{2},\frac{1}{2}\right)^{|\cM|}: \bz \in \bbZ^{|\cM|} \big\} \label{slantedParallelotope}
    % \big\{\bb  + [0,1)^{|\cM|}: \bb \in \bbZ^{|\cM|} \big\}
\end{align}
tessellates $\bbR^{|\cM|}$ (see again Fig. \ref{fig:tessellation}b).
An immediate observation is that $\bz\in\bbZ^{|\cM|}$ satisfying
\begin{equation}
\ba - \bT\bz\in \left[\textstyle - \frac{1}{2},\frac{1}{2}\right)^{|\cM|}
\label{Valencia}
\end{equation}
uniquely exists. And, since $\bT$ is upper unitriangular,
\begin{align}
    [\ba-\bT\bz]_\bell = a_\bell-\sum_{\bmm\succ \bell} t_{\bell,\bmm}z_\bmm - z_\bell.
\end{align}
A recursive relationship emerges from which the entries of $\bz$ satisfying \eqref{Valencia} can be derived, namely
\begin{align}
    z_\bell = \bigg\lfloor a_\bell-\sum_{\bmm\succ \bell} t_{\bell,\bmm}z_\bmm \bigg\rceil,
\end{align}
with $\lfloor \cdot \rceil\equiv \big\lfloor \cdot +\frac{1}{2} \big\rfloor$ indicating rounding. The procedure can be summarized into Alg. \ref{algo:computeLatticePoint}, also available in \cite[Alg. 2.2]{mckilliam2010lattice}.

The existence and uniqueness of $\bz\in\bbZ^{|\cM|}$ satisfying \eqref{Valencia} indicates that 
\begin{align}
    \big\{\bT\bz  + \left[ \textstyle -\frac{1}{2},\frac{1}{2}\right)^{|\cM|}: \bz \in \bbZ^{|\cM|} \big\} \label{slantedCubic}
\end{align}
tessellates $\bbR^{|\cM|}$ \cite[Prop. 2.1]{mckilliam2010lattice} (see Fig. \ref{fig:tessellation}c). 
Accordingly,
\begin{align}
    a_\bmm \in \left[ \textstyle -\frac{1}{2},\frac{1}{2}\right) \label{aUnique}
\end{align}
for all $\bmm$ eliminates the ambiguities. For given $\bb$, one can straightforwardly find $\ba\in [-\frac{1}{2},\frac{1}{2} )^{|\cM|}$ satisfying
\begin{align}
   \exp\!\bigg[j2\pi\bigg(\sum_{\bmm} a_{\bmm} \frac{\bn^\bmm}{\bmm!}\bigg)\bigg] = \exp\!\bigg[j2\pi\bigg(\sum_{\bmm} b_{\bmm} {\bn \choose \bmm}\bigg)\bigg],
\end{align}
or equivalently $\ba-\bT\bb\in\bT\bbZ^{|\cM|}$, through Alg. \ref{algo:TransformBtoA}.

The result naturally generalizes to any other basis satisfying \eqref{goodBasis}. The change-of-basis matrix that turns $\{\!\textstyle{\bn \choose \bmm}\!\} $-coordinates into $\{p_\bmm\}$-coordinates can be seen as a product of
\begin{enumerate}
    \item the matrix that converts $\{\!\textstyle{\bn \choose \bmm}\!\}$-coordinates into $\{\tfrac{\bn^\bmm}{\bmm!}\}$-coordinates; and
    \item the one that converts $\{\tfrac{\bn^\bmm}{\bmm!}\}$-coordinates into $\{p_\bmm\}$-coordinates; this is the inverse of the matrix that converts $\{p_\bmm\}$-coordinates into $\{\tfrac{\bn^\bmm}{\bmm!}\}$-coordinates.
\end{enumerate}
The resultant change-of-basis matrix is upper unitriangular as the set of upper unitriangular matrices is closed under matrix multiplication and inversion. Repeating the argument, $\bd \in [-\frac{1}{2},\frac{1}{2} )^{|\cM|}$ eliminates the ambiguities.

\section{Polynomial Phase Estimation Via Phase Difference Operator}
\label{sec:polynomialPhaseEstimation}

Equipped with the mathematical machinery set forth in the previous sections, an algorithm is next presented that generalizes the one-dimensional estimator in \cite{kitchen1994method} to the multidimensional realm.

\begin{algorithm}[t]
\caption{Computation of $\ba$ satisfying $\ba-\bT\bb \in \bbZ^{|\cM|}$ and $\ba\in[- \frac{1}{2},\frac{1}{2})^{|\cM|}$ for an upper unitriangular matrix $\bT\in\bbR^{|\cM|\times |\cM|}$}\label{algo:TransformBtoA}
\begin{algorithmic}
\Procedure{Compute-New-Coordinate}{$\bb,\bT$}
    \State $\bz \gets \textsc{Compute-Lattice-Point}(\bT\bb,\bT)$
    \State $\ba \gets \bT(\bb-\bz)$
\State \Return $\ba$
\EndProcedure
\end{algorithmic}
\end{algorithm}

% Equipped with the mathematical machinery set forth in the previous section, an algorithm is next presented to estimate polynomial phase coefficients
% from noisy observations, i.e., from
% \begin{align}
%     y(\bn) = e^{j 2\pi x(\bn)}  + w_{\bbC}(\bn).
% \end{align}
% This algorithm generalizes to the multidimensional realm the one-dimensional estimator devised in \cite{kitchen1994method}.

\subsection{Phase Difference Operators and Their Composition}

For subsequent use, let us introduce the so-called phase difference operator \cite{francos1998two}
% \footnote{
% The phase difference operator presented in \cite{peleg1995discrete} (the term seems to be coined in \cite{francos1998two}) is slightly different, namely
% \begin{align}
%     (\cD_d' y)(\bn) = (E_d^{\tau_d} y)(\bn)\overline{(1_d^{\tau_d} y)(\bn)} \nonumber
% \end{align}
% containing an additional parameter $\tau_d$. All the results to be established can be analogously derived, yet in this paper we let $\tau_d = 1$ since $\tau_d > 1$ incurs a loss in identifiability \cite{mckilliam2009identifiability}. That is, even without noise, it can uniquely determine the parameters only if 
% \begin{align}
%     b_\bmm \in {\textstyle\frac{1}{\btau^\bmm}}\left[ \textstyle -\frac{1}{2},\frac{1}{2}\right) \quad \nonumber
% \end{align}
% for all $\bmm$.
% Comparing it with the result in Sec.~\ref{sec:integerValuedPolynomials}, there is an additional multiplicative factor, which makes the algorithm much more restrictive.
% }
\begin{align}
    (\cD_d s)(\bn) &\equiv (E_d s)(\bn)\overline{s(\bn)} \label{phaseDifferenceOperator}
\end{align}
where $\overline{\,\cdot\,}$ denotes complex conjugation. 
Similar to the difference operator, we introduce a compact notation for the composition of phase difference operators, namely
\begin{align}
    \cD^{\bmm} \equiv \cD_0^{m_0}\cdots\cD_{\sfD-1}^{m_{\sfD-1}}.
\end{align}
The above satisfies
\begin{align}
    \cD^\bmm\big(s_1(\bn)s_2(\bn)\big) = (\cD^\bmm s_1)(\bn)(\cD^\bmm s_2)(\bn), \label{cauchyLike}
\end{align}
which can be proved using induction on $|\bmm|$; the one-dimensional version of this result can be found in \cite[Property 1]{peleg1995discrete}. Another useful property is
\begin{align}
    &\cD^{\bmm}\Big[\big[\bn\in[\bN]\big] \exp(j2\pi s(\bn))\Big]\nonumber\\
    & \qquad\qquad = \big[\bn\in[\bN-\bmm]\big]\exp\!\big(j2\pi (\Delta^{\bmm}s)(\bn)\big), \label{operatorRelation}
\end{align}
which relates the phase difference operator with the difference operator.

\begin{figure*}
    \centering
    \subfloat[$\SNR = -10\text{ dB}$]
    {
        \includegraphics[width=0.24\linewidth]{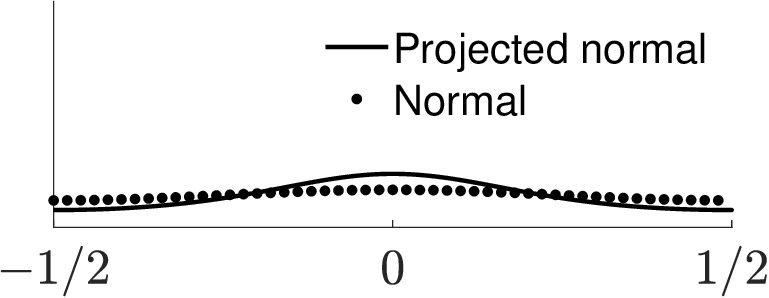}
    }
    \subfloat[$\SNR = -5\text{ dB}$]
    {
        \includegraphics[width=0.24\linewidth]{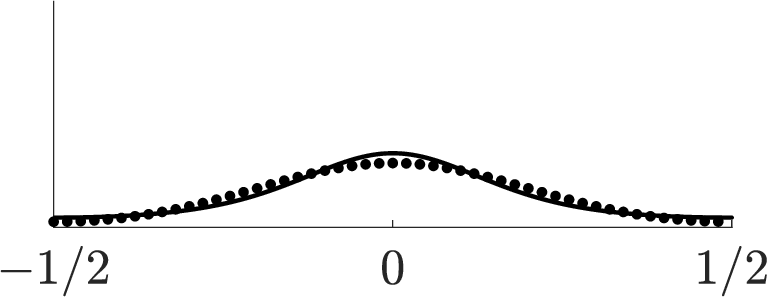}
    }
    \subfloat[$\SNR = 0\text{ dB}$]
    {
        \includegraphics[width=0.24\linewidth]{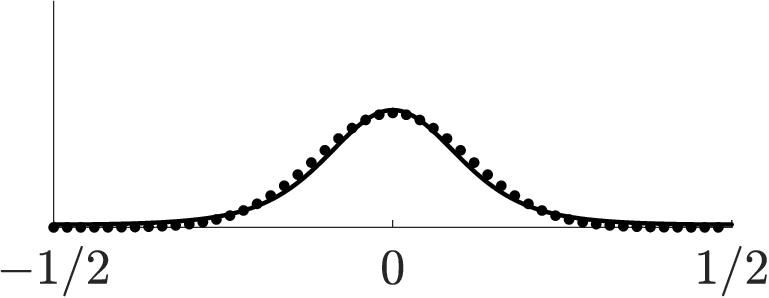}
    }
    \subfloat[$\SNR = 5\text{ dB}$]
    {
        \includegraphics[width=0.24\linewidth]{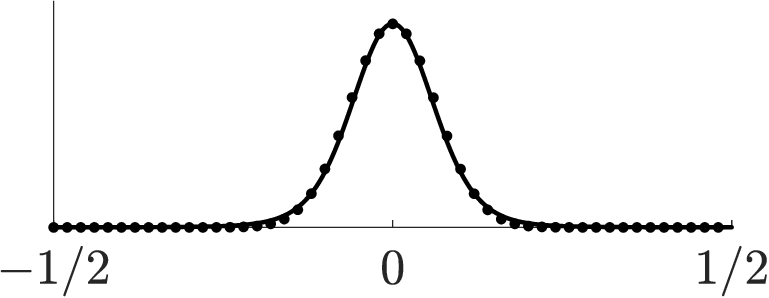}
    }
    \caption{Accuracy of the high-SNR approximation: contrast between the PDFs of $\frac{1}{2\pi}\arg \! \big(1 + \tilde{w}_{\bbC}(\bn)\big)$ and $w(\bn)$.}
    \label{fig:highSnrApproximation}
\end{figure*}

\subsection{The High-SNR Regime}
%\angel{GOOD EFFORT TO JUSTIFY THE HIGH-SNR MODEL WITHOUT MATH, WHICH IS ANYWAY AVAILABLE IN THE REFERENCE. THE ONLY STEP I MISSED IS THE JUSTIFICATION OF WHY IT'S THE NOISE QUADRATURE COMPONENT THAT MATTERS. THAT'S CLEAR WHEN ONE GOES OVER THE MATH, BUT NOT SO EASY TO EXPLAIN IN PLAIN ENGLISH!}
%\heedong{A bit tweaked.}

For the sake of analysis, the high-SNR approximation propounded in \cite[Sec.~II]{tretter1985estimating} is most useful. The signal model in \eqref{signalModel} can be rewritten as
\begin{align}
    % y(\bn) = \big[\bn\in[\bN]\big] e^{j2\pi x(\bn)}\big(1  + e^{-j2\pi x(\bn)}w_{\bbC}(\bn)\big). \label{signalModelMultiplicative}
    y(\bn) = \big[\bn\in[\bN]\big] e^{j2\pi x(\bn)}\big(1  + \tilde{w}_{\bbC}(\bn)\big), \label{signalModelMultiplicative}
\end{align}
where $\tilde{w}_{\bbC}(\bn) \equiv e^{-j2\pi x(\bn)}w_{\bbC}(\bn) \overset{\mathrm{iid}}{\sim} \cN_{\bbC}(0,\frac{1}{\SNR})$.
Letting 
\begin{align}
    (\Pi s)(\bn) \equiv 
    % \equiv \exp(j\arg(s(\bn)))
    % \equiv \frac{s(\bn)}{|s(\bn)|}
    \begin{cases}
        \exp(j\arg(s(\bn))) & \quad s(\bn) \neq 0\\
        % \frac{s(\bn)}{|s(\bn)|} & \qquad s(\bn) \neq 0\\
        0 & \quad \text{otherwise}
    \end{cases}    
\end{align}
be a projection onto the unit circle, which is where $y(\bn)$ would live without noise,
with probability 1 it holds that
\begin{align}
    (\Pi y)(\bn) &= \big[\bn\in[\bN]\big] e^{j2\pi x(\bn)}\Pi\big(1  + \tilde{w}_{\bbC}(\bn)\big) \label{unitCircleProjection}\\
    &=\big[\bn\in[\bN]\big] \exp\!\bigg[j2\pi \bigg(\!x(\bn)\nonumber\\
    &\qquad\qquad+\frac{1}{2\pi}\arg \! \big(1 + \tilde{w}_{\bbC}(\bn)\big)\bigg)\!\bigg]
\end{align}
where $\arg(\cdot)\in[-\pi,\pi)$ returns the argument of a complex number, and with the convention $\arg(0)=0$.
By definition, $\frac{1}{2\pi}\arg \! \big(1 + \tilde{w}_{\bbC}(\bn)\big)$
% \begin{align}
%     \arg\Big(\Pi\big(1 + e^{-j2\pi x(\bn)}\tilde{w}_{\bbC}(\bn)\big)\Big)    
% \end{align}
follows an i.i.d. projected normal distribution whose closed-form probability density function (PDF) is furnished in \cite[Ch. 3.5.6]{mardia2009directional}. 
At high SNR, this argument can be approximated by another normal distribution \cite[Lemma 4.5]{jung2021geodesic}. Precisely, this approximation, whose accuracy is illustrated in
Fig. \ref{fig:highSnrApproximation},
results in
\begin{align}
    (\Pi y)(\bn) & \approx \big[\bn\in[\bN]\big]e^{  j2\pi ( x(\bn)+w(\bn)) }
    \label{highSnrApproximation}
\end{align}
with $w(\bn) \overset{\mathrm{iid}}{\sim} \cN(0,\frac{1}{8\pi^2\SNR})$. This approximation 
can also be obtained via the concatenated Taylor expansions
\begin{align}
    \Pi\big(1 + \tilde{w}_{\bbC}(\bn)\big)
    &= 1+ j\Im \! \big\{\tilde{w}_{\bbC}(\bn)\big\} + o(|\tilde{w}_\bbC(\bn)|)\\
    &= \exp \! \big(j\underbrace{\Im \! \big\{\tilde{w}_{\bbC}(\bn)\big\}}_{\overset{\mathrm{iid}}{\sim} \cN(0,\frac{1}{2\SNR})} \! \big) +  o(|\tilde{w}_\bbC(\bn)|),
\end{align}
which reveal that, at high SNR, 
%can be understood through the following handwavy argument \cite{tretter1985estimating}. At high SNR,
the projection onto the unit circle can be approximated by a projection on the line \cite{tretter1985estimating}.
%Both approximations follow from first-order Taylor expansions.

Every symbol $\approx$ in the sequel indicates the high-SNR approximation embodied by (\ref{highSnrApproximation}), whereby the additive noise in (\ref{signalModel}) has been converted into phase noise. Readers are referred to \cite{baggenstoss1991estimating, quinn2010phase, fu2013phase} for further discussion on the impact of the discarded amplitude information.

\subsection{Estimation of the Highest-Order Coefficient}
\label{sec:estimationOfHighestOrderCoefficient}

%Let us consider high SNR scenario %where \eqref{highSnrApproximation} holds.
Let $\bk$ be the largest element of $\cM$ (with respect to any total order compatible with the partial order defined in Sec.~\ref{sec:order}) and consider the estimation of $b_{\bk}$ at high SNR.
Applying \eqref{operatorRelation} to \eqref{highSnrApproximation} gives 
\begin{align}
    (\cD^{\bk} \Pi y)(\bn) & \approx \big[\bn\in[\bN-\bk]\big] \label{dptComposition} \\
    &\quad \cdot\exp \! \Big[j2\pi \big( (\Delta^{\bk} x)(\bn) +(\Delta^{\bk} w)(\bn)\big)\Big] \nonumber
\end{align}
Plugging \eqref{polynomialDifference} into \eqref{dptComposition},
\begin{align}
    &(\cD^{\bk} \Pi y)(\bn)  \approx\big[\bn\in[\bN-\bk]\big] \label{dptCompositionResult}\\
    & \qquad \cdot \exp \!\bigg[ j2\pi \bigg(  \sum_{\bmm} b_{\bmm} {\bn \choose \bmm-\bk} + (\Delta^{\bk} w)(\bn) \bigg)\bigg] \nonumber\\
     & \quad= \big[\bn\in[\bN-\bk]\big]\exp\!\big[ j2\pi ( b_{\bk} + (\Delta^{\bk}w)(\bn)) \big], \label{extractingHighestOrder}
\end{align}
% where $w_{\bmm'}\equiv \Delta^{\bmm'}w$.
where, in light of \eqref{EOS}, the last step follows from $\bk$ being the largest element in $\cM$.

% \begin{align}
%     (\cD_0^{m_0'}\ldots\cD_D^{m_{\sfD-1}'} y)(\bn) \approx \exp(j (2\pi b_{\bmm'} + w_{\bmm'})) \label{extractingHighestOrder}
% \end{align}
% where
% \begin{align}
%     w_{\bmm'}(\bn) \equiv (\Delta_0^{m_0'}\ldots\Delta_{\sfD-1}^{m_{\sfD-1}'}w)(\bn).
% \end{align}
% The last step \angel{WE SHOULD BE MORE PRECISE} follows from 
% \begin{align}
%     [\forall d\quad m_d \geq m_d'] =[\bmm=\bmm'] ,
% \end{align}
% which can be readily proved from \eqref{orderCondition} and from $\bmm'$ being the largest element:
% \begin{align}
%     \forall d\quad m_d \geq m_d' &\Rightarrow \bmm\geq \bmm'\\
%     &\Rightarrow \bmm=\bmm' ,
% \end{align}
% which also trivially holds in the opposite direction.
% Note that $\cD_0^{m_0'}\ldots\cD_D^{m_{\sfD-1}'} y$ is defined over $[N_0-m_0']\times\ldots\times[N_{\sfD-1}-m_{\sfD-1}']$.

% \begin{figure*}
% \begin{align}
%     (\cD_0^{m_0'}\ldots\cD_D^{m_{\sfD-1}'} y)(\bn)
%     & \approx \exp\big(j \big(2\pi (\Delta_0^{m_0'}\ldots\Delta_{\sfD-1}^{m_{\sfD-1}'} x)(\bn) +(\Delta_0^{m_0'}\ldots\Delta_{\sfD-1}^{m_{\sfD-1}'} w)(\bn)\big)\big)\\
%     & = \exp \!\bigg(j \bigg(2\pi  \sum_{\bmm\in M}\! b_{\bmm} {n_0 \choose m_0-m_0'}\ldots {n_{\sfD-1} \choose m_{\sfD-1}-m_{\sfD-1}'} + w_{\bmm'}(\bn) \bigg)\bigg)\\
%     & = \exp \!\bigg(j \bigg(2\pi  \sum_{\bmm\in M}\! b_{\bmm} {n_0 \choose m_0-m_0'}\ldots {n_{\sfD-1} \choose m_{\sfD-1}-m_{\sfD-1}'}[m_0\geq m_0']\ldots [m_{\sfD-1}\geq m_{\sfD-1}'] + w_{\bmm'}(\bn) \bigg)\bigg)\\
%     & = \exp(j (2\pi b_{\bmm'} + w_{\bmm'}(\bn))).
% \end{align}
% \hrulefill
% \end{figure*}

At high SNR, \eqref{extractingHighestOrder} does not undergo phase wrappings, hence it is tantamount to
\begin{align}
    &\frac{1}{2\pi}\arg \! \big( (\cD^{\bk} y)(\bn) \big) \label{highSnrNoWrapping}\\
    &\qquad \approx  \big[\bn\in[\bN-\bk]\big] \big( b_{\bk} + (\Delta^{\bk}w)(\bn)\big). \nonumber
\end{align}
As per \eqref{binomialExpansion}, $(\Delta^{\bk}w)(\bn)$ is colored, but Gaussian; precisely,
\begin{align}
    (\Delta^{\bk}w)(\bn)
    &=\sum_{\bell} (-1)^{|\bk+\bell|}{\bk \choose \bell} w(\bn+\bell) \\
    &=\sum_{\bell} (-1)^{|\bk+\bell+\bn|}{\bk \choose \bell-\bn} w(\bell). \label{coloredNoise}
\end{align}

The minimum-variance unbiased (and also maximum-likelihood) estimate of $b_{\bk}$ can thus be obtained via noise whitening followed by least squares \cite[Thm. 7.5]{kay1993fundamentals}, namely 
% With a lexicographic ordering on $[\bN-\bmm']$, we can vectorize \eqref{highSnrNoWrapping} as
% \begin{align}
%     \arg(\by_{\bmm'}) \approx 2\pi b_{\bmm'} \one +\bw_{\bmm'}.
% \end{align}
\begin{align}
    \hat{b}_{\bk} = \frac{1}{2\pi}\sum_\bn u_{\bk}(\bn)\arg \! \big((\cD^{\bk} y)(\bn)\big) . \label{mvue}
\end{align}
Here,
\begin{align}
    u_{\bk}(\bn) = \begin{cases}
        \bigg[\frac{\bC_{\bk}^{-1}\one}{\one^\top \bC_{\bk}^{-1}\one}\bigg]_\bn & \quad \bn\in[\bN-\bk]\\
        0 & \quad \text{otherwise}
    \end{cases}
    \label{AmuntValencia}
\end{align}
where $\bC_{\bk}\in \bbR^{|[\bN-\bk]|\times|[\bN-\bk]|}$ is the covariance matrix of $\big[\bn\in[\bN-\bk]\big](\Delta^{\bk}w)(\bn)$, with entries
\begin{align}
    [\bC_{\bk}]_{\bn,\bn'}
    % &=(-1)^{|\bn+\bn'|}  \sum_{\bk\in[\bN]}  {\bmm' \choose \bk-\bn} {\bmm' \choose \bk-\bn'}\\
    &=\frac{(-1)^{|\bn+\bn'|}}{8\pi^2\SNR}  \sum_{\bell}  {\bk \choose \bell-\bn} {\bk \choose \bell-\bn'} \\
    &=\frac{(-1)^{|\bn+\bn'|}}{8\pi^2\SNR}  {2\bk \choose \bn -\bn'+\bk}. \label{coloredNoiseCovariance}
\end{align}
% We can decompose it as
% \begin{align}
%     \bC_{\bmm'} = \bC_{m_0'} \otimes \ldots \otimes \bC_{m_{\sfD-1}'}
% \end{align}
% where $\bC_{m_d'}\in \bbR^{(N_d-m_d')\times(N_d-m_d')}$ is a matrix with entries
% \begin{align}
%     [\bC_{m_d'}]_{n,n'}
%     &=(-1)^{n+n'}  \sum_{k\in[N_d]}  {m_d' \choose k-n} {m_d' \choose k-n'}.
% \end{align}
% Altogether, in the estimation of $b_\bk$ in \eqref{mvue} we finally have
By capitalizing on the structure of $\bC_\bk$, App. \ref{app:weightProof} derives a closed form for the weights in \eqref{AmuntValencia}. Free of matrix inversions, this expression is
\begin{align}
    u_{\bk}(\bn) = \big[\bn\in[\bN-\bk]\big]\frac{{\bn+\bk \choose \bk}{\bN-\bn-\one \choose \bk}}{{\bN+\bk\choose 2\bk+\one}}, \label{kayWeight}
\end{align}
which extends existing results for $\cM=\{0,1\}$ \cite{kay1989fast}, $\cM=\{0,1,2\}$ \cite{djuric1990parameter}, $\cM=\{0,1,\ldots,M\}$ \cite{kitchen1994method}, and $\cM=\{(0,0),(0,1),(1,0)\}$ \cite{kay1990efficient}.
For later use, we note that
\begin{align}
    \sum_\bn u_\bk(\bn) &= \sum_{\bn\in[\bN-\bk]} \bigg[\frac{\bC_{\bk}^{-1}\one}{\one^\top \bC_{\bk}^{-1}\one}\bigg]_\bn\\
    &=\frac{\one^\top\bC_{\bk}^{-1}\one}{\one^\top \bC_{\bk}^{-1}\one}\\
    &= 1. \label{weightProperty}
\end{align}

\subsection{Sequential Estimation of Coefficients}

% \angel{HERE $\bmm'$ IS NO LONGER THE HIGHEST ORDER, BUT THE HIGHEST UNESTIMATED ORDER. PERHAPS WORTH MAKING IT EXPLICIT?}
Let us posit that accurate estimates of $b_{\bmm}$ for $\bmm \succ \bk$, denoted by $\{\hat{b}_{\bmm}\}$, are available. Thus, $\bk$ is the highest unestimated degree and the estimated terms can be cancelled from the observation $y(\bn)$ to yield 
\begin{align}
    y_{\bk}(\bn) \equiv y(\bn)\exp \! \left(-j2\pi \!\sum_{\bmm \succ \bk} \hat{b}_{\bmm}{\bn \choose \bmm}\right).
    \label{Lamine}
\end{align}
Using \eqref{polynomialDifference}, \eqref{cauchyLike}, and \eqref{operatorRelation},
\begin{align}
    &(\cD^{\bk} y_{\bk})(\bn)\nonumber\\
    &= (\cD^{\bk} y)(\bn)  \cD^{\bk}\bigg[\exp\!\bigg(-j2\pi \!\sum_{\bmm \succ \bk}\hat{b}_{\bmm}{\bn \choose \bmm}\bigg)\bigg]\\
    &= (\cD^{\bk} y)(\bn)  \exp\!\bigg(\Delta^{\bk}\bigg[ -j2\pi \!\sum_{\bmm \succ \bk}\hat{b}_{\bmm}{\bn \choose \bmm}\bigg]\bigg)\\
    &= (\cD^{\bk} y)(\bn)  \exp\!\bigg(-j2\pi \!\sum_{\bmm \succ \bk}\hat{b}_{\bmm}{\bn \choose \bmm-\bk}\bigg), \label{reuseComputation}
\end{align}
which further simplifies into
\begin{align}
    (\cD^{\bk} y)(\bn)  \exp\!\bigg( \! -j2\pi \!\! \sum_{\bmm > \bk}\hat{b}_{\bmm}{\bn \choose \bmm-\bk}\bigg) \label{independentToTotalOrder}
\end{align}
by virtue of
\begin{align}
    [\bmm \succ \bk] {\bn \choose \bmm-\bk} & = [\bmm \succ \bk] [\bmm \geq \bk] {\bn \choose \bmm-\bk} \nonumber \\
    & = [\bmm \succ \bk] [\bmm > \bk] {\bn \choose \bmm-\bk} \nonumber \\
    & = [\bmm > \bk] {\bn \choose \bmm-\bk}. \label{compatability}
\end{align}
The first identity above follows from the definition of binomial coefficient and the last identity holds because the total order is compatible with the partial order; this implies that $(\cD^{\bk} y_{\bk})(\bn)$ does not depend on the choice of total order.

The next task is to estimate $b_{\bk}$. Applying the high-SNR approximation in \eqref{dptCompositionResult} to \eqref{independentToTotalOrder}, what emerges is \eqref{sequentialEstimationEnd} atop the next page.
\begin{figure*}
\begin{align}
    (\Pi\cD^{\bk}y_{\bk})(\bn)
    & \approx \big[\bn\in[\bN-\bk]\big] \exp \! \bigg[j2\pi \bigg( b_{\bk} + \sum_{\bmm > \bk} (b_{\bmm}-\hat{b}_{\bmm}) {\bn \choose \bmm-\bk} + \sum_{\bell} (-1)^{|\bk+\bell+\bn|}{\bk \choose \bell-\bn} w(\bell)\bigg)\bigg] \label{sequentialEstimationEnd}
\end{align}
\hrulefill
\end{figure*}
Compared to the estimation of the highest-order coefficient---recall \eqref{extractingHighestOrder}---there is an additional term in \eqref{sequentialEstimationEnd} brought about by the propagation of cancellation errors, namely the term
\begin{align}
    \sum_{\bmm > \bk} (b_{\bmm}-\hat{b}_{\bmm}) {\bn \choose \bmm-\bk}.
\end{align} 
Neglecting this term, we would again be faced with 
\begin{align}
   \hat{b}_{\bk} = \frac{1}{2\pi}\sum_\bn u_{\bk}(\bn)\arg \! \big((\cD^{\bk} y_{\bk})(\bn)\big). \label{sequentialEstimation}
\end{align}
Recalling that $(\cD^{\bk} y_{\bk})(\bn)$ is independent of the choice of total order, the same holds for \eqref{sequentialEstimation}.
Remarkably, the neglect of error propagation does not hamper the high-SNR performance---more on this in Sec.~\ref{sec:performanceEvaluation}---and the sequential estimation process crystallizes into Alg.~\ref{algo:estimateCoefficients} with the averaging operator
\begin{align}
    \mu_{\bmm}(s) = \exp \! \left(j\sum_\bn u_{\bmm}(\bn)\arg \! \big(s(\bn)\big)\right). \label{linearAverager}
\end{align}
Later, in Sec. \ref{sec:accountingForTheCircularNature}, other averaging operators will be considered. 
\begin{algorithm}[t]
\caption{Estimation of the polynomial coefficients in a binomial representation by means of the averaging operator $\bmu_{\bmm}(\cdot)$. % in \eqref{linearAverager}
}
\label{algo:estimateCoefficients}
\begin{algorithmic}
\Procedure{Estimate-Coefficients}{$y, \cM$}
\For{$\bmm\in \cM$ (in descending order)}
    \State $\hat{b}_{\bmm} \gets \frac{1}{2\pi}\arg(\mu_{\bmm}(\cD^{\bmm} y))$
    \State $y(\bn) \gets y(\bn)\exp\!\big(\!-j2\pi \hat{b}_{\bmm}{\bn \choose \bmm} \big)$
\EndFor
\State \Return $\{\hat{b}_\bmm\}$
\EndProcedure
\end{algorithmic}
\end{algorithm}

\subsection{Computational Complexity}

Let us proceed to gauge the complexity of the proposed algorithm in number of multiply-and-accumulate operations.
\begin{itemize}
\item The calculation of the weight $u_\bmm(\bn)$ for each $\bn$ is $\cO(|\bmm|)$. For example, let us consider 
\begin{align}
    {\bn+\bmm \choose \bmm} = {n_0+m_0 \choose m_0}\cdots{n_{\sfD-1}+m_{\sfD-1} \choose m_{\sfD-1}} \label{binomialExample}
\end{align}
and invoke the formula
\begin{align}
    {n_d+m_d \choose m_d} = \frac{(n_d+m_d)\cdots(n_d+1)}{m_d!},
\end{align}
which is $\cO(m_d)$ \cite{rolfe2001binomial}. Running over the dimension $d$ then compounds the complexity to $\cO(|\bmm|)$.
Further running over the index $\bn$ brings it up to $\cO \big( |\bmm|\, |[\bN]| \big)$.
%(This could be alleviated at the cost of additional memory; one may compute the weights once, store them in a lookup table, and reuse them.)
\item The computation of $\cD^{\bmm}y$ is $\cO( |\bmm|\, |[\bN]|)$ as each operator $\cD_d$ is $\cO(|[\bN]|)$.
\end{itemize}
The overall complexity is therefore
\begin{align}
    \cO\bigg(|[\bN]|\sum_{\bmm\in\cM} |\bmm|\bigg) \label{complexity},
\end{align}
which depends linearly on the number of observations, $|[\bN]|$.

\section{Cramer-Rao Bound and Signal Reconstruction Performance}
\label{sec:crb}

\subsection{The Cramer-Rao Bound}

For an unbiased\footnote{The notion of unbiasedness is somewhat complicated for the problem at hand, because of the ambiguity expounded in Sec. \ref{sec:integerValuedPolynomials}. 
As discussed, a resolution is the constraint $\bb\in [-\frac{1}{2},\frac{1}{2} )^{|\cM|}$, which leads to the notion of constrained CRB \cite{gorman1990lower, marzetta1993simple, stoica1998cramer}. Unfortunately, an unbiased estimator does not exist when
there is an extreme point in the set of parameters \cite[Lemma 1]{somekh2017non}.
And, when the distribution is periodic with respect to the parameters, relaxing the range of the estimator (while constraining the set of parameters) does not resolve the issue \cite[Prop. 1]{todros2015limitations} \cite[Sec. III-A]{routtenberg2014cyclic}.
The issue could be fixed by redefining the unbiasedness to account for the periodic nature \cite{smith2005covariance, xavier2005intrinsic, routtenberg2012non} (see \cite{jackson1978frequency, kay1979comments, jackson1979reply} for a related discussion on frequency estimation). 
The set $[-\frac{1}{2},\frac{1}{2} )^{|\cM|}$ could be regarded as the quotient space $(\bbR/\bbZ)^{|\cM|}$, which can be identified with a multidimensional torus.
%it essentially identifies $-\frac{1}{2}$ and $\frac{1}{2}$.
This would come at the cost of additional mathematical and computational complexity.
% The coefficients lie in a manifold (specifically the multidimensional torus) rather than the Euclidean space---into account
As it turns out, though, this additional complexity can be circumvented because, in our problem, the high-SNR estimation errors are small enough---intuitively, the manifold where the parameters lie locally resembles the Euclidean space---that the CRB %in Sec. \ref{sec:crb}
becomes fully pertinent. 
% Particularly, it is stated in \cite[p. 1618]{smith2005covariance} that
% ``The significance of the sectional and Riemannian curvature terms is an open question that depends upon the specific application; however, as noted earlier, these terms become negligible for small errors and biases.'' 
}
estimator, the CRB is a lower bound on the variance.
The CRB for the one-dimensional polynomial estimation is analyzed in \cite{peleg1991cramer} and a computational method is developed in \cite{mckilliam2014cramer}.
With a view to its multidimensional generalization, the %$(\bmm,\bmm')$th entry of the
Fisher information matrix $\bJ\in\bbR^{|\cM|\times|\cM|}$ can be derived (see App.~\ref{fisherProof}) as
\begin{align}
[\bJ]_{\bmm,\bmm'} =
    8\pi^2 \SNR \! \sum_{\bn \in [\bN]}  {\bn \choose \bmm}{\bn \choose \bmm'}. \label{fisherEntry}
\end{align}
The CRB on the unbiased estimation of $\bb$ is then $\bJ^{-1}$ \cite[Thm. 3.2]{kay1993fundamentals}.

\subsection{High-SNR Performance for Signal Reconstruction}

For a collection of unbiased estimates $\{\hat{b}_{\bmm}\}$, let us introduce
%$\bK\in\bbR^{|\cM|\times|\cM|}$ be the covariance matrix of a collection of unbiased estimates $\{\hat{b}_{\bmm}\}$, namely
$
\bK = \bbE \big[(\hat{\bb}-\bb) (\hat{\bb}-\bb)^\top \big].
$
From these estimated coefficients, one can then reconstruct the polynomial phase signal $\big[\bn\in[\bN]\big]e^{j2\pi \hat{x}(\bn)}$, where
\begin{align}
    \hat{x}(\bn) = \sum_{\bmm} \hat{b}_{\bmm} {\bn \choose \bmm}.
\end{align}
The mean-square error (MSE) on that reconstruction is
\begin{align}
    \bbE\Bigg[\sum_{\bn\in[\bN]} \left| e^{j2\pi \hat{x}(\bn)}-e^{j2\pi x(\bn)} \right|^2\Bigg] \label{reconstructionMse},
\end{align}
which at high SNR satisfies
\begin{align}
    & \!\!\!\!\!\!\!\!\!\!\!  4\pi^2 \!\! \sum_{\bn\in[\bN]} \bbE\big[(\hat{x}(\bn)-x(\bn))^2\big]\\
    &\approx 4\pi^2 \!\! \sum_{\bn\in[\bN]} \bbE\Bigg[\bigg(\sum_{\bmm} (\hat{b}_{\bmm}-b_{\bmm}) {\bn \choose \bmm}\bigg)^{\! 2} \Bigg]\\
    &= \frac{1}{2 \, \SNR} \sum_{\bmm}\sum_{\bmm'} \underbrace{\bbE\big[(\hat{b}_{\bmm}-b_{\bmm})(\hat{b}_{\bmm'}-b_{\bmm'})\big]}_{(\bmm,\bmm')\text{th entry of }\bK}\nonumber\\
    &\qquad\qquad\qquad \cdot \underbrace{8\pi^2\SNR\sum_{\bn\in[\bN]}{\bn \choose \bmm}{\bn \choose \bmm'}}_{(\bmm,\bmm')\text{th entry of }\bJ} \\
    &= \frac{1}{2 \, \SNR}\tr (\bK\bJ) \label{usingMainResult}.
\end{align}

\subsection{Remarks} 

For any unbiased estimator, which necessarily satisfies $\bK\geq \bJ^{-1}$ with $\geq$ the Loewner order for positive-semidefinite matrices, we have that \cite[Thm. 7.7.2(a)]{horn2012matrix}
\begin{align}
    \bJ^{\frac{1}{2}}\bK\bJ^{\frac{1}{2}}- \bI \geq \zero. \label{psd}
\end{align}
This implies that
\begin{align}
    \tr (\bK\bJ) &= \tr (\bK^{\frac{1}{2}}\bJ\bK^{\frac{1}{2}})\\
    &= \tr (\underbrace{\bK^{\frac{1}{2}}\bJ\bK^{\frac{1}{2}}-\bI}_{\geq \zero}) + \underbrace{\tr(\bI)}_{=|\cM|}\\
    &\geq|\cM|,
\end{align}
with equality condition
\begin{align}
    \tr (\bK^{\frac{1}{2}}\bJ\bK^{\frac{1}{2}}-\bI) = 0.
\end{align}

Let us simplify the above equality condition.
From \eqref{psd}, the eigenvalues of $\bJ^{\frac{1}{2}}\bK\bJ^{\frac{1}{2}}- \bI$ are nonnegative. Since their sum (the trace) equals zero, all of the eigenvalues should be zero. The condition is thus $\bK^{\frac{1}{2}}\bJ\bK^{\frac{1}{2}} - \bI= \zero$, i.e., $\bK=\bJ^{-1}$. 

Returning to the inequality, the slightly stronger version
\begin{align}
    \tr(\bK\bJ) \geq |\cM|+\underbrace{\ln\det(\bK\bJ)}_{\geq 0}
\end{align}
can be obtained from the Kullback-Leibler divergence between $\cN(\zero,\bK)$ and $\cN(\zero,\bJ^{-1})$ being \cite[Eq. 1]{zhang2024properties}
\begin{equation}
\frac{1}{2}\big(\tr(\bK\bJ)- |\cM|-\ln\det(\bK\bJ)\big) ,
\end{equation}
which is nonnegative. An alternative derivation of this stronger inequality is provided in
App. \ref{app:traceProof}.

In summary, 
% provided that $\bK\geq\bJ^{-1}$, 
\begin{align}
    \bbE\Bigg[\sum_{\bn\in[\bN]} \left| e^{j2\pi \hat{x}(\bn)}-e^{j2\pi x(\bn)} \right|^2\Bigg]
    &\approx \frac{1}{2 \, \SNR}\tr (\bK\bJ)\\
    &\geq \frac{|\cM|}{2 \, \SNR}, \label{reconstructionErrorCrb}
\end{align}
with equality if and only if $\bK = \bJ^{-1}$. Therefore:
\begin{enumerate}
    \item When the estimator achieves the CRB, the reconstruction MSE does not depend on the number of observations, $\bN$. In contrast, without any processing, that MSE would be
    \begin{align}
        &\bbE\Bigg[\sum_{\bn\in[\bN]} \left| y(\bn)-e^{j2\pi x(\bn)} \right|^2\Bigg]\\
        &\qquad\qquad =\bbE\Bigg[\sum_{\bn\in[\bN]} |w_\bbC(\bn)|^2\Bigg]= \frac{|[\bN]|}{\SNR}.
    \end{align}
    Recalling \eqref{moreSampleThanDegree}, the high-SNR performance is always superior with the proposed estimator. In this respect, one can think of it as a denoising method that harnesses the signal structure.
    \item Since the equality holds if and only if the estimator attains the CRB, we can assess how close the estimator's covariance is to the CRB by comparing \eqref{reconstructionMse} with $\frac{|\cM|}{2 \, \SNR}$.
    %Also, the high-SNR power offset can be expressed as 
    %\heedong{What I wanted to say here is quite simple. For an estimator which does not attain the CRB at high SNR, its performance can be quantified by the high-SNR power offset:
    %\begin{align*}
    %    \frac{|\cM|}{\SNR} = \frac{\tr(\bK\bJ)}{\SNR \cdot \SNR_{\sf offset}}.
    %\end{align*}
    %We shall consider such inefficient estimator in Sec. \ref{sec:introducingLagParameters}, which is still in progress (for motivation, please see Fig. \ref{fig:lagTest}; better low-SNR performance at the cost of high-SNR performance loss).
    %}
    %\begin{align}
    %    \frac{\tr(\bK\bJ)}{|\cM|}.
    %\end{align}
\end{enumerate}

The general characterization of the accuracy of the estimates of $\bb$ is
    \begin{align}
        \left\{ [\bK]_{\bmm,\bmm} : \bmm\in\cM \right\} ,
    \end{align}
with convenient scalar measures being the total variance and the generalized variance, which are respectively the trace and determinant of the covariance matrix of those estimates, $\bK$; these scalar measures satisfy $\tr(\bK)\geq \tr(\bJ^{-1})$ and $\det(\bK)\geq \det(\bJ^{-1})$. %with equality if and only if $\bK=\bJ^{-1}$.
This paper, though, favors $\tr(\bK\bJ)$ as scalar measure, given its direct relevance---recall \eqref{usingMainResult}---to the MSE on the signal reconstruction.
%which is a convenient scalar summary of the estimator's performance
%This scalar measure is the one used for numerical evaluations.

\section{Performance Evaluation}
\label{sec:performanceEvaluation}

In this section, the high-SNR behavior of the proposed estimator's covariance matrix is appraised in relation to the CRB. To set the stage for this assessment, a family of orthogonal polynomials is first introduced \cite{mckilliam2014cramer}.
%We commence with introducing a mathematical tool---orthogonal polynomials---to set the stage for the comparison. 

\subsection{Orthogonal Polynomials}
\label{sec:orthogonalPolynomials}
%This subsection introduces a family of orthogonal polynomials which has been found to be useful \cite{mckilliam2014cramer}.
The polynomial of degree $k$,
\begin{align}
    q_{k}(n) &\equiv \Delta^k\bigg[{n \choose k}{n-N \choose k}\bigg] 
\end{align}
satisfies the orthogonality condition (see \cite[Prop. 5.7]{hirvensalo2003studies} and \cite[Prop. 2]{eisinberg2007discrete})
\begin{align}
    \langle q_k,q_{k'} \rangle &\equiv \sum_{n\in[N]} q_k(n)q_{k'}(n)\\ 
    &= [k=k']{N+k \choose 2k+1} {2k \choose k} \label{PSG}
\end{align}
for $k,k' \in [N]$.
For the sake of completeness, a self-contained proof of (\ref{PSG}) is provided in App.~\ref{app:orthogonalityProof}.
This result has a straightforward multidimensional counterpart; the polynomials
\begin{align}
    q_{\bk}(\bn) &\equiv \Delta^\bk\bigg[{\bn \choose \bk}{\bn-\bN \choose \bk}\bigg] \label{orthogonalPolynomial}
\end{align}
satisfy the orthogonality condition 
\begin{align}
    \langle q_\bk,q_{\bk'} \rangle &\equiv \sum_{\bn\in[\bN]} q_\bk(\bn)q_{\bk'}(\bn)\\ 
    &= [\bk=\bk']{\bN+\bk \choose 2\bk+\one} {2\bk \choose \bk} \label{distributiveLawExample}
\end{align}
for $\bk,\bk' \in [\bN]$, as follows from $q_{\bk}(\bn) = \prod_d q_{k_d}(n_d)$ and the general distributive law.

Two identities shall come in handy later in this section. % Sec. \ref{sec:highSnrPerformance}.
The first one, proved in App. \ref{app:usefulIdentity1}, is
\begin{align}
    q_{\bk}(\bn) \!=\!\! \sum_{\bell\in [\bN-\bk]}\!\! (-1)^{|\bk+\bn+\bell|}\!{\bell+\bk \choose \bk}\!{\bN-\bell-\one \choose \bk}\!{\bk \choose \bn-\bell}.\label{usefulIdentity1}
\end{align}
The second one, proved in App. \ref{app:usefulIdentity2}, is
\begin{align}
    \bigg\langle {\bn\choose \bmm}, q_\bk \bigg\rangle \!=\!\! \sum_{\bn\in[\bN-\bk]} {\bn \choose \bmm-\bk}{\bn+\bk \choose \bk}{\bN-\bn-\one \choose \bk}. \label{usefulIdentity2}
\end{align}
% \subsection{Other Representations}
% Also, using Leibniz's rule, we have
% \begin{align}
%     q_{\ell}(n) &= (-1)^{\ell} \sum_k {\ell \choose k} \Delta^{k}{n \choose \ell}\Delta^{\ell-k}{n+k-N \choose \ell}\\
%     &= (-1)^{\ell} \sum_k {\ell \choose k} {n \choose \ell-k}{n+k-N \choose k}
% \end{align}
% The last one can be obtained from representing ${n \choose \ell}{n-N \choose \ell}$ in terms of binomial basis:
% \begin{align}
%     {n \choose \ell}{n-N \choose \ell}
%     &={n \choose \ell} \sum_k {\ell-N \choose 2\ell-k}{n-\ell \choose k-\ell}\\
%     &=\sum_k {\ell-N \choose 2\ell-k}{n-\ell \choose k-\ell}{n \choose \ell}\\
%     &=\sum_k {\ell-N \choose 2\ell-k}{k \choose \ell}{n \choose k},
% \end{align}
% where Vandermonde convolution \cite[Eqn. 5.22]{graham1994concrete} and trinomial revision \cite[Table 174]{graham1994concrete} are used in the first and the last identities. We therefore obtain
% \begin{align}
%     q_\ell(n) &= (-1)^\ell\sum_k {k \choose \ell}{\ell-N \choose 2\ell-k}{n \choose k-\ell}\\
%     &=(-1)^\ell\sum_k {k+\ell \choose \ell}{\ell-N \choose \ell-k}{n \choose k}. \label{orthogonalBinomialRepresentation}
% \end{align}

% \heedong{DUAL REPRESENTATION}
% \begin{align}
%     \sum_{\ell \in [N]} \frac{q_\ell(n) q_\ell(n')}{{N+\ell \choose 2\ell+1} {2\ell \choose \ell}} = [n = n']
% \end{align}

\subsection{Decomposition of the Cramer-Rao Bound}

There are $|[\bN]|$ orthogonal polynomials in $\{q_\bk(\bn):\bk\in[\bN]\}$. Recalling that $q_\bk$ is of degree $\bk$, the $|[\bN]|$-dimensional vector space of polynomials whose degrees are in $[\bN]$ is spanned by $\{q_\bk(\bn):\bk\in[\bN]\}$ \cite[Thm. 1.10]{friedberg2019linear}.
Recalling that ${\bn \choose \bmm}$ is a polynomial of degree $\bmm\in\cM \subset[\bN]$,
we can decompose it as \cite[Thm. 6.3]{friedberg2019linear}
\begin{align}
    {\bn \choose \bmm} &= \sum_{\bk\in[\bN]} \frac{\langle {\bn \choose \bmm}, q_\bk \rangle}{\langle q_\bk,q_\bk \rangle} q_\bk(\bn)
\end{align}
thanks to the orthogonality. Plugged into \eqref{fisherEntry}, this gives
\begin{align}
   [\bJ]_{\bmm,\bmm'} & = 8\pi^2 \SNR \sum_{\bn \in [\bN]}  \Bigg[\sum_{\bk\in[\bN]} \frac{\langle {\bn \choose \bmm}, q_\bk \rangle}{\langle q_\bk,q_\bk \rangle} q_\bk(\bn)\Bigg] \nonumber\\
    & \quad \cdot\Bigg[\sum_{\bk'\in[\bN]} \frac{\langle {\bn \choose \bmm'}, q_{\bk'} \rangle}{\langle q_{\bk'},q_{\bk'} \rangle} q_{\bk'}(\bn)\Bigg]\\
    & \!\!\!\!\!\!\! =8\pi^2 \SNR \! \sum_{\bk\in[\bN]}\sum_{\bk'\in[\bN]} \! \frac{\langle {\bn \choose \bmm}, q_\bk \rangle}{\langle q_\bk,q_\bk \rangle} \frac{\langle {\bn \choose \bmm'}, q_{\bk'} \rangle}{\langle q_{\bk'},q_{\bk'} \rangle} \langle q_\bk,q_{\bk'} \rangle\\
    & \!\!\!\!\!\!\! =8\pi^2 \SNR \! \sum_{\bk\in[\bN]} \! \frac{\langle {\bn \choose \bmm}, q_\bk \rangle \langle {\bn \choose \bmm'}, q_{\bk} \rangle}{\langle q_\bk,q_\bk \rangle}
\end{align}
or, in matrix form,
\begin{align}
    \bJ = 8\pi^2\SNR \cdot \bS^\top(\bQ\bQ^\top)^{-1}\bS \label{fisherDecomposition}
\end{align}
where $\bS\in\bbR^{|\cM|\times |\cM|}$ and $\bQ\in\bbR^{|\cM|\times|[\bN]|}$ are matrices whose $(\bk,\bmm)$th and $(\bk,\bn)$ entries are $\langle {\bn \choose \bmm}, q_\bk \rangle$ and $q_\bk(\bn)$, respectively.
% \begin{align}
%     [\bS]_{\bmm,\bell} = \bigg\langle {\bn \choose \bmm}, q_\bell \bigg\rangle\qquad
%     [\bQ]_{\bell,\bn} = q_\bell(\bn).
% \end{align}
Although the decomposition using orthogonal polynomials appears to complicate the formulation, it shall prove useful in the sequel.

\subsection{High-SNR Performance of the Coefficient Estimation}
\label{sec:highSnrPerformance}

The high-SNR regime is often the only one where an estimator's performance lends itself to analysis \cite{stoica1998asymptomania}, an observation that guides the remainder of the section.

Let us denote $b_\bk$'s estimation error by $v(\bk)$, whose covariance matrix is $\bK$. Thus,
\begin{align}
    v(\bk)& \equiv \hat{b}_{\bk}-b_{\bk}\label{estimationError}\\
    &= \frac{1}{2\pi}\sum_\bn u_{\bk}(\bn)\arg \! \big((\cD^{\bk} y_{\bk})(\bn)\big) - b_{\bk}\\
    &= \frac{1}{2\pi}\sum_\bn u_{\bk}(\bn) \Big( \! \arg \! \big((\cD^{\bk} y_{\bk})(\bn)\big) - b_{\bk}\Big), \label{estimationErrorExpanded}
\end{align}
where the last step follows from \eqref{weightProperty}. 
Recasting \eqref{sequentialEstimationEnd} as
\begin{align}
    &(\Pi\cD^{\bk}y_{\bk})(\bn)\approx \big[\bn\in[\bN-\bk]\big]\exp \! \bigg[j2\pi \bigg( b_{\bk} \\
    &- \sum_{\bmm > \bk} {\bn \choose \bmm-\bk} v(\bmm) + \sum_{\bell} (-1)^{|\bk+\bell+\bn|}{\bk \choose \bell-\bn} w(\bell)\bigg)\bigg] \nonumber
\end{align}
and plugging it into \eqref{estimationErrorExpanded} yields a recurrence relation for $v(\bk)$, namely
\begin{align}
    v(\bk) 
    &\approx \sum_\bn u_{\bk}(\bn)\Bigg[-\sum_{\bmm>\bk}{\bn \choose \bmm-\bk} v(\bmm) \label{boltana}\\
    &\qquad\qquad\qquad\qquad +\sum_{\bell} (-1)^{|\bk+\bell+\bn|}{\bk \choose \bell-\bn} w(\bell)\Bigg]. \nonumber 
\end{align}
From this recurrence, $v(\bk)$ can be seen to be asymptotically Gaussian at high SNR.
Indeed, starting from the maximal element $\bk$, $v(\bk)$ is a summation of Gaussian noise terms, and the Gaussian nature of $v(\bk)$ follows by induction.

Using \eqref{weightProperty} to rewrite the left-hand side of \eqref{boltana} as
\begin{align}
    \sum_\bn u_\bk(\bn) v(\bk)
\end{align}
and rearranging terms,
\begin{align}
    & \!\!\!\! \sum_{\bmm} \sum_\bn u_{\bk}(\bn){\bn \choose \bmm-\bk} v(\bmm) \nonumber \\
    &\qquad \approx\sum_\bn \sum_{\bell} (-1)^{|\bk+\bell+\bn|} u_{\bk}(\bn) {\bk \choose \bell-\bn} w(\bell)\\
    &\qquad =\sum_\bn \sum_{\bell} (-1)^{|\bk+\bn+\bell|} u_{\bk}(\bell) {\bk \choose \bn-\bell} w(\bn),
\end{align}
where the last step is a mere relabelling. Plugging in \eqref{kayWeight},
\begin{align}
    &\sum_\bmm\! \sum_{\bn\in[\bN-\bk]} \!\!{\bn \choose \bmm\!-\!\bk}\!{\bn\!+\!\bk \choose \bk}\!{\bN\!-\!\bn\!-\!\one \choose \bk} v(\bmm) \\
    & \!\approx \! \sum_\bn\! \sum_{\bell\in[\bN-\bk]}\!\!\! (-1)^{|\bk+\bn+\bell|} {\bell\!+\!\bk \choose \bk}\!{\bN\!-\!\bell\!-\!\one \choose \bk}\! {\bk \choose \bn\!-\!\bell} w(\bn). \nonumber 
\end{align}
Recalling \eqref{usefulIdentity1} and \eqref{usefulIdentity2}, the above boils down to
\begin{align}
    \sum_\bmm \bigg\langle {\bn \choose \bmm}, q_\bk \bigg\rangle v(\bmm) \approx \sum_\bn q_\bk(\bn) w(\bn). \nonumber    
\end{align}
From the observation that the left- and right-hand sides are
\begin{align}
    [\bS\bv]_\bk
    &=\sum_\bmm [\bS]_{\bk,\bmm}[\bv]_\bmm\\
    [\bQ\bw]_\bk
    &=\sum_\bn [\bQ]_{\bk,\bn}[\bw]_\bn,
\end{align}
with $\bS$ and $\bQ$ as introduced in the previous subsection, we obtain the vectorized form
\begin{align}
    \bS \bv \approx \bQ\bw. \label{noiseRelation}
\end{align}
As the covariance of $\bv$ is $\bK$, it follows that
\begin{align}
    \bS\bK\bS^\top \approx \frac{1}{8\pi^2\SNR}\bQ\bQ^\top, 
\end{align}
or, equivalently,
\begin{align}
    \bK \big(8\pi^2\SNR\cdot \bS^\top (\bQ\bQ^\top)^{-1}\bS\big) \approx \bI. 
\end{align}
Recalling \eqref{fisherDecomposition}, we obtain the remarkable relationship
\begin{align}
    \bK\approx\bJ^{-1}. \label{mainResult}
\end{align}
Therefore, with the error propagation neglected, the proposed estimator attains the CRB at high SNR. This greatly generalizes \cite{kay1989fast}, \cite{djuric1990parameter}, \cite{kitchen1994method}, and \cite{kay1990efficient} in two ways:
\begin{enumerate}
    \item In those references, only $\cM=\{0,1\}$, $\cM = \{0,1,2\}$, $\cM=\{0,1,\ldots,M\}$, and $\cM=\{(0,0),(0,1),(1,0)\}$ are considered.
    \item Only the variances of the maximal degrees are examined, meaning that only $[\bK]_{\bmm,\bmm}=[\bJ^{-1}]_{\bmm,\bmm}$ is established for a maximal element $\bmm\in\cM$.
\end{enumerate}
% which only consider the cases $\cM=\{0,1\}$, $\cM = \{0,1,2\}$, and $\cM=\{(0,0),(0,1),(1,0)\}$. 

\begin{figure*}
    \centering
    \begin{tabular}{@{}c@{}}
    \subfloat[Reconstruction MSE vs SNR. Besides the MSE over all noise instances, the plot also shows the MSE conditioned on the existence or absence of phase wrappings. \label{fig:mseWrapping}]{%
          \includegraphics[width=.45\linewidth]{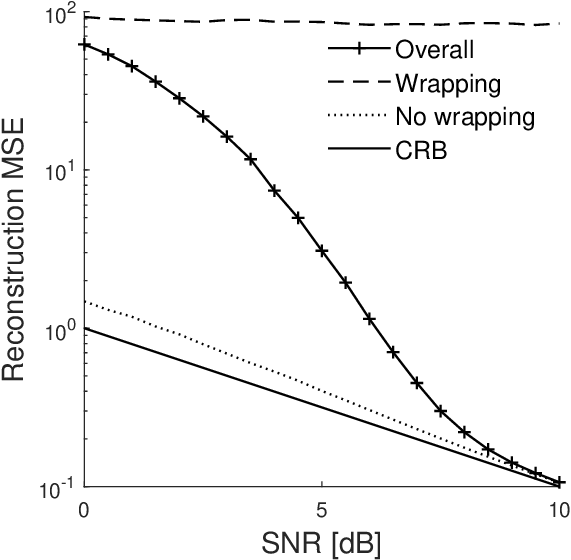}}
    \end{tabular}\quad      
    \begin{tabular}{@{}c@{}}
    \subfloat[Probability of phase wrapping. \label{fig:probabilityWrapping}]{%
          \includegraphics[width=.45\linewidth]{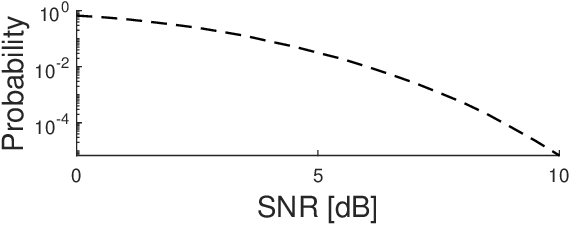}} \\
    \subfloat[Contribution to the reconstruction MSE. \label{fig:mseProportion}]{%
          \includegraphics[width=.45\linewidth]{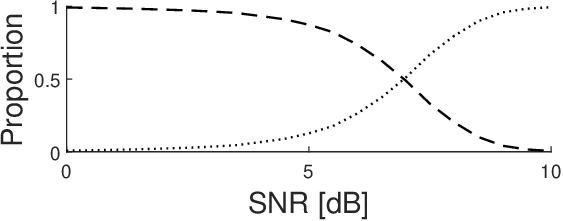}}
    \end{tabular}
    \caption{Evaluation of Alg. \ref{algo:estimateCoefficients} for $\sfD=1$ with $\cM=\{0,1\}$ and $N=64$. The zero polynomial (i.e., $\bb=\zero$) is considered. 
    % The reconstruction MSE in \eqref{reconstructionMse} is obtained by averaging over noise instances;
    The reconstruction MSE is obtained by Monte-Carlo runs with noise instances generated until a thousand wrapping events are collected. For example, at $\SNR = 10$ dB, 147 million noise instances were generated.
    }
    \label{fig:snrThreshold}
\end{figure*}
% \begin{figure*}
%     \centering
%     \begin{tabular}{cc}
%     \adjustbox{valign=t}{\subfloat[Reconstruction MSE vs SNR. Besides the MSE over all noise instances, the plot also shows the MSE conditioned on the existence or absence of phase wrappings. \label{fig:mseWrapping}]{%
%           \includegraphics[width=.45\linewidth]{images/mseWrapping.eps}}}
%     &      
%     \adjustbox{valign=t}{\begin{tabular}{@{}c@{}}
%     \subfloat[Probability of phase wrapping. \label{fig:probabilityWrapping}]{%
%           \includegraphics[width=.45\linewidth]{images/probabilityWrapping.eps}} \\
%     \subfloat[Contribution to the reconstruction MSE. \label{fig:mseProportion}]{%
%           \includegraphics[width=.45\linewidth]{images/mseProportion.eps}}
%     \end{tabular}}
%     \end{tabular}
%     \caption{Evaluation of Alg. \ref{algo:estimateCoefficients} for $\sfD=1$ with $\cM=\{0,1\}$ and $N=64$. The zero polynomial (i.e., $\bb=\zero$) is considered. 
%     % The reconstruction MSE in \eqref{reconstructionMse} is obtained by averaging over noise instances;
%     The reconstruction MSE is obtained by Monte-Carlo runs with noise instances generated until a thousand wrapping events are collected. For example, at $\SNR = 10$ dB, 147 million noise instances were generated.
%     }
%     \label{fig:snrThreshold}
% \end{figure*}

\subsection{Threshold Behavior}

Supporting the foregoing high-SNR analysis, Fig. \ref{fig:mseWrapping} illustrates how the proposed estimator (curve ``overall'') attains the CRB at high SNR in a simple one-dimensional setting ($\cM=\{0,1\}$, $N=64$).
%for a simple one-dimensional setting with $\cM=\{0,1\}$ and $N=64$, the proposed estimator  attains the CRB at high SNR, confirming the efficacy of the foregoing high-SNR analysis.
This welcome behavior, however, breaks down and the estimates degrade rapidly as the SNR dips below a threshold. Such a threshold effect is prevalent in many estimation problems, and can be attributed to outliers \cite{vertatschitsch1991impact, thomas1995probability, athley2005threshold}, meaning severely unreliable estimates whose occurrence decays swiftly with the SNR. Characterizing outlier events is contingent on both the specific estimation problem and the estimator of interest, and is somewhat of an art.

In our case, an outlier corresponds to a phase-wrapping situation, i.e., to
\begin{align}
     b_\bk + \Delta^\bk\bigg(\frac{1}{2\pi}\arg \! \big(1 + \tilde{w}_{\bbC}(\bn)\big)\bigg) \notin \bigg[-\frac{1}{2},\frac{1}{2}\bigg)  \label{outlierCondition}
\end{align}
for some $\bn\in[\bN-\bk]$ and $\bk\in\cM$. In Fig. \ref{fig:mseWrapping}, the reconstruction MSE is further shown when conditioned on \eqref{outlierCondition}, with ``wrapping'' and ``no wrapping'' indicating whether the segregated noise instances trigger (or not) a phase wrapping.
%\begin{enumerate}
%    \item \eqref{outlierCondition} holding (labelled ``wrapping''); and
%    \item otherwise (labelled ``no wrapping''). \heedong{I can't decide between ``wrapping'' and ``outlier''...}
% \end{enumerate}
\begin{figure}
    \centering
    \includegraphics[width=0.9\linewidth]{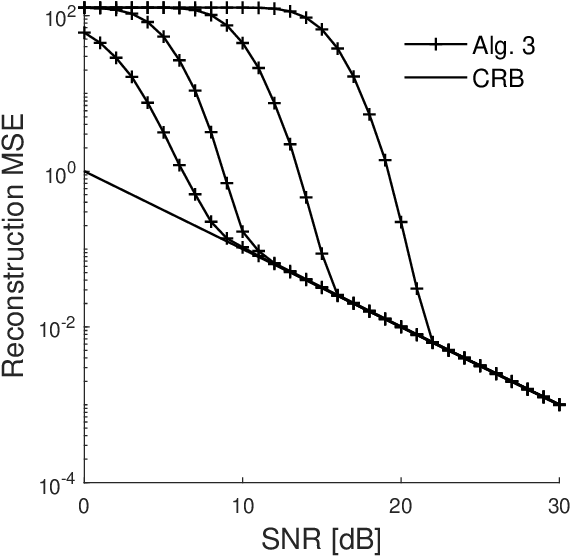}
    \caption{Evaluation of Alg. \ref{algo:estimateCoefficients} for $\sfD=1$ with $\cM=\{0,1\}$ and $N=64$. The reconstruction MSE in \eqref{reconstructionMse} is obtained by averaging over 10000 noise instances. The curves from left to right correspond to $(b_0,b_1) = \big(0,\frac{1}{2}-\frac{1}{2}\big), \big(0,\frac{1}{2}-\frac{1}{2^2}\big),\big(0,\frac{1}{2}-\frac{1}{2^3}\big)$, and $\big(0,\frac{1}{2}-\frac{1}{2^4}\big)$.}
    \label{fig:dependenceOnParameter}
\end{figure}

In the absence of wrappings, the reconstruction MSE exhibits only a small shortfall from the CRB below the threshold, attributable to the discarded amplitude information \cite{baggenstoss1991estimating, quinn2010phase, fu2013phase} (see also \cite[Thm. 1]{mckilliam2014polynomial}).
With phase wrappings, conversely, the MSE is far removed from the CRB and does not improve with the SNR; its contribution to the overall MSE, however, does decline sharply (Fig. \ref{fig:probabilityWrapping}).

\section{Accounting for the Circular Nature}
\label{sec:accountingForTheCircularNature}

A drawback of the averaging operator in \eqref{linearAverager} is that the performance of the resultant estimator at a given SNR depends on the parameters being estimated---recall the characterization of a phase-wrapping situation in \eqref{outlierCondition}.
% in relation to the SNR. 
% \angel{Wouldn't the drawback be of the averaging operator, rather than the algorithm?} \heedong{You're totally right. The culprit is the averaging operator, not the algorithm. Incidentally, I don't get what ``in relation to the SNR'' means...}
% \angel{Ok, then let's rephrase. I wanted to inject the SNR into the consideration, since the tolerace of proximity to a branch cut is dictated by the SNR. If the SNR increases, we can get closer to the branch cut without running into problems, as the figure nicely shows graphically.} \heedong{Oh, I now fully understand what it means and I think there's no need to rephrase it.}
This dependence is demonstrated in Fig. \ref{fig:dependenceOnParameter} for a simple one-dimensional setting with $\cM=\{0,1\}$ and $N=64$, whereby Alg. \ref{algo:estimateCoefficients} with \eqref{linearAverager} becomes Kay's weighted phase averager \cite[Eq. 16]{kay1989fast}. Hereafter, ``CRB'' in the legends corresponds to \eqref{reconstructionErrorCrb}, the high-SNR reconstruction MSE derived from the CRB. As expected (and as recognized in \cite[Figs. 6-7]{fowler2002phase}), the performance at any given SNR worsens as $b_1$ approaches the cell boundary at $\frac{1}{2}$.
Particularly, the step,
\begin{align}
    \hat{b}_{\bmm} &\gets \frac{1}{2\pi}\arg \! \big(\mu_\bmm(\cD^{\bmm} y)\big)\\
    &= \frac{1}{2\pi}\sum_\bn u_{\bmm}(\bn) \arg\!\big((\cD^{\bmm} y)(\bn)\big), \label{branchCutAtPi}
\end{align}
is problematic. This step computes the average of the argument of a complex signal in a linear sense, but, as that argument is circular in nature, it should be processed in a circular fashion.
% For circular data, averaging in a linear sense is inappropriate. 

\begin{figure*}
\centering
    \begin{tikzpicture} 
    \fill (0,0) circle (0.05cm);
    \draw (0,0) circle [radius=2cm];
    
    \def\angleOne{110};  % 11:20
    \def\angleTwo{100};  % 11:40
    \def\angleThree{75}; % 00:30
    \def\angleCircular{95};
    \def\angleLinear{215};
    \coordinate (pointOne) at ({2*cos(\angleOne)},{2*sin(\angleOne)});
    \coordinate (pointTwo) at ({2*cos(\angleTwo)},{2*sin(\angleTwo)});
    \coordinate (pointThree) at ({2*cos(\angleThree)},{2*sin(\angleThree)});
    \coordinate (pointCircular) at ({2*cos(\angleCircular)},{2*sin(\angleCircular)});
    \coordinate (pointLinear) at ({2*cos(\angleLinear)},{2*sin(\angleLinear)});
    
    % circular clock
    \foreach \angle [count=\xi] in {60,30,...,-270}
    {
      \draw[line width=1pt] (\angle:1.9cm) -- (\angle:2cm);
      \node at (\angle:1.6cm) {\textsf{\xi}};
    }
    \foreach \angle in {0,90,180,270}
      \draw[line width=2pt] (\angle:1.8cm) -- (\angle:2cm);
    
    \node[draw,circle,fill=black,scale=0.5] at (pointOne) [] {};
    \node[draw,circle,fill=black,scale=0.5] at (pointTwo) [] {};
    \node[draw,circle,fill=black,scale=0.5] at (pointThree) [] {};
    \node[draw,circle,fill=white,scale=0.5] at (pointCircular) [label=above:{Circular}] {};
    \node[draw,rectangle,fill=white,scale=0.5] at (pointLinear) [label= left:{Linear}] {};

    % linear clock
    \draw[line width=1pt] (0,2.8) -- ({4*pi},2.8);
    \node at (0,3.3) {\small\textsf{12}};
    \foreach \xi in {1,2,...,12}
    {
      \draw[line width=1pt] ({4*pi/12*\xi},2.8) -- ({4*pi/12*\xi},2.9);
      \node at ({4*pi/12*\xi},3.3) {\small \textsf{\xi}};
    }
    \foreach \xi in {0,3,...,12}
    \draw[line width=2pt] ({4*pi/12*\xi},2.8) -- ({4*pi/12*\xi},3);
    \node[draw,circle,fill=black,scale=0.5] at ({4*pi*(360+90-\angleOne)/360},2.8) [] {};
    \node[draw,circle,fill=black,scale=0.5] at ({4*pi*(360+90-\angleTwo)/360},2.8) [] {};
    \node[draw,circle,fill=black,scale=0.5] at ({4*pi*(90-\angleThree)/360},2.8) [] {};
    \node[draw,rectangle,fill=white,scale=0.5] at ({4*pi*((360+90-\angleOne)+(360+90-\angleTwo)+(90-\angleThree))/3/360},2.8) [label=below:{Linear}] {};
    \draw[line width=1pt, ->] (2.5,0) -- (7,0) node[midway, above, align=center] {\small Rotate the clock by\\ \small 90$^\circ$ counterclockwise};
    \draw[line width=1pt, ->] (1.7,1.5) -- (5.2,1.5) -- (5.2,2.5);
    \node[align=center] at (3.4,2) {\small Unfold the clock with\\ \small boundary at 12 o'clock};

    \begin{scope}[shift={(0,-5.6)}]
        % Another linear clock
        \draw[line width=1pt] (0,2.8) -- ({4*pi},2.8);
        \node at (0,2.3) {\small\textsf{6}};
        \foreach \xi in {1,2,...,6}
        {
          \draw[line width=1pt] ({4*pi/12*\xi},2.8) -- ({4*pi/12*\xi},2.7);
          \node at ({4*pi/12*(\xi+6)},2.3) {\small \textsf{\xi}};
        }
        \foreach \xi in {7,8,...,12}
        {
          \draw[line width=1pt] ({4*pi/12*\xi},2.8) -- ({4*pi/12*\xi},2.7);
          \node at ({4*pi/12*(\xi-6)},2.3) {\small \textsf{\xi}};
        }
        \foreach \xi in {0,3,...,12}
        \draw[line width=2pt] ({4*pi/12*\xi},2.8) -- ({4*pi/12*\xi},2.6);
        \node[draw,circle,fill=black,scale=0.5] at ({4*pi*(360+90-\angleOne-180)/360},2.8) [] {};
        \node[draw,circle,fill=black,scale=0.5] at ({4*pi*(360+90-\angleTwo-180)/360},2.8) [] {};
        \node[draw,circle,fill=black,scale=0.5] at ({4*pi*(90-\angleThree+180)/360},2.8) [] {};
        \node[draw,circle,fill=white,scale=0.5] at ({4*pi*((360+90-\angleOne-180)+(360+90-\angleTwo-180)+(90-\angleThree+180))/3/360},2.8) [label=above:{Circular}] {};
        \draw[line width=1pt, ->,] (1.7,4.1) -- (5.2,4.1) -- (5.2,3.1);
        \node[align=center] at (3.4,3.6) {\small Unfold the clock with\\ \small boundary at 6 o'clock};
    \end{scope}

    \begin{scope}[shift={(10,0)}]
    \draw[line width=1pt, ->, opacity=0.5] (-2.5,0) -- (2.5,0);
    \draw[line width=1pt, ->, opacity=0.5] (0,-2.5) -- (0,2.5);
    \begin{scope}[rotate=90]
    \fill (0,0) circle (0.05cm);
    \draw (0,0) circle [radius=2cm];
    
    \def\angleOne{110};  % 11:20
    \def\angleTwo{100};  % 11:40
    \def\angleThree{75}; % 00:30
    \def\angleCircular{95};
    \def\angleLinear{215};
    \coordinate (pointOne) at ({2*cos(\angleOne)},{2*sin(\angleOne)});
    \coordinate (pointTwo) at ({2*cos(\angleTwo)},{2*sin(\angleTwo)});
    \coordinate (pointThree) at ({2*cos(\angleThree)},{2*sin(\angleThree)});
    \coordinate (pointCircular) at ({2*cos(\angleCircular)},{2*sin(\angleCircular)});
    \coordinate (pointLinear) at ({2*cos(\angleLinear)},{2*sin(\angleLinear)});
    
    % circular clock
    \foreach \angle [count=\xi] in {60,30,...,-270}
    {
      \draw[line width=1pt] (\angle:1.9cm) -- (\angle:2cm);
      \node[rotate=90] at (\angle:1.6cm) {\textsf{\xi}};
    }
    \foreach \angle in {0,90,180,270}
      \draw[line width=2pt] (\angle:1.8cm) -- (\angle:2cm);
    
    \node[draw,circle,fill=black,scale=0.5] at (pointOne) [] {};
    \node[draw,circle,fill=black,scale=0.5] at (pointTwo) [] {};
    \node[draw,circle,fill=black,scale=0.5] at (pointThree) [] {};
    \node[draw,circle,fill=white,scale=0.5] at (pointCircular) [] {};
    \node[draw,rectangle,fill=white,scale=0.5] at (pointLinear) [] {};
    \end{scope}
    \end{scope}
    \end{tikzpicture}
\caption{Circular data and their averages in the linear and circular senses.}
% \heedong{Please look into the additional linear clock. If you think it's more than enough, I'll undo the change. And I find ``unfolding the clock with boundary at'' is verbose and unnatural. I was wondering if you have better wording...} \angel{Great pic! Do keep it. As of the text, it's fine. Perhaps just change "Unfolding" to "Unfold" and "Rotating" to "Rotate"}
\label{fig:clockExample}
\end{figure*}
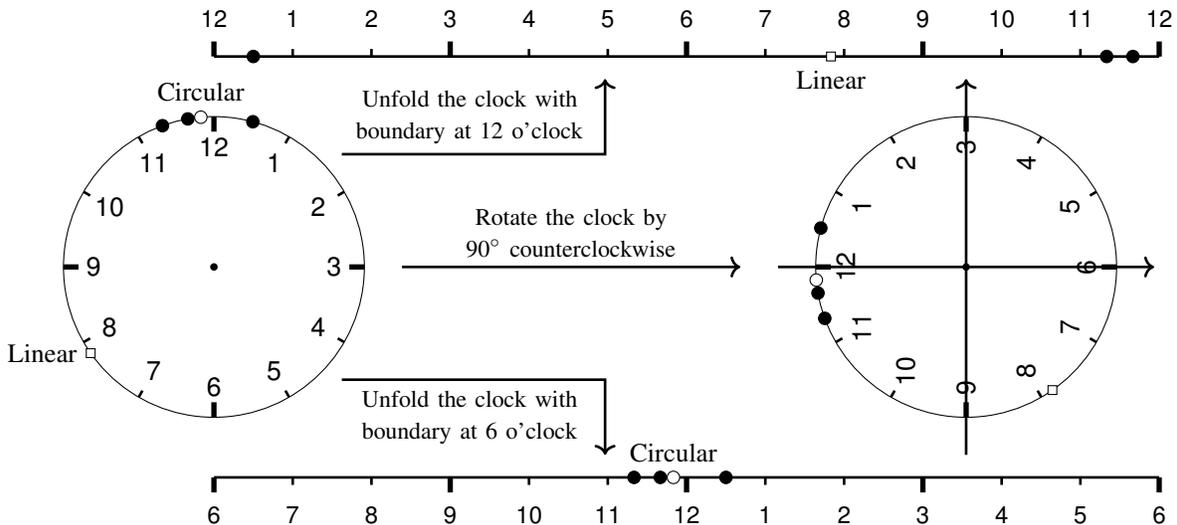

An example of circular data would be the clock in Fig.~\ref{fig:clockExample}, which, rotated by 90$^\circ$ counterclockwise, can be identified with the unit circle on the complex plane.
What should the mean of 11:20, 11:40, and 00:30 be? Averaging in a linear sense, it is $(\text{11:20}+\text{11:40}+\text{00:30})/3 = \text{07:50}$, %which is not sensible. It
while it rather should be $(\text{11:20}+\text{11:40}+\text{12:30})/3=11:50$; this requires unfolding the clock with boundary other than 12 o'clock, say 6 o'clock. 
%Rotated by 90$^\circ$ counterclockwise, the clock can be identified with the unit circle on the complex plane: 11:20, 11:40, and 00:30 correspond to $e^{-j2\pi\frac{320}{720}}$, $e^{-j2\pi\frac{340}{720}}$, and $e^{j2\pi\frac{330}{720}}$, respectively. This rotation aligns the branch cuts, 12 o'clock on the clock and $-1$ on the complex plane.

% \begin{algorithm}[t]
% \caption{Modification of Algorithm \ref{algo:estimateCoefficients}}\label{algo:circular}
% \begin{algorithmic}
% \Procedure{Estimate-Coefficients-Circular}{$y, \cM, \mu$}
% \While{$\cM \neq \emptyset$}
%     \State Pick a maximal element $\bmm\in \cM$
%     \State $u_{\bmm}(\bn) \!\gets\! \big[\bn\!\in\![\bN-\bmm]\big]\!{\bn+\bmm \choose \bmm}\!{\bN-\bn-\one \choose \bmm}/{\bN+\bmm\choose 2\bmm+\one}$
%     \State $\tau \gets \frac{1}{2\pi}\arg\!\big(\sum_\bn (\Pi\cD^{\bmm} y)(\bn)\big)$
%     \State $\hat{b}_{\bmm} \gets \big\{\sum_\bn u_{\bmm}(\bn) \big\{ \frac{1}{2\pi} \arg\!\big((\cD^{\bmm} y)(\bn)\big)\big\}_{\!\tau}\big\}_{\!0}$
%     \State $y(\bn) \gets y(\bn)\exp\!\big(\!-j2\pi \hat{b}_{\bmm}{\bn \choose \bmm} \big)$
%     \State $\cM \gets \cM \setminus \{\bmm\}$
% \EndWhile
% \State \Return $\{\hat{b}_\bmm\}$
% \EndProcedure
% \end{algorithmic}
% \end{algorithm}

\subsection{Averaging in a Circular Sense}
\label{sec:averagingInCircularSense}
To account for the circular nature, the averaging operator should exhibit a certain property, namely
\begin{align}
    \mu_{\bk}\big( e^{j2\pi\phi} s \big) = e^{j2\pi\phi} \mu_{\bk}(s) \label{circularMean}
\end{align}
for any $\phi\in\bbR$.
It is shown in App.~\ref{app:circularProof} that Alg. \ref{algo:estimateCoefficients} with an averaging operator abiding by \eqref{circularMean} satisfies
\begin{align}
\hat{b}_\bk(y(\bn)) - b_\bk-\hat{b}_\bk\big(y(\bn)e^{-j2\pi x(\bn)}\big) \in \bbZ \label{circularProperty}
\end{align}
for all $\bk\in\cM$. 
% From $\hat{b}_\bk\big(y(\bn)e^{-j2\pi x(\bn)}\big)\in \big[-\frac{1}{2},\frac{1}{2}\big)$, the above can be recast as
% \begin{align}
%     \hat{b}_\bk(y(\bn)) - b_\bk = \hat{b}_\bk\big(y(\bn)e^{-j2\pi x(\bn)}\big) + \big\lfloor \hat{b}_\bk(y(\bn)) - b_\bk \big\rceil. \label{circularPropertyRestated}
% \end{align}
% or, equivalently,
% \begin{align}
%     \exp\!\Big(j2\pi\Big(\hat{b}_\bmm\big(y(\bn)\big)-b_\bmm-\hat{b}_\bmm\big(y(\bn)e^{-j2\pi x(\bn)}\big)\Big)\Big) = 1.
% \end{align}
Recalling \eqref{signalModelMultiplicative}, we have that
\begin{align}
    y(\bn)e^{-j2\pi x(\bn)} = \big[\bn\in[\bN]\big] \big(1  + \tilde{w}_{\bbC}(\bn)\big),
\end{align}
whose distribution does not depend on the parameters being estimated.
% The same holds for the distribution of
% \begin{align}
%     \exp\!\big(j2\pi\big(\hat{b}_\bk(y(\bn))-b_\bk\big)\big) = \exp\!\big(j2\pi\hat{b}_\bk\big(y(\bn)e^{-j2\pi x(\bn)}\big)\big) .
% \end{align}
%That is, the performance of the estimator does not depend on the parameters, which is desirable.
The reconstruction error
\begin{align}
    &\sum_{\bn\in[\bN]} \! \big| e^{j2\pi \hat{x}(\bn)}-e^{j2\pi x(\bn)} \big|^2  \\
    &=\!\sum_{\bn\in[\bN]} \bigg| \exp\!\bigg(j2\pi\! \sum_\bmm (\hat{b}_\bmm(y(\bn))-b_\bmm){\bn\choose \bmm}\!\bigg)-1 \bigg|^2\!, \nonumber
\end{align}
which can be expanded as \eqref{reconstructionErrorReformulated} atop the next page,
\begin{figure*}
\begin{align}
    &\sum_{\bn\in[\bN]} \bigg| \exp\!\bigg[j2\pi \bigg(\sum_\bmm \hat{b}_\bmm\big(y(\bn)e^{-j2\pi x(\bn)}\big){\bn\choose \bmm}+\sum_\bmm \big( \hat{b}_\bmm(y(\bn)) - b_\bmm - \hat{b}_\bmm\big(y(\bn)e^{-j2\pi x(\bn)}\big) \big) {\bn\choose \bmm}\bigg)\bigg]-1 \bigg|^2 \label{reconstructionErrorReformulated}
\end{align}
\hrulefill
\end{figure*}
reduces, because  binomial coefficients are integer-valued, to
\begin{align}
    \sum_{\bn\in[\bN]} \!\bigg|\! \exp\!\bigg(\!j2\pi\! \sum_\bmm \hat{b}_\bmm\big(y(\bn)e^{-j2\pi x(\bn)}\big){\bn\choose \bmm}\!\bigg)\!-\!1 \bigg|^2\!. 
\end{align}
Consequently, the reconstruction error exhibits a distribution that does not depend on the parameters, as desired. 

For one-dimensional linear polynomials, two averaging operations satisfying \eqref{circularMean}, precisely
\begin{align}
    \mu_\bmm(s) = \sum_\bn u_{\bmm}(\bn)s(\bn) \label{Kay}
\end{align}
and
\begin{align}
    \mu_\bmm(s) = \sum_\bn u_{\bmm}(\bn)(\Pi s)(\bn), \label{LW}
\end{align}
are proposed in \cite[Eq. 18]{kay1989fast} and \cite[Sec. VI-E]{lovell1992statistical}, respectively.
Recalling that $(\Pi s)(\bn) = \exp\!\big(j \arg( s(\bn))\big)$, the latter
%For the latter operation, we can write
%\begin{align}
%    \arg(\mu_\bmm(s)) = \arg\!\bigg(\sum_\bn u_{\bmm}(\bn)\exp\!\big(j \arg( s(\bn))\big)\bigg).
%\end{align}
projects the data onto the unit circle and averages it thereon,
%and then turns the result back to the angular domain,
which aligns with the notion of circular mean \cite[Ch. 2]{mardia2009directional}.

Next, a new averaging operator is set forth that can handle arbitrary dimensionalities and orders, and that outperforms \eqref{Kay} and \eqref{LW} in one-dimensional settings.

%\heedong{A subtle point is that, for random variables $X$ and $Y$, it is not generally true that $\exp(j2\pi a X)$ and $\exp(j2\pi a Y)$ are identically distributed even if $\exp(j2\pi X)$ and $\exp(j2\pi Y)$ are identically distributed. The simplest example would be $X \sim {\sf Uniform}([0,1])$ and $Y \sim {\sf Uniform}([0,2])$. For $a=1.5$, $\exp(j2\pi a X)$ is not uniformly distributed over the circle while $\exp(j2\pi a Y)$ is. This is because, in general, $\exp(j2\pi a X) \neq (\exp(j2\pi X))^a$ unless $a$ is integer. In this regard, I think the last step may help the readers although seasoned readers may not need it. Perhaps, we need to stress that the last step holds because binomial coefficients are integer-valued.}

\begin{figure*}
\centering
\subfloat[Compute the boundary]{
\begin{tikzpicture}
    \begin{scope}[shift={(10,0)},scale=0.8]
    \draw[line width=1pt, ->, opacity=0.5] (-2.5,0) -- (2.5,0);
    \draw[line width=1pt, ->, opacity=0.5] (0,-2.5) -- (0,2.5);
    \begin{scope}[rotate=90]
    \fill (0,0) circle (0.05cm);
    \draw (0,0) circle [radius=2cm];
    
    \def\angleOne{110};  % 11:20
    \def\angleTwo{100};  % 11:40
    \def\angleThree{75}; % 00:30
    \def\angleCircular{95};
    \def\angleLinear{215};
    \coordinate (pointOne) at ({2*cos(\angleOne)},{2*sin(\angleOne)});
    \coordinate (pointTwo) at ({2*cos(\angleTwo)},{2*sin(\angleTwo)});
    \coordinate (pointThree) at ({2*cos(\angleThree)},{2*sin(\angleThree)});
    \coordinate (pointCircular) at ({2*cos(\angleCircular)},{2*sin(\angleCircular)});
    \coordinate (pointBranchCut) at ({-(2*cos(\angleOne)+2*cos(\angleTwo)+2*cos(\angleThree))/sqrt((cos(\angleOne)+cos(\angleTwo)+cos(\angleThree))^2+(sin(\angleOne)+sin(\angleTwo)+sin(\angleThree))^2)},{-(2*sin(\angleOne)+2*sin(\angleTwo)+2*sin(\angleThree))/sqrt((cos(\angleOne)+cos(\angleTwo)+cos(\angleThree))^2+(sin(\angleOne)+sin(\angleTwo)+sin(\angleThree))^2)});
    
    % circular clock
    \foreach \angle [count=\xi] in {60,30,...,-270}
    {
      \draw[line width=1pt] (\angle:1.9cm) -- (\angle:2cm);
      \node[rotate=90] at (\angle:1.6cm) {\textsf{\xi}};
    }
    \foreach \angle in {0,90,180,270}
      \draw[line width=2pt] (\angle:1.8cm) -- (\angle:2cm);
    
    \node[draw,circle,fill=black,scale=0.5] at (pointOne) [] {};
    \node[draw,circle,fill=black,scale=0.5] at (pointTwo) [] {};
    \node[draw,circle,fill=black,scale=0.5] at (pointThree) [] {};
    \draw[white, fill=white] (0.15,-0.5) rectangle (0.65,0.5);
    \node[draw,diamond,fill=white,scale=0.5] at (pointBranchCut) [label= {[xshift=0.05cm, yshift=-0.15cm]above left:{\small New boundary}}] {};
    \end{scope}
    \end{scope}
\end{tikzpicture}
}\hspace{1mm}
\subfloat[Rotate so the boundary is at $-1$]{
\begin{tikzpicture}
    \def\angleRotate{174.9219}
    \begin{scope}[shift={(10,0)},scale=0.8]
    \draw[line width=1pt, ->, opacity=0.5] (-2.5,0) -- (2.5,0);
    \draw[line width=1pt, ->, opacity=0.5] (0,-2.5) -- (0,2.5);
    \begin{scope}[rotate={90+\angleRotate}]
    \fill (0,0) circle (0.05cm);
    \draw (0,0) circle [radius=2cm];
    
    \def\angleOne{110};  % 11:20
    \def\angleTwo{100};  % 11:40
    \def\angleThree{75}; % 00:30
    \def\angleCircular{95};
    \def\angleLinear{215};
    \coordinate (pointOne) at ({2*cos(\angleOne)},{2*sin(\angleOne)});
    \coordinate (pointTwo) at ({2*cos(\angleTwo)},{2*sin(\angleTwo)});
    \coordinate (pointThree) at ({2*cos(\angleThree)},{2*sin(\angleThree)});
    \coordinate (pointCircular) at ({2*cos(\angleCircular)},{2*sin(\angleCircular)});
    \coordinate (pointBranchCut) at ({-(2*cos(\angleOne)+2*cos(\angleTwo)+2*cos(\angleThree))/sqrt((cos(\angleOne)+cos(\angleTwo)+cos(\angleThree))^2+(sin(\angleOne)+sin(\angleTwo)+sin(\angleThree))^2)},{-(2*sin(\angleOne)+2*sin(\angleTwo)+2*sin(\angleThree))/sqrt((cos(\angleOne)+cos(\angleTwo)+cos(\angleThree))^2+(sin(\angleOne)+sin(\angleTwo)+sin(\angleThree))^2)});
    
    % circular clock
    \foreach \angle [count=\xi] in {60,30,...,-270}
    {
      \draw[line width=1pt] (\angle:1.9cm) -- (\angle:2cm);
      \node[rotate={90+\angleRotate}] at (\angle:1.6cm) {\textsf{\xi}};
    }
    \foreach \angle in {0,90,180,270}
      \draw[line width=2pt] (\angle:1.8cm) -- (\angle:2cm);
    
    \node[draw,circle,fill=black,scale=0.5] at (pointOne) [] {};
    \node[draw,circle,fill=black,scale=0.5] at (pointTwo) [] {};
    \node[draw,circle,fill=black,scale=0.5] at (pointThree) [] {};
    \node[draw,diamond,fill=white,scale=0.5] at (pointBranchCut) [] {};
    \end{scope}
    \end{scope}
\end{tikzpicture}
}\hspace{1mm}
\subfloat[Average in the linear sense]{
\begin{tikzpicture}
    \def\angleRotate{174.9219}
    \begin{scope}[shift={(10,0)},scale=0.8]
    \draw[line width=1pt, ->, opacity=0.5] (-2.5,0) -- (2.5,0);
    \draw[line width=1pt, ->, opacity=0.5] (0,-2.5) -- (0,2.5);
    \begin{scope}[rotate={90+\angleRotate}]
    \fill (0,0) circle (0.05cm);
    \draw (0,0) circle [radius=2cm];
    
    \def\angleOne{110};  % 11:20
    \def\angleTwo{100};  % 11:40
    \def\angleThree{75}; % 00:30
    \def\angleCircular{95};
    \def\angleLinear{215};
    \coordinate (pointOne) at ({2*cos(\angleOne)},{2*sin(\angleOne)});
    \coordinate (pointTwo) at ({2*cos(\angleTwo)},{2*sin(\angleTwo)});
    \coordinate (pointThree) at ({2*cos(\angleThree)},{2*sin(\angleThree)});
    \coordinate (pointCircular) at ({2*cos(\angleCircular)},{2*sin(\angleCircular)});
    \coordinate (pointBranchCut) at ({-(2*cos(\angleOne)+2*cos(\angleTwo)+2*cos(\angleThree))/sqrt((cos(\angleOne)+cos(\angleTwo)+cos(\angleThree))^2+(sin(\angleOne)+sin(\angleTwo)+sin(\angleThree))^2)},{-(2*sin(\angleOne)+2*sin(\angleTwo)+2*sin(\angleThree))/sqrt((cos(\angleOne)+cos(\angleTwo)+cos(\angleThree))^2+(sin(\angleOne)+sin(\angleTwo)+sin(\angleThree))^2)});
    
    % circular clock
    \foreach \angle [count=\xi] in {60,30,...,-270}
    {
      \draw[line width=1pt] (\angle:1.9cm) -- (\angle:2cm);
      \node[rotate={90+\angleRotate}] at (\angle:1.6cm) {\textsf{\xi}};
    }
    \foreach \angle in {0,90,180,270}
      \draw[line width=2pt] (\angle:1.8cm) -- (\angle:2cm);
    
    \node[draw,circle,fill=black,scale=0.5] at (pointOne) [] {};
    \node[draw,circle,fill=black,scale=0.5] at (pointTwo) [] {};
    \node[draw,circle,fill=black,scale=0.5] at (pointThree) [] {};
    \node[draw,circle,fill=white,scale=0.5] at (pointCircular) [] {};
    \node[draw,diamond,fill=white,scale=0.5] at (pointBranchCut) [] {};
    \end{scope}
    \end{scope}
\end{tikzpicture}
}\hspace{1mm}
\subfloat[Undo the rotation]{
\begin{tikzpicture}
    \begin{scope}[shift={(10,0)},scale=0.8]
    \draw[line width=1pt, ->, opacity=0.5] (-2.5,0) -- (2.5,0);
    \draw[line width=1pt, ->, opacity=0.5] (0,-2.5) -- (0,2.5);
    \begin{scope}[rotate=90]
    \fill (0,0) circle (0.05cm);
    \draw (0,0) circle [radius=2cm];
    
    \def\angleOne{110};  % 11:20
    \def\angleTwo{100};  % 11:40
    \def\angleThree{75}; % 00:30
    \def\angleCircular{95};
    \def\angleLinear{215};
    \coordinate (pointOne) at ({2*cos(\angleOne)},{2*sin(\angleOne)});
    \coordinate (pointTwo) at ({2*cos(\angleTwo)},{2*sin(\angleTwo)});
    \coordinate (pointThree) at ({2*cos(\angleThree)},{2*sin(\angleThree)});
    \coordinate (pointCircular) at ({2*cos(\angleCircular)},{2*sin(\angleCircular)});
    \coordinate (pointBranchCut) at ({-(2*cos(\angleOne)+2*cos(\angleTwo)+2*cos(\angleThree))/sqrt((cos(\angleOne)+cos(\angleTwo)+cos(\angleThree))^2+(sin(\angleOne)+sin(\angleTwo)+sin(\angleThree))^2)},{-(2*sin(\angleOne)+2*sin(\angleTwo)+2*sin(\angleThree))/sqrt((cos(\angleOne)+cos(\angleTwo)+cos(\angleThree))^2+(sin(\angleOne)+sin(\angleTwo)+sin(\angleThree))^2)});
    
    % circular clock
    \foreach \angle [count=\xi] in {60,30,...,-270}
    {
      \draw[line width=1pt] (\angle:1.9cm) -- (\angle:2cm);
      \node[rotate=90] at (\angle:1.6cm) {\textsf{\xi}};
    }
    \foreach \angle in {0,90,180,270}
      \draw[line width=2pt] (\angle:1.8cm) -- (\angle:2cm);
    
    \node[draw,circle,fill=black,scale=0.5] at (pointOne) [] {};
    \node[draw,circle,fill=black,scale=0.5] at (pointTwo) [] {};
    \node[draw,circle,fill=black,scale=0.5] at (pointThree) [] {};
    \node[draw,circle,fill=white,scale=0.5] at (pointCircular) [] {};
    \node[draw,diamond,fill=white,scale=0.5] at (pointBranchCut) [] {};
    \end{scope}
    \end{scope}
\end{tikzpicture}
}
\caption{Illustration of how the proposed averaging operator in \eqref{proposedAverager} processes circular data. For this example, uniform weights $u_\bk(\bn)$ are considered.}
\label{fig:proposedAverager}
\end{figure*}
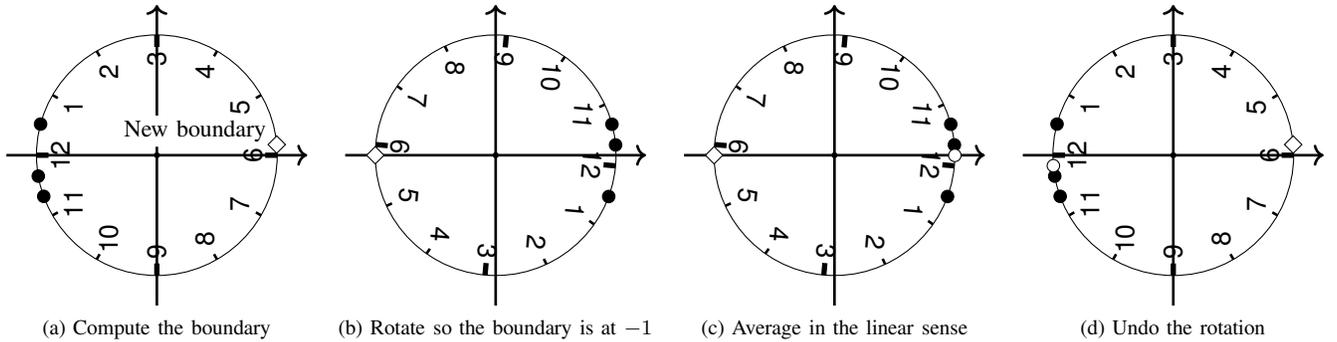

\begin{figure*}
    \centering
    \subfloat[$M=1$]
    {
        \includegraphics[width=0.29\linewidth]{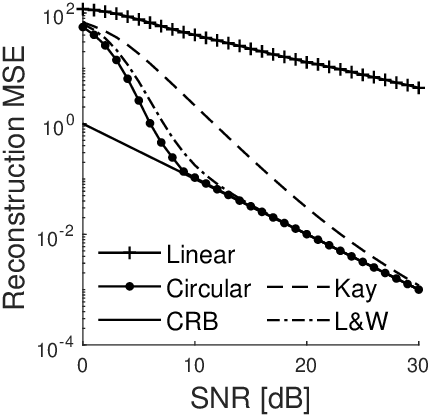}
    }
    \subfloat[$M=2$]
    {
        \includegraphics[width=0.29\linewidth]{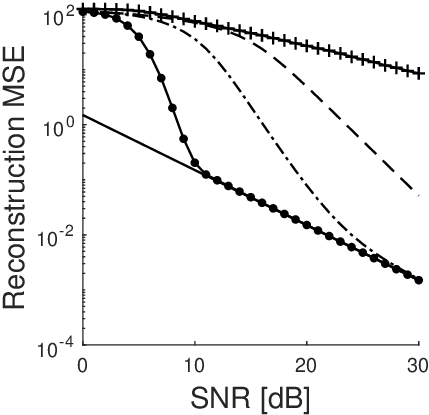}
    }
    \subfloat[$M=3$]
    {
        \includegraphics[width=0.29\linewidth]{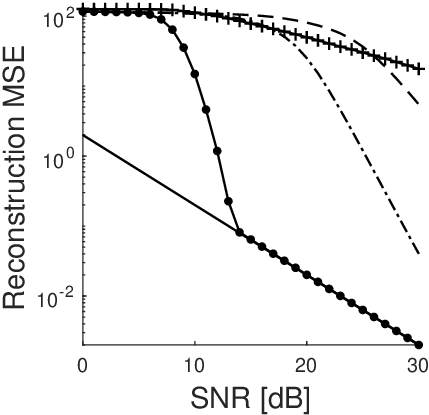}
    }
    \caption{Evaluation of Alg. \ref{algo:estimateCoefficients} with \eqref{proposedAverager} (labelled ``circular'') for $\sfD=1$ with $\cM=[M+1]$ and $N=64$. Also included are Alg. \ref{algo:estimateCoefficients} with \eqref{linearAverager} (labelled ``linear''), \eqref{Kay} (labelled ``Kay''), and \eqref{LW} (labelled ``L\&W'').
    For each one, the reconstruction error in \eqref{reconstructionMse} is averaged over ten thousands instances of noise and parameters with $\bb$ drawn uniformly over $[-\frac{1}{2},\frac{1}{2} )^{M+1}$.}
    \label{fig:modification}
\end{figure*}

\begin{figure*}
    \centering
    \subfloat[$\sfD=1, M=1, N\in\{16,64,256,1024\}$]
    {
        \includegraphics[width=0.29\linewidth]{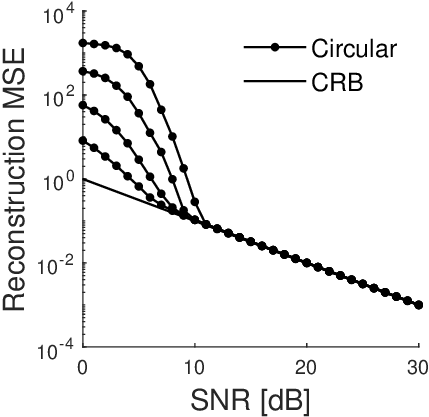}
    }
    \subfloat[$\sfD=1, M=2, N\in\{16,64,256,1024\}$]
    {
        \includegraphics[width=0.29\linewidth]{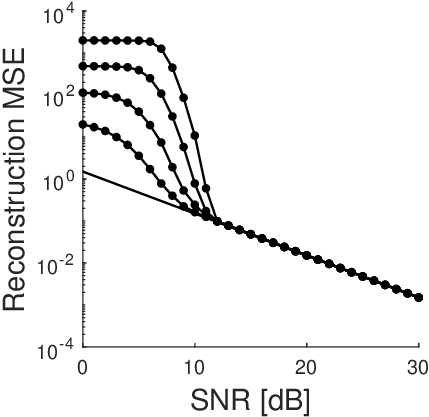}
    }
    \subfloat[$\sfD=1, M=3, N\in\{16,64,256,1024\}$]
    {
        \includegraphics[width=0.29\linewidth]{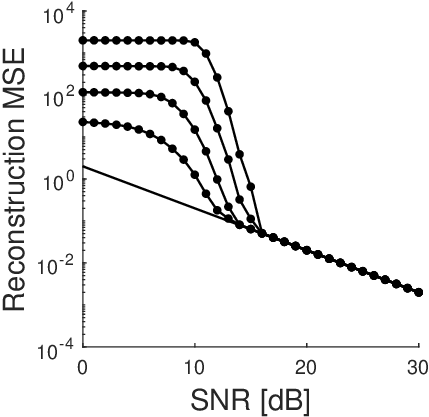}
    }\\
    \subfloat[$\sfD=2, M=1, N\in\{4,8,16,32\}$]
    {
        \includegraphics[width=0.29\linewidth]{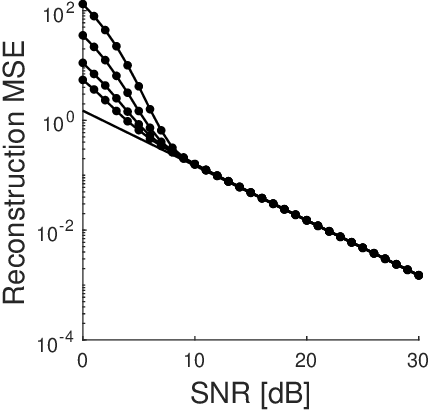}
    }
    \subfloat[$\sfD=2, M=2, N\in\{4,8,16,32\}$]
    {
        \includegraphics[width=0.29\linewidth]{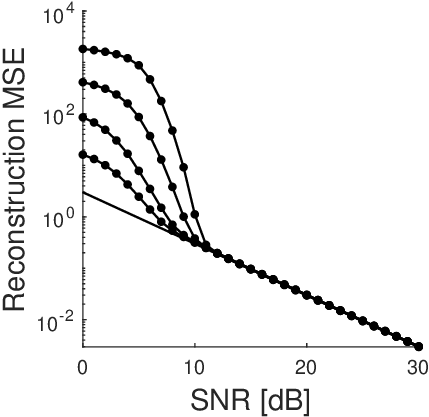}
    }
    \subfloat[$\sfD=2, M=3, N\in\{4,8,16,32\}$]
    {
        \includegraphics[width=0.29\linewidth]{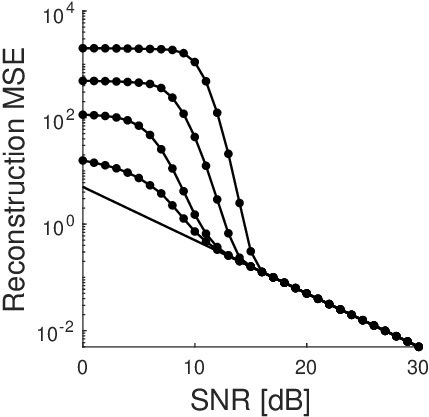}
    }\\
    \subfloat[$\sfD=3, M=1, N\in\{4,8,16,32\}$]
    {
        \includegraphics[width=0.29\linewidth]{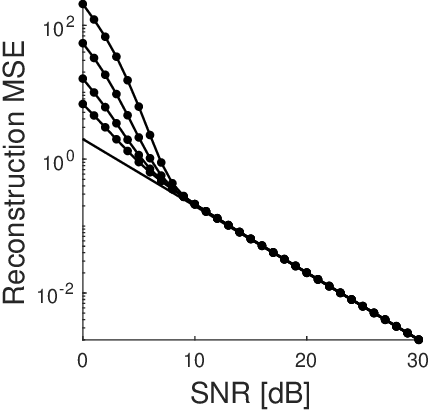}
    }
    \subfloat[$\sfD=3, M=2, N\in\{4,8,16,32\}$]
    {
        \includegraphics[width=0.29\linewidth]{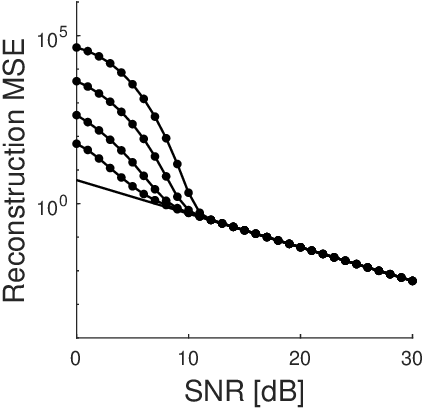}
    }
    \subfloat[$\sfD=3, M=3, N\in\{4,8,16,32\}$]
    {
        \includegraphics[width=0.29\linewidth]{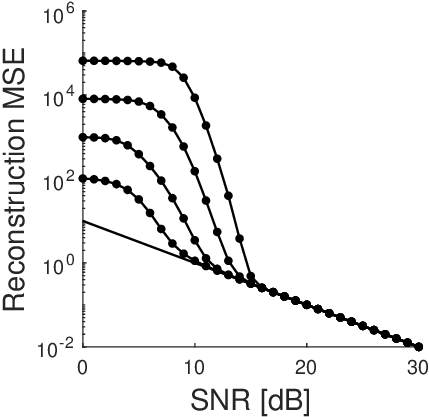}
    }\\
    \subfloat[$\sfD=4, M=1, N\in\{4,8,16,32\}$]
    {
        \includegraphics[width=0.29\linewidth]{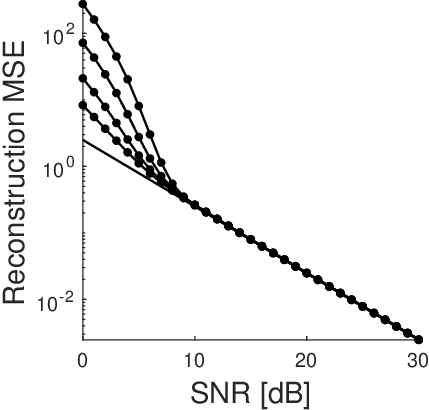}
    }
    \subfloat[$\sfD=4, M=2, N\in\{4,8,16,32\}$]
    {
        \includegraphics[width=0.29\linewidth]{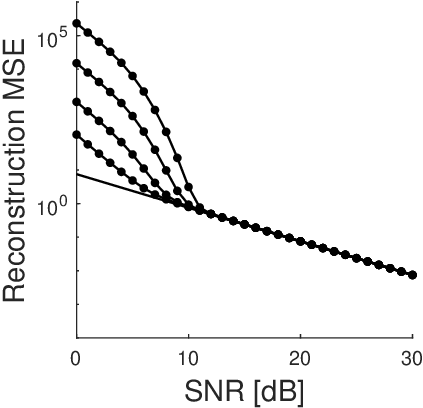}
    }
    \subfloat[$\sfD=4, M=3, N\in\{4,8,16,32\}$]
    {
        \includegraphics[width=0.29\linewidth]{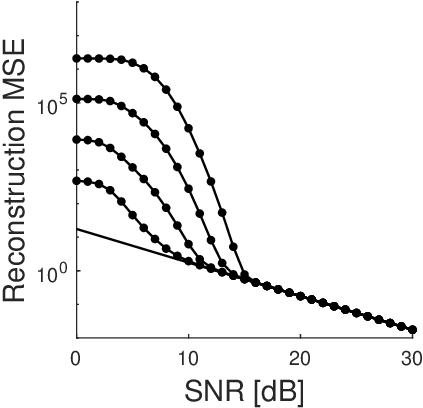}
    }
    \caption{Reconstruction performance of Alg. \ref{algo:estimateCoefficients} with \eqref{proposedAverager} in a variety of setups, $\bN = (N,\ldots,N)$ and $\cM = \{\bmm\in\bbN_0^{\sfD}: |\bmm|\leq M\}$. For each figure, the reconstruction MSE is averaged over ten thousands instances of noise and parameters with $\bb$ drawn uniformly over $[-\frac{1}{2},\frac{1}{2})^{M+1}$. In every case, the MSE worsens as $N$ increases.}
    \label{fig:numericalResults}
\end{figure*}

\subsection{Proposed Averaging Operator}

%\angel{The two new figures are great, really illuminating. Now let's see to the text. Before jumping into the proposed branch cut, perhaps we should introduce the idea of having a branch cut that is not fixed, but rather a function of the data, followed by some intuition that a desirable cut would be in some sense "away" from the data.}
Recalling the clock example in Fig. \ref{fig:clockExample}, intuition suggests that the boundary should be pushed away from the data. Thus, rather than being fixed, the boundary should be a function of the data itself.
One sensible choice for that function is the additive inverse of the %unweighted
circular mean,
\begin{align}
    -\arg\!\bigg(\sum_\bell (\Pi s)(\bell)\bigg).
\end{align}
This leads to the averaging operator
\begin{align}
    \mu_\bk(s) & = \Pi\bigg[\sum_\bell (\Pi s)(\bell)\bigg] \label{proposedAverager}\\
    &\quad \cdot \exp\!\bigg(j  \sum_\bn u_{\bk}(\bn)  \arg\!\bigg(s(\bn)\overline{\sum_\bell (\Pi s)(\bell)}\bigg)\bigg) , \nonumber 
\end{align}
which centers $\arg\!\big(s(\bn)\big)$ around $\arg\!\big(\sum_\bell (\Pi s)(\bell)\big)$ 
%\angel{We again have a bit of back-and-forth between the indices $\bell$ and $\bn$ here...} \heedong{I think, for the latter, $\bell$ seems to be better choice than $\bn$. In \eqref{twoIndices}, there's a nested summation which requires two indices. The index for the outer summation should be $\bn$. We can choose the inner one freely, and I just chose $\bell$.}
instead of zero, identically to \cite[Alg. 1]{madsen2019finite}.
Resorting again to the clock analogy, the functioning of the above averaging operator is illustrated in Fig. \ref{fig:proposedAverager}.

%  \heedong{I have also tested weighted circular mean but they performed almost identically, which is not unexpected. Consider 6 o'clock and 7 o'clock (as boundaries) in the clock example, which produces the identical result despite the one-hour difference. Generalizing the idea slightly, we can construct another averaging operator $\tilde{\mu}_\bk(s)$ from an averaging operator $\mu_\bk(s)$ satisfying \eqref{circularMean}:
% \begin{align*}
%     \tilde{\mu}_\bk(s) & = \mu_\bk(s) \exp\!\bigg(j  \sum_\bn u_{\bk}(\bn)  \arg\!\big(s(\bn)\overline{\mu_\bk(s)}\big)\bigg). 
% \end{align*}
% This corresponds to unfolding the clock with boundary determined in a circular sense.
% The resultant operation also satisfies \eqref{circularMean} as
% \begin{align*}
%     &\tilde{\mu}_\bk(e^{j2\pi\phi} s)\\
%     &= \mu_\bk(e^{j2\pi\phi}s) \exp\!\bigg(\!j  \sum_\bn u_{\bk}(\bn)  \arg\!\big(e^{j2\pi\phi}s(\bn)\overline{\mu_\bk(e^{j2\pi\phi}s)}\big)\!\bigg) \\
%     &=e^{j2\pi\phi}\tilde{\mu}_\bk( s),
% \end{align*}
% where the last step follows from the assumption that $\mu_\bk(s)$ satisfies \eqref{circularMean}.
% The proposed estimator is a special case! Is this generalization more of a digression?
% } \angel{It's both a nice generalization and a nice digression :-) In the balance, I'd say we skip it, for the sake of conciseness only}

The desired property in \eqref{circularMean} is satisfied as
\begin{align}
    \mu_\bmm(e^{j2\pi\phi} s)&=  \Pi\bigg[e^{j2\pi\phi}\sum_\bell  (\Pi s)(\bell)\bigg] \label{twoIndices}\\
    &\!\!\!\!\!\!\!\!\!\!\!\!\!\!\!\!\!\!\!\! \cdot \exp\!\bigg(j\sum_\bn u_{\bmm}(\bn)  \arg\!\bigg(e^{j2\pi\phi} s(\bn) \overline{e^{j2\pi\phi}\sum_\bell (\Pi s)(\bell)}\bigg)\bigg) \nonumber\\
    % &=e^{j2\pi\phi} \Pi\bigg[\sum_\bell  (\Pi s)(\bell)\bigg] \exp\bigg(j  \sum_\bn u_{\bmm}(\bn)  \arg\!\bigg( s(\bn) \overline{\sum_\bell (\Pi s)(\bell)}\bigg)\bigg)\\
    &=e^{j2\pi\phi}\mu_\bmm( s)
\end{align}
and it is clear that the computation entails linear time.

% The resulting modified estimator is embodied by Algorithm~\ref{algo:circular}. 

In Fig. \ref{fig:modification}, the proposed averaging operator in \eqref{proposedAverager}
is shown to compare favorably with its counterparts in \eqref{linearAverager}, \eqref{Kay}, and \eqref{LW}, especially when the polynomial degree is high.
For a more comprehensive evaluation of this proposed operator in a variety of setups,
readers are referred to Fig. \ref{fig:numericalResults}.

\subsection{Change of Basis Yet Again}
\label{sec:changeOfBasisYetAgain}

Throughout the paper, the binomial representation has been seen to be analytically most convenient. In most instances though, there is interest in mapping the estimates back to
%the signal model in \eqref{signalModel} with
\eqref{polynomialMonomial}.
%That said, in most applications, at the end of the day, one need to map it back. 
%Let us consider the estimation of $\{a_\bmm\}$ under the signal model \eqref{signalModel} with \eqref{newBasis}.

Recalling %the result in
Sec. \ref{sec:changeOfBasisRevisited}, a two-stage approach materializes: obtaining $\{\hat{b}_\bmm\}$ with Alg. \ref{algo:estimateCoefficients}, and then invoking the change-of-basis formula described in Alg.~\ref{algo:TransformBtoA} to reach $\{\hat{a}_\bmm\}$. 
% This procedure can be summarized into Algorithm \ref{algo:twoStage}.
Recalling Sec. \ref{sec:changeOfBasisRevisited}, we have that, by construction,
\begin{align}
    \hat{\ba}-\bT\hat{\bb}\in\bT\bbZ^{|\cM|}
\end{align}
and the reconstructed signal is therefore invariant under the change of basis
\begin{align}
   \exp\!\bigg[j2\pi\bigg(\sum_{\bmm} \hat{a}_{\bmm} \frac{\bn^\bmm}{\bmm!}\bigg)\bigg] = \exp\!\bigg[j2\pi\bigg(\sum_{\bmm} \hat{b}_{\bmm} {\bn \choose \bmm}\bigg)\bigg]. \label{trivialIdentity}
\end{align}
% Although the CRB depends on the basis \cite{kay1993fundamentals}, the reconstruction error does not.

Another natural approach is a direct one, summarized into Alg. \ref{algo:oneStage}. It is shown in App.~\ref{app:twoAlgorithmIdenticalProof} that both approaches, two-stage and direct, output identical results provided that the averaging operator satisfies \eqref{circularMean}, and this naturally generalizes to any other basis satisfying \eqref{goodBasis}.

% \begin{algorithm}[t]
% \caption{Estimating the polynomial coefficients for any basis with the two-stage approach}\label{algo:twoStage}
% \begin{algorithmic}
% \Procedure{Estimate-Two-Stage}{$y,\cM,\bT$}
% \State $\hat{\bb}\gets\textsc{Estimate-Coefficients}(y,\cM)$
% \vspace{1mm}
% \State $\hat{\ba}\gets\textsc{Compute-New-Coordinate}(\hat{\bb},\bT)$
% \State \Return $\{\hat{a}_\bmm\}$
% \EndProcedure
% \end{algorithmic}
% \end{algorithm}

\begin{algorithm}[t]
\caption{Direct approach to estimate polynomial coefficients on monomial basis}\label{algo:oneStage}
\begin{algorithmic}
\Procedure{Estimate-Coefficients-Direct}{$y, \cM$}
\For{$\bmm\in\cM$ (in descending order)}
    \State $\hat{a}_{\bmm} \gets \frac{1}{2\pi}\arg\!\big(\mu_{\bmm}(\cD^{\bmm} y)\big)$
    \vspace{1mm}
    \State $y(\bn) \gets y(\bn)\exp\!\big(\!-j2\pi \hat{a}_{\bmm}\frac{\bn^\bmm}{\bmm!} \big)$
\EndFor
\State \Return $\{\hat{a}_\bmm\}$
\EndProcedure
\end{algorithmic}
\end{algorithm}

\section{Lifting the Technical Condition in \eqref{containingAllLowerDegrees}}
\label{sec:LiftingTechnicalCondition}

When \eqref{containingAllLowerDegrees} does not hold, Alg. \ref{algo:estimateCoefficients} can be still applied with
\begin{align}
    \cM' = \bigcup_{\bmm\in\cM} [\bmm] \label{newSetOfDegrees}
\end{align}
in lieu of $\cM$. The set $\cM'$ satisfies \eqref{containingAllLowerDegrees} by construction, and $\cM' = \cM$ if and only if $\cM$ satisfies \eqref{containingAllLowerDegrees}.
After the application of Alg. \ref{algo:estimateCoefficients}, the additional step of enforcing $b_\bmm = 0$ for each $\bmm\in\cM'\setminus\cM$ could be taken, but
this section evidences that this na{\"i}ve approach can be highly suboptimal.
With a simple modification, though, this suboptimality can be corrected and the CRB attained at high SNR.
% \heedong{I find the notation $\cM'$ and $\bJ'$ a bit inconvenient (especially an extra parenthesis for $(\bJ')^{-1}$ is annoying). How about $\bar{\cM}$ and $\bar{\bJ}$?}

% All of the results presented in this section can be easily generalized to any polynomial basis. 

\subsection{Suboptimality of the Na{\"i}ve Approach}

On the one hand, the Fisher information matrix pertaining to $\{b_\bmm: \bmm\in\cM'\}$ is 
$\bJ'\in \bbR^{|\cM'|\times|\cM'|}$ with entries
\begin{align}
    [\bJ']_{\bmm,\bmm'} &= 8\pi^2 \SNR \! \sum_{\bn \in [\bN]}  {\bn \choose \bmm}{\bn \choose \bmm'}.
\end{align}
At high SNR, with the na{\"i}ve approach, the covariance matrix of the estimator becomes a submatrix of $(\bJ')^{-1}$, namely 
$
    \bE^\top(\bJ')^{-1}\bE,
$
where $\bE \in \bbR^{|\cM'|\times|\cM|}$ has entries
\begin{align}
    [\bE]_{\bmm,\bmm'} = [\bmm = \bmm' \in\cM].
\end{align}
On the other hand, the Fisher information matrix for $\{b_\bmm: \bmm\in\cM\}$ is $\bE^\top\bJ'\bE$, a submatrix of $\bJ'$. The CRB is thus
\begin{align}
    (\bE^\top\bJ'\bE)^{-1}. \label{newCrb}    
\end{align}
As per \cite[Thm. 7.7.15]{horn2012matrix}, 
\begin{align}
    \bE^\top(\bJ')^{-1}\bE \geq (\bE^\top\bJ'\bE)^{-1}, 
\end{align}
and the inequality is strict unless $\cM'=\cM$. In terms of the high-SNR reconstruction error,
recalling \eqref{usingMainResult},
the inequality becomes
\begin{align}
    \frac{1}{2\,\SNR}\tr(\bE^\top(\bJ')^{-1}\bE(\bE^\top\bJ'\bE))\geq \frac{|\cM|}{2\,\SNR}. \label{reconstructionErrorConstrained}
\end{align}

\subsection{A Concrete Example}

The inequality in \eqref{reconstructionErrorConstrained} is not only strict, but possibly loose.
Let us take the simplest example of a one-dimensional monomial, i.e., $\sfD =1$ and $\cM = \{M\}$, where \eqref{reconstructionErrorConstrained} is simply
\begin{align}
    \frac{[(\bJ')^{-1}]_{M,M} [\bJ']_{M,M}}{2 \, \SNR} \geq \frac{1}{2\, \SNR}.
    \label{YL}
\end{align}
% We have that
% \begin{align}
%     [\bJ^{-1}]_{\bmm,\bmm}[\bJ]_{\bmm,\bmm} &=\big(\bee_{\bmm}^\top\bJ^{-1}\bee_{\bmm}\big)\big(\bee_{\bmm}^\top\bJ\bee_{\bmm}\big)\\
%     &=\|\bJ^{-\frac{1}{2}}\bee_\bmm\|^2\|\bJ^{\frac{1}{2}}\bee_\bmm\|^2\\
%     &\geq \big((\bJ^{-\frac{1}{2}}\bee_\bmm)^\top(\bJ^{\frac{1}{2}}\bee_\bmm)\big)^2 =  1,
% \end{align}
% where the Cauchy-Schwarz inequality is applied in the last step.
% The equality holds if and only if $\bee_\bmm$ is an eigenvalue of $\bJ$, which is not the case. We therefore have
% \begin{align}
%     [\bJ^{-1}]_{\bmm,\bmm} > [\bJ]_{\bmm,\bmm}^{-1},
% \end{align}
% implying that we cannot attain the CRB with Algorithm \ref{algo:circular}. It essentially says that the CRB gets smaller as we have additional information that all lower-order coefficients are zero.

For this particular case, a quantitative analysis is available in the large-$N$ regime \cite{peleg1991cramer}.
From
\begin{align}
    &[\bJ']_{m,m'} = 8\pi^2\SNR \sum_{n \in [N]}  {n \choose m}{n \choose m'} \\
    &\,\, = 8\pi^2\SNR\bigg(\frac{N^{m+m'+1}}{m!m'!(m+m'+1)} +\cO \big( N^{m+m'} \big)\bigg), \label{hilbertForm}
\end{align}
we have that
\begin{align}
    [\bJ']_{M,M} = 8\pi^2\SNR \bigg(\frac{N^{2M+1}}{(M!)^2(2M+1)} + \cO \big( N^{2M} \big) \bigg). \label{JEntry}
\end{align}
Also from \eqref{hilbertForm},
\begin{align}
    \bJ' &= 8\pi^2\SNR \cdot N\diag\bigg(\frac{1}{0!},\frac{N}{1!},\frac{N^2}{2!},\ldots,\frac{N^{M}}{M!}\bigg) \label{largeN}\\
    &\qquad \cdot\big(\bH+\cO(N^{-1})\big)\diag\bigg(\frac{1}{0!},\frac{N}{1!},\frac{N^2}{2!},\ldots,\frac{N^{M}}{M!}\bigg),\nonumber
\end{align}
where $\bH$ is the Hilbert matrix whose entries are given by $[\bH]_{m,m'}=\frac{1}{m+m'+1}$.
Inverting \eqref{largeN} gives
\begin{align}
    (\bJ')^{-1} &= \frac{1}{8\pi^2\SNR}\!\cdot\! \frac{1}{N}\diag\bigg(\frac{0!}{1},\frac{1!}{N},\frac{2!}{N^2},\ldots,\frac{M!}{N^{M}}\bigg)\\
    &\quad \cdot\big(\bH^{-1}+\cO(N^{-1})\big)\diag\bigg(\frac{0!}{1},\frac{1!}{N},\frac{2!}{N^2},\ldots,\frac{M!}{N^{M}}\bigg) \nonumber
\end{align}
and, plugging in the closed form for $\bH^{-1}$ \cite{choi1983tricks},
\begin{align}
    &\big[\bH^{-1}\big]_{m,m'} = (-1)^{m+m'}(m+m'+1)\\
    &\qquad\qquad\quad \cdot{M+m+1 \choose M-m'}{M+m'+1 \choose M-m}{m+m' \choose m}^2, \nonumber
\end{align}
finally gives
\begin{align}
    &[(\bJ')^{-1}]_{M,M}  = \frac{1}{8\pi^2\SNR} \label{JinvEntry} \\
    &\qquad\qquad \cdot \bigg(\frac{(M!)^2(2M+1)}{N^{2M+1}}{2M \choose M}^2 \!\! + \cO(N^{-2M-2}) \!
 \bigg).\nonumber
\end{align}
Combining \eqref{JEntry} and \eqref{JinvEntry}, the reconstruction error in the left-hand side of \eqref{YL} converges, asymptotically in $N$, to
\begin{align}
    \frac{1}{2 \, \SNR}{2M \choose M}^2, \label{msenaive}
\end{align}
while the CRB-induced bound on the right-hand side is $\frac{1}{2\,\SNR}$. The factor
\begin{align}
{2M \choose M}^{\! 2}
\end{align}
% \begin{equation}
% {2M \choose M}^{\! 2}
% \end{equation}
quantifies the suboptimality of the na{\"i}ve approach; for $M=3$, it amounts to ${6 \choose 3}^{\scriptscriptstyle 2}=400$, which represents a 26-dB penalty.
% Interestingly, without the additional step enforcing $b_\bmm = 0$ for each $\bmm\in\cM'\setminus\cM$, the reconstruction error is only $\frac{|\cM'|}{2\,\SNR}=\frac{M+1}{2\,\SNR}$, which is a 6-dB penalty for $M=3$. It implies that simply discarding the information from $\bmm\in\cM'\setminus\cM$ is inadvisable, and we need to make better use of this information.

Rather unintuitively, without the additional step of forcing $b_\bmm = 0$ for each $\bmm\in\cM'\setminus\cM$, the reconstruction error is more modest, namely $\frac{|\cM'|}{2\,\SNR}=\frac{M+1}{2\,\SNR}$, which represents a 6-dB penalty for $M=3$. The intuition behind the na{\"i}ve approach is decidedly deceiving.

\begin{algorithm}[t]
\caption{Extending Alg. \ref{algo:estimateCoefficients} to the cases where 
\eqref{containingAllLowerDegrees} does not hold}\label{algo:conditionNotHold}
\begin{algorithmic}
\Procedure{Estimate-Coefficients}{$y, \cM$}
\If{$\cM$ satisfies \eqref{containingAllLowerDegrees}}
    \For{$\bmm\in\cM$ (in descending order)}
    \State $\hat{b}_{\bmm} \gets \frac{1}{2\pi}\arg(\mu_{\bmm}(\cD^{\bmm} y))$
    \State $y(\bn) \gets y(\bn)\exp\!\big(\!-j2\pi \hat{b}_{\bmm}{\bn \choose \bmm} \big)$
    \EndFor
    \State \Return $\{\hat{b}_\bmm\}$
\Else
    \State $\cM' \gets \bigcup_{\bmm\in\cM} [\bmm]$
    \State $\hat{\bb}\gets\textsc{Estimate-Coefficients}(y,\cM')$
    \State $\hat{\bb}\gets (\bE^\top\bJ'\bE)^{-1}\bE^\top\bJ'\hat{\bb}$
    \State \Return $\{\hat{b}_\bmm\}$
\EndIf
\EndProcedure
\end{algorithmic}
\end{algorithm}

% We can close the remaining gap from the following high-SNR behavior:
% \begin{align}
%     \hat{\bb}\equiv (\hat{b}_0, \ldots, \hat{b}_{M-1}, \hat{b}_M)
%     \sim
%     \cN\big((0,\ldots,0,b_M),\bJ^{-1}\big).
% \end{align}
% The minimum-variance unbiased estimator for $b_M$ is then \cite[Thm. 7.5]{kay1993fundamentals}
% \begin{align}
%     \frac{\bee_M^\top\bJ}{\bee_M^\top\bJ\bee_M}\hat{\bb},
% \end{align}
% and its variance is approximately
% \begin{align}
%     \frac{1}{\bee_M^\top\bJ\bee_M} = [\bJ]_{M,M}^{-1},
% \end{align}
% which coincides with the CRB.

\subsection{Proposed Approach}

Let us return to the general problem. The na{\"i}ve approach (application of Alg. \ref{algo:estimateCoefficients} with $\cM'$ followed by the additional step of forcing $b_\bmm = 0$ for each $\bmm\in\cM'\setminus\cM$) is not satisfactory. To do better, let us start by recalling the CRB attainment of Alg. \ref{algo:estimateCoefficients} at high SNR and
the asymptotically Gaussian nature of the ensuing error
on the estimation of $\bb\in[-\frac{1}{2},\frac{1}{2})^{|\cM'|}$,
whereby the vectorization of $\hat{b}_\bk (y(\bn),\cM')$, denoted by $\hat{\bb}\big(y(\bn),\cM'\big)$, is asymptotically distributed as
\begin{equation}
     \cN\big(\bE \bb,(\bJ')^{-1}\big)
\end{equation}
%$\hat{\bb}\big(y(\bn),\cM'\big)$, follows $\cN\big(\bE \bb,(\bJ')^{-1}\big)$
at high SNR.
The corresponding minimum-variance unbiased estimator for $\{b_\bmm: \bmm\in\cM\}$ is \cite[Thm. 7.5]{kay1993fundamentals}
\begin{align}
    \hat{\bb}\big(y(\bn),\cM\big)\equiv (\bE^\top\bJ'\bE)^{-1}\bE^\top\bJ' 
\, \hat{\bb}\big(y(\bn),\cM'\big), \label{modifiedEstimator}
\end{align}
and its covariance is $(\bE^\top\bJ'\bE)^{-1}$ at high SNR,
which does coincide with the corresponding CRB---recall \eqref{newCrb}. This proposed approach is condensed into Alg. \ref{algo:conditionNotHold}. Note that \eqref{modifiedEstimator} may not lie in $[-\frac{1}{2},\frac{1}{2})^{|\cM|}$, but one can use 
\begin{align}
    \hat{\bb}\big(y(\bn),\cM\big)-\lfloor \hat{\bb}\big(y(\bn),\cM\big) \rceil
\end{align}
in its stead, with rounding applied component-wise.

With an averaging operator abiding by \eqref{circularMean}, it is shown in App. \ref{app:modifiedEstimatorProof} that, for all $\bk\in\cM'$,
\begin{align}
    & \!\!\!\!\!\!\!\!\!\!\! \hat{b}_\bk\big(y(\bn),\cM\big)-b_\bk-\hat{b}_\bk\big(y(\bn)e^{-j2\pi x(\bn)},\cM\big) \label{modificationCircular}\\
    &=\hat{b}_\bk\big(y(\bn),\cM'\big)-b_\bk-\hat{b}_\bk\big(y(\bn)e^{-j2\pi x(\bn)},\cM'\big), \nonumber
\end{align}
which is an integer---recall \eqref{circularProperty}. Repeating the argument in Sec. \ref{sec:averagingInCircularSense}, the distribution of reconstruction error can be seen not to depend on the parameters being estimated. 
Altogether then, \eqref{modifiedEstimator} is highly superior to the na{\"i}ve approach and behaves as desired.

\begin{figure}
    \centering
    \includegraphics[width=0.9\linewidth]{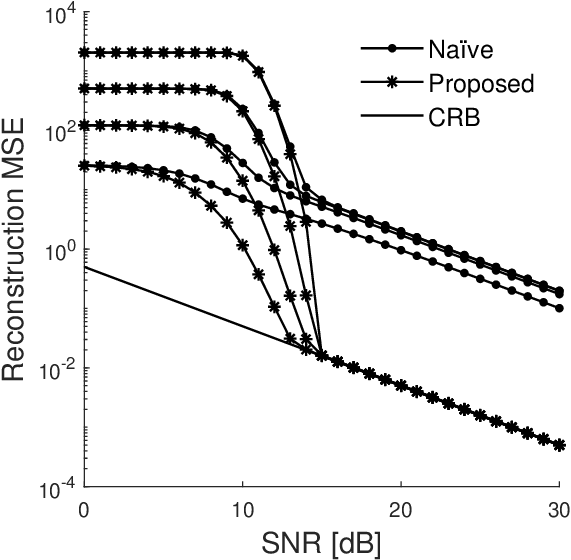}
    \caption{Evaluation of the na{\"i}ve approach and the proposed approach for $\sfD=1$ with $\cM=\{3\}$ and $N\in\{16,64,256,1024\}$. The reconstruction error in \eqref{reconstructionMse} is averaged over ten thousand noise and parameter instances with $b_3$ uniform over $[-\frac{1}{2},\frac{1}{2})$. In every case, the MSE worsens as $N$ increases.}
    \label{fig:conditionNotHold}
\end{figure}

Fig. \ref{fig:conditionNotHold} compares the na{\"i}ve approach, 
the proposed approach,
% Alg.~\ref{algo:conditionNotHold},
%Alg. \ref{algo:estimateCoefficients}, Alg. \ref{algo:conditionNotHold},
and the high-SNR reconstruction error derived from the CRB.

\section{Incorporating Lag Parameters}
\label{sec:introducingLagParameters}

Presented in \cite{peleg1995discrete} is a phase difference operator that is
slightly different to the one in \eqref{phaseDifferenceOperator}, namely
\begin{align}
    (\cD_d y)(\bn) \equiv (E_d^{\tau_d} y)(\bn)\overline{ y(\bn)}, \nonumber
\end{align}
which features an additional integer parameter $\tau_d>0$ termed \textit{delay} or \textit{lag}. In this section, Alg. \ref{algo:estimateCoefficients} is extended by incorporating this lag parameter.

Let us define $\btau = (\tau_0,\ldots,\tau_{\sfD-1})\in\bbN^\sfD$ and 
\begin{align}
    \cD_{\btau}^\bmm \equiv \cD_0^{m_0}\cdots\cD_{\sfD-1}^{m_{\sfD-1}},
\end{align}
where the notation $\cD_{\btau}^\bmm$ is used, to distinguish from $\cD^\bmm$.
The results established in Sec. \ref{sec:polynomialPhaseEstimation} can be analogously derived. In particular, \eqref{highSnrNoWrapping} becomes
\begin{align}
    & \!\!\!\! \frac{1}{2\pi}\arg \! \big( (\cD_{\btau}^{\bk} y)(\bn) \big) \label{lagSystemModel}\\
    &\quad \approx  \big[\bn\in[\bN-\btau\circ\bk]\big] \big( \btau^{\bk} b_{\bk} + ((E^{\btau}-1)^{\bk}w)(\bn)\big), \nonumber 
\end{align}
where $\circ$ denotes element-wise multiplication.
From \eqref{lagSystemModel}, a loss in identifiability is incurred for $\tau_d > 1$ \cite{mckilliam2009identifiability}. Precisely, even without noise, there is no ambiguity only if 
\begin{align}
    b_\bmm \in \tfrac{1}{\btau^\bmm}\left[ -\tfrac{1}{2},\tfrac{1}{2}\right) \label{lagIdentifiability}
\end{align}
for all $\bmm$.
In comparison with the result in Sec.~\ref{sec:integerValuedPolynomials}, there is an additional multiplicative factor that shrinks the cell and renders the algorithm more restrictive. 

\begin{figure*}
    \centering
    \subfloat[$\cM=\{0,1\}$]
    {
        \includegraphics[width=0.29\linewidth]{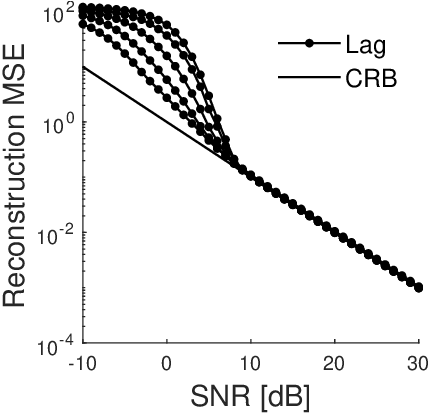}
    }
    \subfloat[$\cM=\{0,1,2\}$]
    {
        \includegraphics[width=0.29\linewidth]{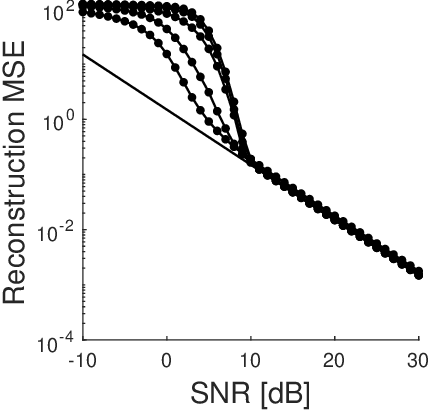}
    }
    \subfloat[$\cM=\{0,1,2,3\}$]
    {
        \includegraphics[width=0.29\linewidth]{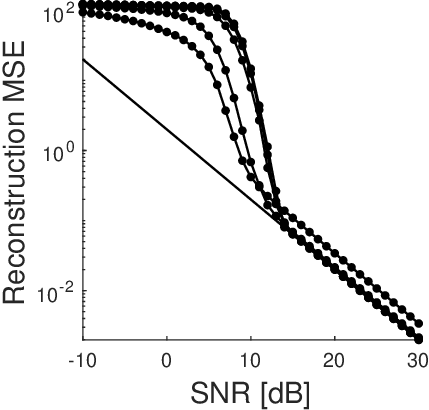}
    }
    \caption{Evaluation of Alg. \ref{algo:lagTest} for $\sfD=1$ with $N=64$. The zero polynomial (i.e., $\bb=\zero$) is considered. The reconstruction MSE is averaged over ten thousands noise instances for each $\tau \in \{1, 2, 4, 8, 16\}$. In every case, the MSE at low SNR improves as $\tau$ increases.}
    \label{fig:lagTest}
\end{figure*}

\begin{algorithm}[t]
\caption{Extension of Alg. \ref{algo:estimateCoefficients} to any lag parameter}\label{algo:lagTest}
\begin{algorithmic}
\Procedure{Estimate-Coefficients}{$y, \cM,\btau$}
\For{$\bmm\in\cM$ (in descending order)}
    \State $\hat{b}_{\bmm} \gets \frac{1}{2\pi \btau^\bmm}\arg(\mu_{\bmm}(\cD_{\btau}^{\bmm} y))$
    \State $y(\bn) \gets y(\bn)\exp\!\big(\!-j2\pi \hat{b}_{\bmm}{\bn \choose \bmm} \big)$
\EndFor
\State \Return $\{\hat{b}_\bmm\}$
\EndProcedure
\end{algorithmic}
\end{algorithm}

Provided that \eqref{lagIdentifiability} holds, the correspondence to Alg. \ref{algo:estimateCoefficients} with the averaging operator in \eqref{proposedAverager} is Alg. \ref{algo:lagTest} with the averaging operator
\begin{align}
    \mu_{\bk,\btau}(s) & = \Pi\bigg[\sum_\bell (\Pi s)(\bell)\bigg]\\
    &\quad \cdot \exp\!\bigg(j  \sum_\bn u_{\bk,\btau}(\bn)  \arg\!\bigg(s(\bn)\overline{\sum_\bell (\Pi s)(\bell)}\bigg)\bigg), \nonumber 
\end{align}
where
\begin{align}
    u_{\bk,\btau}(\bn) = \begin{cases}
        \bigg[\frac{\bC_{\bk,\btau}^{-1}\one}{\one^\top \bC_{\bk,\btau}^{-1}\one}\bigg]_\bn & \quad \bn\in[\bN-\btau\circ\bk]\\
        0 & \quad \text{otherwise}
    \end{cases}
    \label{weightLag}
\end{align}
with $\bC_{\bk,\btau}\in\bbR^{|[\bN-\btau\circ\bk]|\times|[\bN-\btau\circ\bk]|}$ the covariance matrix of
\begin{align}
    \big[\bn\in[\bN-\btau\circ\bk]\big]((E^{\btau}-1)^{\bk}w)(\bn). \label{noiseLag}
\end{align}
% The closed-form expression of the weights in \eqref{weightLag} can be found in App. \ref{app:weightComputation}.
As detailed in App. \ref{app:weightComputation} this variance can be expressed in a manner that enables its inversion, whereby \eqref{weightLag} unravels into
\begin{align}
    u_{\bk,\btau}(\bn) = \prod_{d=0}^{\sfD-1} u_{k_d,\tau_d}(n_d),
\end{align}
where
\begin{align}
    u_{k_d,\tau_d}(n_d) &\propto [n_d\in[N_d-\tau_d k_d]]\\
    &\quad \cdot {\lfloor \frac{n_d}{\tau_d} \rfloor + k_d\choose k_d} {\lceil \frac{N_d-n_d}{\tau_d} \rceil-1 \choose k_d} \nonumber
\end{align}
with the normalizing factor determined from $\sum_n u_{k,\tau}(n) = 1$. 

\begin{algorithm}[t]
\caption{Extension of Alg. \ref{algo:lagTest} to multiple lags}\label{algo:multiLag}
\begin{algorithmic}
\Procedure{Estimate-Coefficients}{$y, \cM, \cT$}
\For{$\bmm\in\cM$ (in descending order)}
    \State $\hat{b}_{\bmm} \gets 0$
    \For{$\btau\in\cT$ (in ascending order)}
        \State $\delta \gets \frac{1}{2\pi \btau^\bmm}\arg(\mu_{\bmm}(\cD_{\btau}^{\bmm} y))$
        \State $\hat{b}_{\bmm} \gets \hat{b}_{\bmm} + \delta$
        \State $y(\bn) \gets y(\bn)\exp\!\big(\!-j2\pi \delta{\bn \choose \bmm} \big)$
    \EndFor
\EndFor
\State \Return $\{\hat{b}_\bmm\}$
\EndProcedure
\end{algorithmic}
\end{algorithm}

\begin{figure*}
    \centering
    \subfloat[$\cM=\{0,1\}$]
    {
        \includegraphics[width=0.29\linewidth]{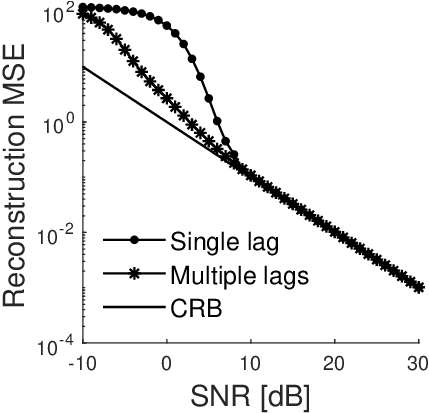}
    }
    \subfloat[$\cM=\{0,1,2\}$]
    {
        \includegraphics[width=0.29\linewidth]{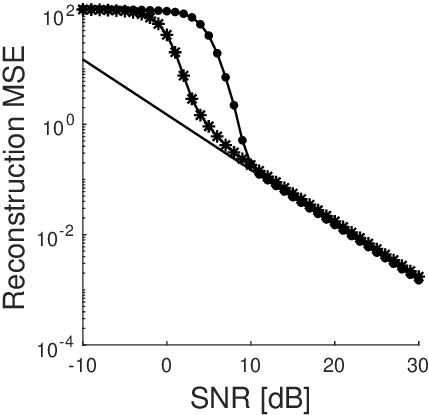}
    }
    \subfloat[$\cM=\{0,1,2,3\}$]
    {
        \includegraphics[width=0.29\linewidth]{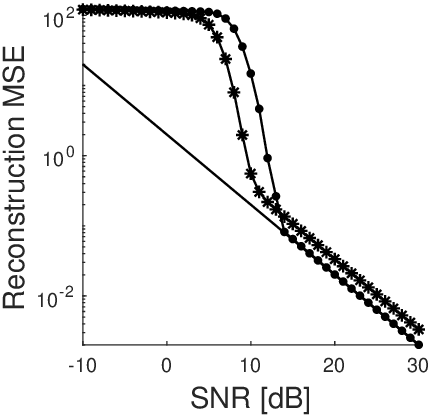}
    }
    \caption{Evaluation of Alg. \ref{algo:multiLag} for $\sfD=1$ with $N=64$ and $\cT = \{1, 2, 4, 8, 16\}$. The reconstruction MSE is averaged over ten thousands instances of noise and parameters with $\bb$ drawn uniformly over $[-\frac{1}{2},\frac{1}{2})^{M+1}$. Also shown is the reconstruction MSE with $\cT=\{1\}$.}
    \label{fig:multiLag}
\end{figure*}

% Before delving into how to efficiently compute the weights in \eqref{weightLag}, let us exemplify the reconstruction performance of Alg.~\ref{algo:lagTest}.

% Although non-unity lag parameter incurs loss in identifiability as well as a minute loss at high SNR, the resultant estimator compares favorably with the vanilla estimator at low to medium SNR, which is the main motivation of this modification.
% Even better, a repeated application with multiple lag parameters, which will be demonstrated at the end of the section, (i) removes the ambiguity originating from the additional multiplicative factor, and (ii) makes the algorithm's performance independent to the parameters being estimated. 

\begin{figure*}
    \centering
    \subfloat[$\sfD=2$]
    {
        \includegraphics[width=0.29\linewidth]{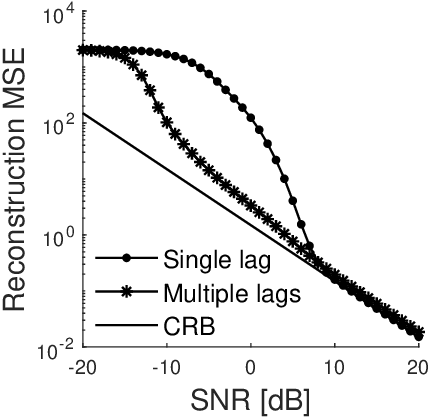}
    }
    \subfloat[$\sfD=3$]
    {
        \includegraphics[width=0.29\linewidth]{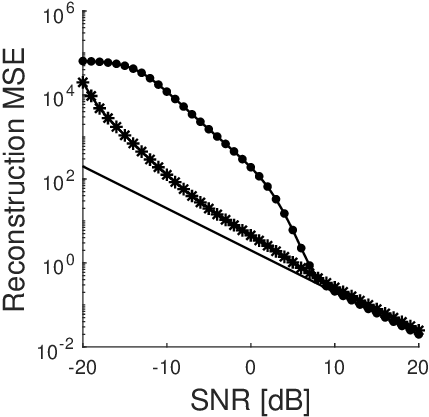}
    }
    \subfloat[$\sfD=4$]
    {
        \includegraphics[width=0.29\linewidth]{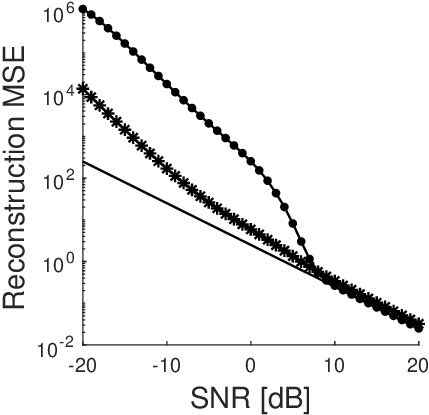}
    }
    \caption{Reconstruction performance of Alg. \ref{algo:multiLag} in multidimensional setups: $\bN=(32,\ldots,32)$, $\cM = \{\bmm\in\bbN_0^{\sfD}: |\bmm|\leq 1\}$, and $\cT = \{(1,\ldots,1), (2,\ldots, 2), (4,\ldots,4), (8,\ldots,8), (16,\ldots,16)\}$. The reconstruction MSE is averaged over ten thousands instances of noise and parameters with $\bb$ drawn uniformly over $[-\frac{1}{2},\frac{1}{2})^{|\cM|}$. Also shown is the reconstruction MSE with $\cT=\{(1,\ldots,1)\}$.}
    \label{fig:multiLagMultidimension}
\end{figure*}

Shown in  Fig.~\ref{fig:lagTest} is how $\tau_d > 1$ outperforms $\tau_d = 1$ for the zero polynomial ($\bb=\boldsymbol{0}$) at low-to-medium SNR, which is what motivates this extension.
This suggests potentially substantial improvements, albeit with \eqref{lagIdentifiability} standing in the way of a more general applicability.
% \angel{I've edited this, check it out}
% As mentioned, Alg. \ref{algo:lagTest} is contingent upon \eqref{lagIdentifiability}.
This hindrance can be sidestepped through repeated application of the procedure with multiple lags, starting with the all-ones lag.
The intuition is that the polynomial coefficient after cancelling out the estimate obtained with $\btau=\one$ is not large, and most often abides by \eqref{lagIdentifiability}.
%in the rare occasions in which it does not, a larger error is incurred, but in the mean-square sense the use of multiple lags is decidedly advantageous.
%\heedong{One thing I have to make sure is that ``in the rare occasions in which it does not, a larger error is incurred''; I'll check it out numerically.} \angel{Ok, maybe I jumped the gun. Let's see... We'll rephrase if it's not the case} \heedong{It's not the case; even if the coefficient lies outside of $\frac{1}{\btau^\bmm}[ -\tfrac{1}{2},\tfrac{1}{2})$, the additional error is bounded by $\frac{1}{\btau^\bmm}$, which is not catastrophic.}
The ensuing procedure is summarized in Alg.~\ref{algo:multiLag}, where $\cT\subset \bbN^{\sfD}$ is the ordered set of lags with minimum element $\one$.

Furthermore, denoting the estimate of $\hat{b}_\bk$ by $\hat{b}_\bk(y(\bn),\cT)$, 
it is shown in App. \ref{app:repeatedApplicationProof} that
\begin{align}
    \hat{b}_\bk(y(\bn),\cT) - b_\bk-\hat{b}_\bk\big(y(\bn)e^{-j2\pi x(\bn)},\cT\big) \in \bbZ
\end{align}
as long as \eqref{circularMean} holds.
That is, just like Alg. \ref{algo:estimateCoefficients}, the proposed estimator with multiple lags performs identically for every parameter. Figs. \ref{fig:multiLag} and \ref{fig:multiLagMultidimension} compare this multi-lag algorithm with its single-lag brethren (i.e., Alg. \ref{algo:estimateCoefficients}). For these numerical examples, which merely seek to evince the potential of the idea, the lags have been set heuristically. 
%which is enough for evidencing the potential of the idea.
Note that only a fraction of the gain observed in Fig. \ref{fig:lagTest} is realizable
once multiple lags are applied to ensure lack of ambiguity via \eqref{lagIdentifiability}.

%\angel{Readers may get confused about the difference between Fig. 10 and Figs. 11-12, which I guess is the issue of ambiguity} \heedong{Right. The result in Fig. 10 shows the potential of non-unity lag and it cannot be used in general case. In Fig. 11 (c), at SNR $>$ 10 dB, the absolute value of highest-order coefficient after passing $\btau=\one$ is less than $\frac{1}{16^3}$ with high probability, so the curve is almost identical to that of Fig. 10 (c) with $\tau=16$. But at SNRs around 7-10 dB, the absolute value does not reside between $\pm \frac{1}{16^3}$, but resides between $\pm \frac{1}{8^3}$ with high probability. So it follows the curve in Fig. 10 (c) with $\tau=8$.}

%\angel{If you wanna keep it short, we can simply content ourselves with evidencing the potential of the idea, indicating that we have picked the lags heuristically for the examples, posing as future work the development of a systematic procedure to determine $\cT$. Alternatively, we can think about providing some guidelines, if you have some ideas for that. This may be simpler for specific applications, such as channel estimation in LOS MIMO, so it is tempting to keep things short here and devote ourselves to finding a systematic procedure for that problem, on the basis of the SNR, numbers of antennas, etc.} \heedong{I'd go for the first option.}

% The rest of the section is devoted to the computation of the weights in \eqref{weightLag}.
% \heedong{Perhaps, it'd be better to make it as an appendix?} \angel{Yes, I agree} \heedong{Done!}

\section{Conclusion}
\label{sec:conclusion}

An estimation method for polynomial phase signals has been set forth that can handle an arbitrary number of dimensions and polynomial degrees, with the sole requirement that the number of observations along each dimension exceeds the highest degree thereon.
Embodied by Algs. \ref{algo:oneStage} and \ref{algo:conditionNotHold}, which feature an original circular averaging operator,
the estimator exhibits a linear complexity in the number of observations and attains the CRB at high SNR.
To reinforce the performance at low and medium SNRs, the estimates can be progressively refined through repeated application with varying lags. 
%where any phase estimator is hampered by the inherent ambiguity caused by phase wrappings, suitable functionalities are incorporated and shown to be highly effective.
Systematic procedures to establish such lags for specific incarnations of the estimation problem are an interesting avenue for follow-up work.

Other potential extensions include assessing the impact of quantization \cite{stoica2021cramer} and
the consideration of a sum of polynomial phase signals in lieu of a single one, i.e., a multi-component signal as opposed to a mono-component one. While the one-dimensional multi-component case is treated in \cite{barbarossa1998product}, the generalization to multiple dimensions remains open.

\appendices

\section{}
\label{app:secondConditionProof}
 From \eqref{containingAllLowerDegrees}, 
for $\bmm\in\cM$ it holds that
\begin{align}
    \{0\cdot \bee_d, 1\cdot \bee_d, \ldots, m_d\cdot \bee_d\} \subset \cM \label{ALS}
\end{align}
for all $d$.
% \angel{Wonder whether the meaning of $\bee_d$ will be obvious to readers} \heedong{It appears in Table \ref{table:priorArt}, so it'd be better to introduce it in Sec. \ref{sec:notations}.} \angel{Ok}
This, in turn, implies that
\begin{align}
    \bigg\{\big[\bn\in[\bN]\big]\frac{n_d^{0}}{0 !}, \ldots, \big[\bn\in[\bN]\big]\frac{n_d^{m_d}}{m_d !} \bigg\} 
\end{align}
are linearly independent, which implies linear independence of the one-dimensional signals
\begin{align}
    \bigg\{\big[n_d\in[N_d]\big]\frac{n_d^{0}}{0 !}, \ldots, \big[n_d\in[N_d]\big]\frac{n_d^{m_d}}{m_d !} \bigg\}.\label{oneDimensionalPolynomial}
\end{align}
As the space spanned by \eqref{oneDimensionalPolynomial} has a dimension of at most $N_d$, which is the dimension of the space of all signals over $[N_d]$, we have that
$N_d \geq m_d+1$ for all $d$.

\section{}
\label{FCB}

To show that \eqref{moreSampleThanDegree} is a sufficient condition for the linear independence of \eqref{monomialBasis},
it suffices to show that $x(\bn) = 0$ for all $ \bn\in[\bN]$ implies
\begin{align}
    b_{\bk} = \sum_{\bell\in[\bN]} (-1)^{|\bk+\bell|}{\bk \choose \bell} x(\bell) =0 
\end{align}
for all $\bk$.
Note that this, which would be rather obvious over $\bbN_0$, need not hold over $[\bN]$ without \eqref{moreSampleThanDegree}. As a one-dimensional example, $x(n) = {n \choose N}$ is a nonzero polynomial that is zero for every $n\in[N]$.
By virtue of \eqref{moreSampleThanDegree}, though, the range of summation in \eqref{binomialTransform} can indeed be restricted to $[\bN]$.

\section{}
\label{app:weightProof}
We can decompose $\bC_\bk$ as
\begin{align}
    [\bC_\bk]_{\bn,\bn'} = \frac{1}{8\pi^2\SNR}\prod_{d=0}^{\sfD-1}[\bC_{k_d}]_{n_d,{n_d}'} \label{essentiallyKronecker}
\end{align}
where $\bC_{k_d}\in\bbR^{(N_d-k_d)\times(N_d-k_d)}$ is given by
\begin{align}
    [\bC_{k_d}]_{n_d,{n_d}'} = (-1)^{n_d+{n_d}'}  {2k_d \choose n_d -{n_d}'+k_d} \label{covarianceEntries}
\end{align}
Provided that the inverse of $\bC_{k_d}$ exists for each $d$, %we have that
\begin{align}
    [\bC_\bk^{-1}]_{\bn,\bn'}=8\pi^2\SNR\prod_{d=0}^{\sfD-1}[\bC_{k_d}^{-1}]_{n_d,{n_d}'},
\end{align}
which follows from
\begin{align}
    &\sum_\bell \bigg([\bC_{\bk}]_{\bn,\bell} \cdot 8\pi^2\SNR\prod_{d=0}^{\sfD-1}[\bC_{k_d}^{-1}]_{\ell_d,{n_d}'}\bigg)\\
    % &\sum_\bell \prod_{d=0}^{\sfD-1}[\bC_{k_d}]_{n_d,{\ell_d}} \prod_{d=0}^{\sfD-1}[\bC_{k_d}^{-1}]_{\ell_d,{n_d}'}\\
    &\qquad =\sum_\bell \prod_{d=0}^{\sfD-1}[\bC_{k_d}]_{n_d,{\ell_d}} [\bC_{k_d}^{-1}]_{\ell_d,{n_d}'} \\
    &\qquad= \prod_{d=0}^{\sfD-1} \sum_{\ell_d} [\bC_{k_d}]_{n_d,\ell_d}[\bC_{k_d}^{-1}]_{\ell_d,{n_d}'} \label{generalDistributiveLaw}\\
    &\qquad= \prod_{d=0}^{\sfD-1} [\bC_{k_d}\bC_{k_d}^{-1}]_{n_d,{n_d}'}\\
    &\qquad= \prod_{d=0}^{\sfD-1} [n_d={n_d}']\\
    &\qquad=[\bn=\bn'] %\\
    %&\qquad=[\bI]_{\bn,\bn'}
\end{align}
where \eqref{generalDistributiveLaw} holds on account of the general distributive law \cite[Eq. 2.28]{graham1994concrete}. 

For $\bn\in[\bN-\bk]$, the application again of the general distributive law gives
\begin{align}
    \big[\bC_{\bk}^{-1}\one\big]_\bn = \prod_{d=0}^{\sfD-1} \big[\bC_{k_d}^{-1}\one\big]_{n_d}
    \label{plou}
\end{align}
and
   \begin{align} 
    \one^\top \bC_{\bk}^{-1}\one = \prod_{d=0}^{\sfD-1}  \one^\top\bC_{k_d}^{-1}\one.
    \label{molt}
\end{align}
% \begin{align}
%     \big[\bC_{\bk}^{-1}\one\big]_\bn &= \sum_\bell[\bC_{\bk}^{-1}]_{\bn,\bell}\\
%     &= \sum_\bell \prod_{d=0}^{\sfD-1} [\bC_{k_d}^{-1}]_{n_d,\ell_d}\\
%     &= \prod_{d=0}^{\sfD-1} \sum_{\ell_d} [\bC_{k_d}^{-1}]_{n_d,\ell_d}\\
%     &= \prod_{d=0}^{\sfD-1} \big[\bC_{k_d}^{-1}\one\big]_{n_d}\\
% \end{align}
% and
% \begin{align}
%     \one^\top \bC_{\bk}^{-1}\one &= \sum_\bn \prod_{d=0}^{\sfD-1} \big[\bC_{k_d}^{-1}\one\big]_{n_d}\\
%     &= \prod_{d=0}^{\sfD-1} \sum_{n_d}  \big[\bC_{k_d}^{-1}\one\big]_{n_d}\\
%     &= \prod_{d=0}^{\sfD-1}  \one^\top\bC_{k_d}^{-1}\one.
% \end{align}
It follows from \eqref{plou} and \eqref{molt} that
\begin{align}
    u_{\bk}(\bn) = \prod_{d=0}^{\sfD-1} u_{k_d}(n_d)
\end{align}
where
\begin{align}
    u_{k_d}(n_d)=
    \begin{cases}
        \bigg[\frac{\bC_{k_d}^{-1}\one}{\one^\top \bC_{k_d}^{-1}\one}\bigg]_{n_d} & \quad n_d\in[N_d-k_d]\\
        0 & \quad \text{otherwise}
    \end{cases}.    
\end{align}
It is stated in \cite{kitchen1994method} that, with the index $d$ omitted for brevity,
\begin{align}
    u_{k}(n) = \big[n\in[N-k]\big]\frac{{n+k \choose k}{N-n-1 \choose k}}{{N+k\choose 2k+1}} , 
\end{align}
with a proof of this result available in the earlier work \cite{hoskins1972some}. Particularly, \cite[Thm. 3]{hoskins1972some} states that
\begin{align}
    [\bC_{k}^{-1}\one]_{n} = \frac{{n+k \choose k}{N-n-1 \choose k}}{{2k\choose k}}. \label{sumOfColumns}
\end{align}
In \cite[Lemma 5]{hoskins1972some}, even an explicit formula for the inverse matrix is presented.
% , namely
% \begin{align}
%     \bC_{k}^{-1} = \bU_{k}\bU_{k}^\top, 
% \end{align}
% where $\bU_{k}\in \bbR^{|[N-k]|\times|[N-k]|}$ is an upper triangular matrix with entries
% \begin{align}
%     [\bU_{k}]_{n,\ell} = \frac{{n+k \choose k}{\ell-n+k-1 \choose k-1}}{\sqrt{{\ell+k\choose k}{\ell+2k\choose k}}} [n \leq \ell].
% \end{align}

\section{}
\label{fisherProof}

The log-likelihood is
\begin{align}
    -\SNR\sum_{\bn} \left| y(\bn)-e^{j2\pi x(\bn)} \right|^2,
\end{align}
and, using the chain rule, its partial derivative emerges as
\begin{align}
    & \!\!\!\!\!\!\!\! \frac{\partial}{\partial b_{\bmm}} \bigg(-\SNR\sum_{\bn} \left| y(\bn)-e^{j 2\pi x(\bn)} \right|^2 \bigg) \nonumber\\
    &=-\SNR \sum_{\bn} \frac{\partial}{\partial b_{\bmm}} \left| y(\bn)-e^{j2\pi x(\bn)} \right|^2 \\
    % &=-\SNR \sum_{\bn} \frac{\partial}{\partial b_{\bmm}}\Big(|y(\bn)|^2+1-2\operatorname{Re}\{y(\bn)\exp(-j2\pi x(\bn))\}  \Big) \nonumber\\
    &=-4\pi \, \SNR \sum_{\bn} \operatorname{Re} \left \{ j y(\bn) e^{-j2\pi x(\bn)} \right \} {\bn\choose \bmm}. 
\end{align}
The calculations in \eqref{fisher1}--\eqref{fisher5} conclude the proof, with \eqref{fisher4} holding because $w_\bbC(\bn)$ is circularly symmetric.

\begin{figure*}
\begin{align}
    [\bJ]_{\bmm,\bmm'}
    &=16\pi^2\SNR^2 \, \bbE\bigg[\sum_{\bn}  \operatorname{Re}\{j y(\bn)\exp(-j2\pi x(\bn))\} {\bn\choose \bmm} \cdot \sum_{\bn'}  \operatorname{Re}\{j y(\bn')\exp(-j2\pi x(\bn'))\} {\bn'\choose \bmm'} \bigg] \label{fisher1}\\
    &=16\pi^2\SNR^2 \, \sum_{\bn}\sum_{\bn'}{\bn\choose \bmm}{\bn'\choose \bmm'}  \bbE\Big[\operatorname{Re}\{j y(\bn)\exp(-j2\pi x(\bn))\}
    \operatorname{Re}\{j y(\bn')\exp(-j2\pi x(\bn'))\} \Big]\\
    &=16\pi^2\SNR^2 \, \sum_{\bn}\sum_{\bn'}{\bn\choose \bmm}{\bn'\choose \bmm'}  \bbE\Big[\operatorname{Re}\{j w_\bbC(\bn)\exp(-j2\pi x(\bn))\}
    \operatorname{Re}\{j w_\bbC(\bn')\exp(-j2\pi x(\bn'))\} \Big]\\
    &=16\pi^2\SNR^2 \, \sum_{\bn}\sum_{\bn'}{\bn\choose \bmm}{\bn'\choose \bmm'}  \frac{\big[\bn=\bn'\in [\bN]\big]}{2 \, \SNR} \label{fisher4}\\
    &=8\pi^2 \SNR \sum_{\bn \in [\bN]}  {\bn \choose \bmm}{\bn \choose \bmm'}. \label{fisher5}
\end{align}
\hrulefill
\end{figure*}

\section{}
\label{app:traceProof}
As $\bK\bJ$ is a product of two positive definite matrices, its eigenvalues are positive and real \cite[Cor. 7.6.2]{horn2012matrix}. Recalling that the trace and determinant of a matrix are the sum and product of its eigenvalues respectively, the inequality between the arithmetic and geometric means gives \cite[1.2.P19]{horn2012matrix}
\begin{align}
    \frac{1}{|\cM|}\tr(\bK\bJ) \geq \big(\det(\bK\bJ)\big)^{\frac{1}{|\cM|}} \label{amgm}
\end{align}
while, applying $t\geq 1+ \ln t$, it is alternatively found that 
\begin{align}
    \big(\det(\bK\bJ)\big)^{\frac{1}{|\cM|}} &\geq 1+ \ln \big( \det(\bK\bJ)\big)^{\frac{1}{|\cM|}} \\
    &=1+ \frac{1}{|\cM|}\ln \det(\bK\bJ) \label{tangent}
\end{align}
% From $\bK\geq \bJ^{-1}$, we have that $\det(\bK)\geq\det(\bJ^{-1})$ \cite[Cor. 7.7.4 (e)]{horn2012matrix}, hence
% \begin{align}
%     \det(\bK\bJ) = \frac{\det(\bK)}{\det(\bJ^{-1})} \geq 1. \label{determinant}
% \end{align}
The combination of \eqref{amgm} and \eqref{tangent} concludes the proof.

\section{}
\label{app:orthogonalityProof}
As derived in \cite[Eqn. 2.55]{graham1994concrete},
\begin{align}
    \Delta (f(n)g(n))
    &= f(n+1)g(n+1)-f(n)g(n)\\
    &=f(n+1)g(n+1)- f(n)g(n+1)\nonumber\\
    &\quad + f(n)g(n+1)-f(n)g(n)\\
    &=(\Delta f)(n) (Eg)(n) + f(n)(\Delta g)(n),
\end{align}
which can be rearranged as
\begin{align}
    f(n)(\Delta g)(n) = \Delta (f(n)g(n)) - (\Delta f)(n) (Eg)(n).
\end{align}
Summing over $[N]$ gives the ``summation by parts'' formula
\begin{align}
    \sum_{n\in[N]} f(n)(\Delta g)(n) = [f(n)g(n)]_{0}^{N} - \!\! \sum_{n\in[N]} (\Delta f)(n)(E g)(n) . \nonumber
    % &= f(\bN)g(\bN)-f(\zero)g(\zero) - \sum_{\bn\in[\bN]} (\Delta_d f)(\bn)(E_d g)(\bn) \nonumber.
\end{align}
Repeated summation by parts yields a formula that is the discrete analogue of tabular integration, namely
\begin{align}
    & \!\!\!\!\!\!\!\!\!\!\!\! \sum_{n\in[N]} f(n)(\Delta^m g)(n) \nonumber\\
    &= \Bigg[\sum_{k\in[m]} (-1)^{k}(\Delta^{k}f)(n) (\Delta^{m-k-1}g)(n) \Bigg]_0^N \nonumber\\
    &\quad +(-1)^{m} \sum_{n\in[N]} (\Delta^m f)(n)(E^m g)(n).\label{summationByParts}
\end{align}

\begin{figure*}
\begin{align}
     & \sum_{n\in[N]} \Delta^k\bigg[{n \choose k}{n-N \choose k}\bigg]\Delta^{k'}\bigg[{n \choose k'}{n-N \choose k'}\bigg] = (-1)^{k'} \sum_{n\in[N]} \Delta^{k+k'}\bigg[{n \choose k}{n-N \choose k}\bigg]E^{k'}\bigg[{n \choose k'}{n-N \choose k'}\bigg] \nonumber\\
    &\qquad\qquad\qquad\qquad\qquad\qquad\qquad +\Bigg[\sum_{\ell\in[k']} (-1)^{\ell}\Delta^{k+\ell}\bigg[{n \choose k}{n-N \choose k}\bigg]\Delta^{k'-\ell-1}\bigg[{n \choose k'}{n-N \choose k'}\bigg] \Bigg]_0
    ^N \label{polynomialInnerProduct}
\end{align}
\hrulefill
\end{figure*} 
Without loss of generality, let $k \leq k'$.
Applying \eqref{summationByParts}, $\langle p_k,p_{k'}\rangle$ becomes \eqref{polynomialInnerProduct}.  
Since the polynomial ${n \choose k'}{n-N \choose k'}$
is zero at $n \in [k']$ and $n \in [N+k']\setminus[N]$,
\begin{align}
    \Delta^{k'-\ell-1}\bigg[{n \choose k'}{n-N \choose k'}\bigg]
\end{align}
is zero at $0$ and $N$. Therefore, the latter term in \eqref{polynomialInnerProduct} equals zero.
As of the former term, from the observation that
\begin{align}
    {n \choose k}{n-N \choose k} 
\end{align}
is a polynomial of degree $2k$, if $k < k'$, then 
\begin{align}
    \Delta^{k+k'}\bigg[{n \choose k}{n-N \choose k}\bigg] = 0.
\end{align}
Otherwise, if $k = k'$, \eqref{polynomialInnerProduct} becomes
\begin{align}
     &(-1)^{k'} \sum_{n\in[N]} \Delta^{2k}\bigg[{n \choose k}{n-N \choose k}\bigg]{n+k \choose k}{n-N+k \choose k} \nonumber \\
     &=(-1)^{k'} {2k \choose k}\sum_{n\in[N]}{n+k \choose k}{n-N+k \choose k}\\
     &= {2k \choose k}\sum_{n\in[N]}{n+k \choose k}{N-n-1 \choose k}\\
     &= {2k \choose k}{N+k \choose 2k+1},
\end{align}
where use was made of summation by parts \eqref{summationByParts}, upper negation \cite[Table 174]{graham1994concrete}, and the Chu-Vandermonde identity \cite[Eqn. 5.26]{graham1994concrete}.

\section{}
\label{app:usefulIdentity1}
It is sufficient to prove a one-dimensional result, which can be done through the series of identities
\begin{align}
    q_{k}(n) & = \sum_{\ell} (-1)^{k+\ell}{k \choose \ell}{n+\ell \choose k}{n+\ell-N\choose k}\\
    &= \sum_{\ell} (-1)^{n+\ell}{k \choose \ell-n+k}{k+\ell \choose k}{k+\ell-N\choose k} \label{usefulIdentity1Step1}\\
    &= \sum_{\ell} (-1)^{n+\ell}{k \choose n-\ell}{k+\ell \choose k}{k+\ell-N\choose k} \label{usefulIdentity1Step2}\\
    &= \sum_{\ell} (-1)^{k+n+\ell}{\ell+k \choose k}{N-\ell-1 \choose k}{k \choose n-\ell} \label{usefulIdentity1Step3}
    % &= \!\sum_{\ell\in [N-k]}\! (-1)^{k+n+\ell}{\ell+k \choose k}\!{N-\ell-1 \choose k}\!{k \choose n-\ell}. \label{usefulIdentity1Step4}
\end{align}
where \eqref{usefulIdentity1Step1} follows from replacing $\ell$ by $\ell-n+k$, \eqref{usefulIdentity1Step2} follows from a symmetry relation \cite[Table 174]{graham1994concrete}, and \eqref{usefulIdentity1Step3} makes use of upper negation \cite[Table 174]{graham1994concrete}. Incorporating the boundary condition, $\ell\in [N-k]$:
\begin{itemize}
    \item For $\ell<-k$, we have that $n-\ell>k$ from $n\geq 0$, leading to ${k \choose n-\ell} = 0$.
    \item For $-k\leq \ell<0$, it is clear that ${\ell+k \choose k} = 0$.
    \item For $\ell \geq N-k$, we have that $N-\ell-1<k$, leading to ${N-\ell-1 \choose k} = 0$.
\end{itemize}

% The decomposition \eqref{rightHandSideDecomposition} break the multi-dimensional problem down into one-dimensional problems. From the symmetry, there is no loss of generality when letting $m\leq m'$.
% For brevity, we hereafter drop the index $d$ as long as no confusion arises. From \eqref{rightHandSideStart}-\eqref{rightHandSideEnd},
% it immediately follows that \eqref{rightHandSideDecomposition} equals
% \begin{align}
%     [\bmm=\bmm']{\bN+\bmm \choose 2\bmm+\one} {2\bmm \choose \bmm}.
% \end{align}
% In \eqref{rightHandSideBoundary1}, we can remove the boundary conditions for $k$ since
% \begin{itemize}
%     \item For $k<0$, ${m' \choose k-n'} = 0$ from $n'\geq 0$.
%     \item For $k\geq N$, ${m' \choose k-n'} = 0$ from $n'<N-m'$. 
% \end{itemize}
% Then, \eqref{rightHandSideVandermonde} follows from Chu-Vandermonde identity \cite[Eqn. 5.23]{graham1994concrete}. We can again remove the boundary condition for $n$ as
% \begin{itemize}
%     \item For $n<-m$, ${m+m' \choose m+n-n'} = 0$ from $n'\geq 0$
%     \item For $-m\leq n<0$, ${n+m \choose m} = 0$
%     \item For $N-m\leq n<N$, ${N-n-1 \choose m} = 0$
%     \item For $n \geq N$, ${m+m' \choose m+n-n'} = 0$ from $n'< N-m'$
% \end{itemize}
% resulting in \eqref{rightHandSideBoundary2}. Shifting the summation index, $n\gets n+n'-m$, gives \eqref{rightHandSideShift}. The inversion formula combined with the assumption $m \leq m'$ then gives \eqref{rightHandSideInversion}. \heedong{PENDING...}

\section{}
\label{app:usefulIdentity2}

Invoking summation by parts as per \eqref{summationByParts},
\begin{align}
    &\bigg\langle {n\choose m}, q_k \bigg\rangle\nonumber\\
    &= \sum_{n\in[N]} {n\choose m} \Delta^k\bigg[{n \choose k}{n-N \choose k}\bigg]\\
    &=\Bigg[\sum_{\ell\in[k]} (-1)^{\ell}{n\choose m-\ell} \Delta^{k-\ell-1}\bigg[{n \choose k}{n-N \choose k}\bigg] \Bigg]_0^N \nonumber\\
    &\quad +(-1)^{k} \sum_{n\in[N]} {n\choose m-k}{n+k \choose k}{n+k-N \choose k}. \label{innerProductSummationByParts}
\end{align}
As argued in App. \ref{app:orthogonalityProof},
\begin{align}
    \Delta^{k-\ell-1}\bigg[{n \choose k}{n-N \choose k}\bigg]
\end{align}
is zero at $0$ and $N$, hence \eqref{innerProductSummationByParts} equals
\begin{align}
    (-1)^{k} \sum_{n\in[N]} {n\choose m-k}{n+k \choose k}{n+k-N \choose k}.
\end{align}
Upper negation \cite[Table 174]{graham1994concrete} turns the above into
\begin{align}
    \sum_{n\in[N]} {n\choose m-k}{n+k \choose k}{N-n-1 \choose k}.
\end{align}
The range of summation $[N]$ can be replaced by $[N-k]$ as ${N-n-1 \choose k}=0$ for $n\in[N]\setminus[N-k]$, concluding the proof.

\section{}
\label{app:circularProof}

This appendix proofs \eqref{circularProperty} by induction on $\bk\in\cM$. Using \eqref{polynomialDifference}, \eqref{cauchyLike}, and \eqref{operatorRelation},
\begin{align}
    & \!\! e^{j2\pi\hat{b}_\bk(y(\bn))}\nonumber\\
    &=  \mu_\bk\bigg(\!\cD^{\bk}\!\bigg[y(\bn)\exp\!\bigg(\!\!-j2\pi \!\!\sum_{\bmm \succ \bk}\!\hat{b}_{\bmm}(y(\bn)){\bn \choose \bmm}\!\bigg)\!\bigg]\bigg) \\
    &=  \mu_\bk\bigg(\!(\cD^{\bk}y)(\bn)\exp\!\bigg(\!\!-j2\pi\!\! \sum_{\bmm \succ \bk}\hat{b}_{\bmm}(y(\bn)){\bn \choose \bmm-\bk}\!\bigg)\!\bigg) \label{recursion1}
\end{align}
and
\begin{align}
    & \!\!\!\!\! e^{j2\pi\hat{b}_\bk(y(\bn)e^{-j2\pi x(\bn)})}\nonumber\\
    &=\mu_\bk\bigg(\!\cD^\bk\bigg[y(\bn)e^{-j2\pi x(\bn)}\\
    &\quad\cdot\exp\!\bigg(\!-j2\pi \sum_{\bmm \succ \bk}\hat{b}_{\bmm}(y(\bn)e^{-j2\pi x(\bn)}){\bn \choose \bmm}\!\bigg)\!\bigg]\bigg)\nonumber\\
    &=\mu_\bk\bigg(\!(\cD^\bk y)(\bn)\exp\!\bigg(\!-j2\pi \sum_\bmm b_\bmm {\bn \choose \bmm-\bk}\!\bigg)\\
    &\quad\cdot\exp\!\bigg(\!-j2\pi \sum_{\bmm \succ \bk}\hat{b}_{\bmm}(y(\bn)e^{-j2\pi x(\bn)}){\bn \choose \bmm-\bk}\!\bigg)\!\bigg) \nonumber\\
    &=\mu_\bk\bigg(\!(\cD^\bk y)(\bn)e^{-j2\pi b_\bk} \exp\!\bigg(\!-j2\pi  \\
    &\quad \cdot \sum_{\bmm \succ \bk}\!\big(b_\bmm + \hat{b}_{\bmm}(y(\bn)e^{-j2\pi x(\bn)})\big){\bn \choose \bmm-\bk}\!\bigg)\!\bigg) \nonumber\\
    &=e^{-j2\pi b_\bk}\mu_\bk\bigg(\!(\cD^\bk y)(\bn)\exp\!\bigg(-j2\pi  \label{recursion2}\\
    &\quad \cdot \sum_{\bmm \succ \bk}\big(b_\bmm + \hat{b}_{\bmm}(y(\bn)e^{-j2\pi x(\bn)})\big){\bn \choose \bmm-\bk}\!\bigg)\!\bigg). \nonumber
\end{align}
Combining \eqref{recursion1} and \eqref{recursion2} gives \eqref{recursionCombined}.
\begin{figure*}
\begin{align}
    &\exp\!\Big(j2\pi\Big(\hat{b}_\bk\big(y(\bn)\big)-b_\bk-\hat{b}_\bk\big(y(\bn)e^{-j2\pi x(\bn)}\big)\Big)\Big) =\frac{\mu_\bk\Big((\cD^{\bk}y)(\bn)\exp\!\Big( \! -j2\pi \sum\limits_{\bmm \succ \bk} \! \hat{b}_{\bmm}(y(\bn)){\bn \choose \bmm-\bk}\Big)\Big)}{\mu_\bk\Big((\cD^\bk y)(\bn)\exp\!\Big( \! -j2\pi    \sum\limits_{\bmm \succ \bk} 
\! \big(b_\bmm + \hat{b}_{\bmm}(y(\bn)e^{-j2\pi x(\bn)})\big){\bn \choose \bmm-\bk}\Big)\Big)}\nonumber\\
    &=\frac{\mu_\bk\Big((\cD^{\bk}y)(\bn)\exp\!\Big(-j2\pi \sum\limits_{\bmm \succ \bk}\hat{b}_{\bmm}(y(\bn)){\bn \choose \bmm-\bk}\Big)\Big)}
    {\!\mu_\bk\Big((\cD^\bk y)(\bn)\exp\!\Big(\!-j2\pi\! \sum\limits_{\bmm \succ \bk}\!\hat{b}_{\bmm}(y(\bn)){\bn \choose \bmm-\bk}\!\Big)\exp\!\Big(j2\pi \!\sum\limits_{\bmm \succ \bk}\!\big(\hat{b}_{\bmm}(y(\bn))-b_\bmm - \hat{b}_{\bmm}(y(\bn)e^{-j2\pi x(\bn)})\big){\bn \choose \bmm-\bk}\!\Big)\!\Big)} \label{recursionCombined}
\end{align}
\hrulefill
\end{figure*}

For the highest order $\bk$, as desired, both the numerator and denominator of \eqref{recursionCombined} equal
\begin{align}
    \mu_\bk\big((\cD^\bk y)(\bn)\big).
\end{align}
Let us now assume that \eqref{circularProperty} holds for all $\bmm \succ \bk$.
From the induction hypothesis, the term
\begin{align}
    \sum\limits_{\bmm \succ \bk} \big(\hat{b}_{\bmm}(y(\bn))-b_\bmm - \hat{b}_{\bmm}(y(\bn)e^{-j2\pi x(\bn)})\big){\bn \choose \bmm-\bk}
\end{align}
in the denominator is integer-valued, as binomial coefficients are integer-valued.
Thus, both the numerator and denominator of \eqref{recursionCombined} reduce to
\begin{align}
    \mu_\bk\bigg(\!(\cD^{\bk}y)(\bn)\exp\!\bigg(\!-j2\pi\! \sum\limits_{\bmm \succ \bk}\!\hat{b}_{\bmm}(y(\bn)){\bn \choose \bmm-\bk}\!\bigg)\!\bigg),
\end{align}
which concludes the proof.

\section{}
\label{app:twoAlgorithmIdenticalProof}

%In this appendix, we will show that Algorithms \ref{algo:estimateCoefficients} and \ref{algo:oneStage} output an identical result in the sense that the reconstructed polynomials are identical.
Denoting the outputs of the two-stage and the direct approaches
by $\hat{\bb}$ and $\hat{\ba}$, respectively, it suffices to show that
\begin{align}
    \bT^{-1}\hat{\ba}-\hat{\bb} \in \bbZ^{|\cM|}
\end{align}
or, equivalently, that, for all $\bmm\in\cM$,
\begin{align}
    \exp \! \left( j2\pi \hat{b}_\bmm \right) = \exp \! \left(j2\pi\sum_{\bell} \tilde{t}_{\bmm,\bell} \hat{a}_\bell \right) ,
\end{align}
% \begin{align}
%     \sum_{\bell} \tilde{t}_{\bmm,\bell} \hat{d}_\bell - \hat{b}_\bmm \in \bbZ,
% \end{align}
where $\tilde{t}_{\bmm,\bell} \equiv [\bT^{-1}]_{\bmm,\bell}$. From this definition of $\tilde{t}_{\bmm,\bell}$,
\begin{align}
    \frac{\bn^\bell}{\bell!} = \sum_\bmm \tilde{t}_{\bmm,\bell}  {\bn \choose \bmm}.
\end{align}
Let us now proceed with induction on $\bmm$. The base case for the largest element of $\cM$ trivially holds. Let us assume that the statement holds for $\bmm \succ \bk$. Then,
\begin{align}
    \exp\!\big(j2\pi \hat{a}_\bk\big) =\mu_\bk\bigg(\cD^{\bk}\bigg[y(\bn)\exp \! \bigg( \! -j2\pi \sum_{\bell\succ\bk}\hat{a}_{\bell}\frac{\bn^\bell}{\bell!} \bigg)\bigg]\bigg), \label{temp1}
\end{align}
which can be recast as
\begin{align}
    &\cD^{\bk}\bigg[y(\bn)\exp \! \bigg( \! -j2\pi \sum_{\bell\succ\bk}\hat{a}_{\bell}\frac{\bn^\bell}{\bell!} \bigg)\bigg]\\
    &=\cD^{\bk}\bigg[y(\bn)\exp \! \bigg( \! -j2\pi  \sum_{\bell\succ\bk} \hat{a}_{\bell} \sum_\bmm \tilde{t}_{\bmm,\bell}  {\bn \choose \bmm} \bigg)\bigg]\\
    &=\cD^{\bk}\bigg[y(\bn)\exp \! \bigg( \! -j2\pi  \sum_\bmm {\bn \choose \bmm} \sum_{\bell\succ\bk}   \tilde{t}_{\bmm,\bell}\hat{a}_{\bell}   \bigg)\bigg]\\
    &= \big(\cD^{\bmm}y\big)(\bn)\exp \! \bigg(\!\! -j2\pi\!  \sum_\bmm {\bn \choose \bmm-\bk}\! \sum_{\bell\succ\bk}   \tilde{t}_{\bmm,\bell}\hat{a}_{\bell}  \!\bigg)\\
    &=\big(\cD^{\bk}y\big)(\bn)
    \exp \! \bigg( \! -j2\pi  \sum_{\bell\succ\bk}   \tilde{t}_{\bk,\bell}\hat{a}_{\bell} \bigg)\\
    &\quad \cdot \exp \! \bigg( \! -j2\pi  \sum_{\bmm\succ \bk} {\bn \choose \bmm-\bk} \sum_{\bell\succ\bk}   \tilde{t}_{\bmm,\bell}\hat{a}_{\bell} \bigg)\nonumber\\
    &=\big(\cD^{\bk}y\big)(\bn)
    \exp \! \bigg(\! -j2\pi  \sum_{\bell\succ\bk}   \tilde{t}_{\bk,\bell}\hat{a}_{\bell} \bigg)\\
    &\quad \cdot \exp \! \bigg(\! -j2\pi  \sum_{\bmm\succ \bk} {\bn \choose \bmm-\bk} \sum_{\bell}   \tilde{t}_{\bmm,\bell}\hat{a}_{\bell} \bigg)\nonumber\\    
    &=\big(\cD^{\bk}y\big)(\bn)
    \exp \! \bigg( \! -j2\pi  \sum_{\bell\succ\bk}   \tilde{t}_{\bk,\bell}\hat{a}_{\bell} \bigg) \label{temp2}\\
    &\qquad \cdot \exp \! \bigg( \! -j2\pi  \sum_{\bmm\succ \bk} \hat{b}_\bmm{\bn \choose \bmm-\bk}  \bigg).\nonumber
\end{align}
The second to last step follows from
\begin{align}
    [\bmm\succ \bk]\sum_{\bell\succ\bk} \tilde{t}_{\bmm,\bell}\hat{a}_{\bell} & =[\bmm\succ \bk]\sum_{\bell} [\bell\succ\bk][\bell \succeq \bmm]\tilde{t}_{\bmm,\bell}\hat{a}_{\bell} \nonumber \\
    &=\sum_{\bell} [\bell \succeq \bmm \succ \bk]\tilde{t}_{\bmm,\bell}\hat{a}_{\bell}\\
    &=[\bmm\succ \bk]\sum_{\bell} [\bell \succeq \bmm]\tilde{t}_{\bmm,\bell}\hat{a}_{\bell}\\
    &=[\bmm\succ \bk]\sum_{\bell} \tilde{t}_{\bmm,\bell}\hat{a}_{\bell},
\end{align}
where the upper triangularity of $\bT^{-1}$ was used. In turn, the last step is a direct consequence of the induction hypothesis.

Plugging \eqref{temp2} into \eqref{temp1}, the desired result is obtained,
\begin{align}
    e^{ j2\pi \hat{a}_\bk } & = \mu_\bk \bigg(\big(\cD^{\bmm}y\big)(\bn)
    \exp \! \bigg( \! -j2\pi  \sum_{\bell\succ\bk}   \tilde{t}_{\bk,\bell}\hat{a}_{\bell} \bigg)\\
    &\quad \cdot\exp \! \bigg( -j2\pi  \sum_{\bmm\succ \bk} \hat{b}_\bmm {\bn \choose \bmm-\bk}  \bigg)\bigg)\nonumber\\
    &= \exp \! \bigg( \! -j2\pi  \sum_{\bell\succ\bk}   \tilde{t}_{\bk,\bell}\hat{a}_{\bell} \bigg) 
    \\
    &\quad \cdot\mu_\bk \bigg(\big(\cD^{\bmm}y\big)(\bn)\exp \! \bigg( \! -j2\pi  \sum_{\bmm\succ \bk} \hat{b}_\bmm {\bn \choose \bmm-\bk}  \bigg)\bigg)\nonumber\\
    &= \exp \! \bigg( \! -j2\pi  \sum_{\bell\succ\bk}   \tilde{t}_{\bk,\bell}\hat{a}_{\bell} \bigg) e^{j2\pi \hat{b}_\bk}.
\end{align}

\section{}
\label{app:modifiedEstimatorProof}

This appendix presents a proof for \eqref{modificationCircular}.
For $\bk\in\cM'\setminus\cM$, \eqref{modificationCircular} holds because the left-hand side equals zero from
\begin{align}
    \hat{b}_\bk\big(y(\bn),\cM\big)=b_\bk=\hat{b}_\bk\big(y(\bn)e^{-j2\pi x(\bn)},\cM\big)=0
\end{align}
while the right-hand side,
\begin{align}
    \hat{b}_\bk\big(y(\bn),\cM'\big)-b_\bk-\hat{b}_\bk\big(y(\bn)e^{-j2\pi x(\bn)},\cM'\big),
\end{align}
is an integer in the interval $(-1,1)$ (recall \eqref{circularProperty} and the range of the estimators), which must be zero. This proves \eqref{modificationCircular} for $\bk\in\cM'\setminus\cM$, and implies that
\begin{align}
    & \hat{\bb}\big(y(\bn),\cM'\big)-\bE\bb-\hat{\bb}\big(y(\bn)e^{-j2\pi x(\bn)},\cM'\big) \label{zeroIndices}\\
    & \quad = \bE\bE^\top \big(\hat{\bb}\big(y(\bn),\cM'\big)-\bE\bb-\hat{\bb}\big(y(\bn)e^{-j2\pi x(\bn)},\cM'\big)\big). \nonumber
\end{align}
For $\bk\in\cM$, in turn, vectorization gives 
\begin{align}
    &\hat{\bb}\big(y(\bn),\cM\big)-\bb-\hat{\bb}\big(y(\bn)e^{-j2\pi x(\bn)},\cM\big) \nonumber\\
    &=(\bE^\top\bJ'\bE)^{-1}\bE^\top\bJ' \hat{\bb}\big(y(\bn),\cM'\big) - \bb \\
    &\quad -(\bE^\top\bJ'\bE)^{-1}\bE^\top\bJ'\hat{\bb}\big(y(\bn)e^{-j2\pi x(\bn)},\cM'\big) \nonumber \\
    &=(\bE^\top\bJ'\bE)^{-1}\bE^\top\bJ' \\
    &\quad \cdot \big(\hat{\bb}\big(y(\bn),\cM'\big)-\bE\bb- \hat{\bb}\big(y(\bn)e^{-j2\pi x(\bn)},\cM'\big)\big)\nonumber\\
    &=\bE^\top \!\big(\hat{\bb}\big(y(\bn),\cM'\big)\!-\!\bE\bb\!-\!\hat{\bb}\big(y(\bn)e^{-j2\pi x(\bn)},\cM'\big)\!\big),
\end{align}
where the last step follows from \eqref{zeroIndices}. It immediately follows from \eqref{circularProperty} that it is in $\bbZ^{|\cM|}$, which concludes the proof.

\section{}
\label{app:weightComputation}

\subsection{Weight Decomposition}

From
\begin{align}
  \!\!\!\!  ((E^{\btau}-1)^{\bk}w)(\bn) =\sum_\bell (-1)^{|\bk+\bell|}{\bk \choose \bell}w(\bn+\btau\circ \bell),
\end{align}
the covariance matrix of \eqref{noiseLag} can be expanded as
\begin{align}
    [\bC_{\bk,\btau}]_{\bn,\bn'}&=\frac{1}{8\pi^2\SNR}  \sum_{\bell}\sum_{\bell'}  (-1)^{|\bell+\bell'|}\\
    & \quad \cdot {\bk \choose \bell}{\bk \choose \bell'}[\bn+\btau\circ \bell = \bn'+\btau\circ \bell'] \nonumber\\
    &=\frac{1}{8\pi^2\SNR}  \prod_{d=0}^{\sfD-1}\sum_{\ell_d}\sum_{{\ell_d}'}  (-1)^{|\ell_d+{\ell_d}'|} \label{distributiveLawAnotherExample}\\
    & \quad \cdot {k_d \choose \ell_d}{k_d \choose {\ell_d}'}[n_d+\tau_d\ell_d = {n_d}'+\tau_d {\ell_d}'] \nonumber\\
    &=\frac{1}{8\pi^2\SNR}\prod_{d=0}^{\sfD-1}[\bC_{k_d}]_{n_d,{n_d}'},
\end{align}
where $\bC_{k_d,\tau_d}\in\bbR^{(N_d-\tau_d k_d)\times(N_d-\tau_d k_d)}$ has entries
\begin{align}
    [\bC_{k_d,\tau_d}]_{n_d,{n_d}'} &= \sum_{\ell_d}\sum_{{\ell_d}'}  (-1)^{|\ell_d+{\ell_d}'|} \label{covarianceEntriesLag}\\
    & \quad \cdot {k_d \choose \ell_d}{k_d \choose {\ell_d}'}[n_d+\tau_d\ell_d = {n_d}'+\tau_d {\ell_d}'] \nonumber
\end{align}
and the general distributive law was invoked in \eqref{distributiveLawAnotherExample}.

Recalling Sec. \ref{sec:estimationOfHighestOrderCoefficient}, one can invert $\bC_{\bk,\btau}$ in a dimension-wise fashion. Thus,
\begin{align}
    u_{\bk,\btau}(\bn) = \prod_{d=0}^{\sfD-1} u_{k_d,\tau_d}(n_d),
\end{align}
where
\begin{align}
    u_{k,\tau}(n) = \begin{cases}
        \bigg[\frac{\bC_{k,\tau}^{-1}\one}{\one^\top \bC_{k,\tau}^{-1}\one}\bigg]_n & \qquad n\in[N-\tau k]\\
        0 & \qquad \text{otherwise}
    \end{cases}.
\end{align}

\subsection{Covariance Computation}

For brevity, let us hereafter drop the index $d$.
Fortuitously, \eqref{covarianceEntriesLag} needs not be computed from scratch thanks to the observation that $[\bC_{k,\tau}]_{n,n'}=0$ unless
\begin{align}
    n \equiv n' \pmod \tau. \label{nonzeroCondition}
\end{align} 
Accordingly, \eqref{covarianceEntriesLag} can be recast as
\begin{align}
    [\bC_{k,\tau}]_{n,n'} &= [n\equiv n'\!\! \pmod \tau]\sum_{\ell}\sum_{\ell'}  (-1)^{\ell+\ell'} \\
    & \quad \cdot {k \choose \ell}{k \choose \ell}\bigg[\ell = \ell + \frac{n-n'}{\tau} \bigg] \nonumber\\
    &\!\!\!\!\!\!\!=[n\equiv n'\!\! \pmod \tau](-1)^{\frac{n-n'}{\tau}}{2k \choose k + \frac{n-n'}{\tau}}.
\end{align}

As a congruence relation is an equivalence relation, it naturally constructs a partition of $[N-\tau k]$. Specifically,
there are $\tau$ equivalence classes, namely,
\begin{align}
    \overline{0}\cap [N- \tau k], \overline{1}\cap [N-\tau k], \ldots, \overline{\tau-1}\cap [N-\tau k], \label{equivalenceClass}
\end{align}
where $\overline{\tau'} \equiv \{n\in \bbZ: n = \tau' \!\!\pmod \tau\}$ is the congruence class of $\tau'$ modulo $\tau$.
This implies that $\bC_{k}$ is essentially a block-diagonal matrix comprising $\tau$ blocks, each of the same form of \eqref{covarianceEntries}.

% Since $\btau|\bn-\bn'$ is an equivalence relation, we can construct a partition of the set $[\bN-\btau\circ\bk]$, namely
% \begin{align}
%     I_{\btau'} &\equiv \{\bn\in[\bN-\btau\circ\bk]: \btau|\bn-\btau'\} \nonumber
% \end{align}
% where $\btau'\in[\btau]$. It means that $\bC_{\bk}$ is essentially a block diagonal matrix. Let us consider the $\btau'$-th block.
% For $\bn,\bn'\in I_{\btau'}$, we can write
% \begin{align}
%     &[\bC_{\bk}]_{\bn,\bn'} \nonumber\\
%     &= \frac{1}{8\pi^2\SNR}  \sum_{\bell}\sum_{\bell'} (-1)^{|\bell+\bell'|} {\bk \choose \bell}{\bk \choose \bell'}\\
%     &\qquad\qquad\qquad\qquad \cdot[\btau\circ (\bell+\bk) = \btau\circ (\bell'+\bk')] \nonumber\\
%     &= \frac{1}{8\pi^2\SNR}  \sum_{\bk}\sum_{\bk'} (-1)^{|\bk+\bk'|}  {\bmm' \choose \bk}{\bmm' \choose \bk'}\\
%     &\qquad\qquad\qquad\qquad\qquad\qquad \cdot[\bell+\bk = \bell'+\bk'] \nonumber\\
%     &= \frac{1}{8\pi^2\SNR} (-1)^{|\bell+\bell'|} \sum_{\bk}  {\bmm' \choose \bk}{\bmm' \choose \bell+\bk-\bell'}\\
%     &= \frac{1}{8\pi^2\SNR} (-1)^{|\bell+\bell'|} \sum_{\bk} {\bmm' \choose \bk-\bell}{\bmm' \choose \bk-\bell'}\\
%     &=\frac{1}{8\pi^2\SNR}{2\bmm' \choose \bmm' +\bell-\bell'}.
% \end{align}

As an example, let us consider $N=16$, $k=2$, and $\tau=3$. The covariance matrix becomes
\begin{equation*}
\arraycolsep=2.5pt\def\arraystretch{1.5}
\scriptstyle
\left[\begin{array}{cccccccccc}
        \gray6 &   &   & \gray-4 &   &   & \gray1 &   &   & \gray  \\
          & \ggray6 &   &   &\ggray-4 &   &   & \ggray1 &   &   \\
          &   & \gggray6 &   &   &\gggray-4 &   &   & \gggray1 &   \\
        \gray-4&   &   & \gray6 &   &   &\gray-4 &   &   & \gray1 \\
          &\ggray-4 &   &   & \ggray6 &   &   &\ggray-4 &   &   \\
          &   &\gggray-4 &   &   & \gggray6 &   &   &\gggray-4 &   \\
        \gray1 &   &   &\gray-4 &   &   & \gray6 &   &   &\gray-4 \\
          & \ggray1 &   &   &\ggray-4 &   &   & \ggray6 &   &   \\
          &   & \gggray1 &   &   &\gggray-4 &   &   & \gggray6 &   \\
         \gray &   &   & \gray1 &   &   &\gray-4 &   &   & \gray6 \\
    \end{array}\right]_{\textstyle.}
\end{equation*}
This matrix is essentially a block matrix in that, after rearranging rows and columns, it becomes
\begin{equation*}
\arraycolsep=2.5pt\def\arraystretch{1.5}
\scriptstyle
    \left[\begin{array}{cccccccccc}
    \gray 6 & \gray-4 & \gray 1 & \gray   &   &   &   &   &   &   \\
    \gray -4& \gray 6 & \gray -4& \gray 1 &   &   &   &   &   &   \\
    \gray 1 &\gray -4 &\gray 6 &\gray -4 &   &   &   &   &   &   \\
    \gray  & \gray1 &\gray-4 &\gray  6 &   &   &   &   &   &   \\
      &   &   &   & \ggray6 &\ggray-4 & \ggray1 &   &   &   \\
      &   &   &   &\ggray-4 & \ggray6 &\ggray-4 &   &   &   \\
      &   &   &   & \ggray1 &\ggray-4 & \ggray6 &   &   &   \\
      &   &   &   &   &   &   & \gggray6 &\gggray-4 & \gggray1 \\
      &   &   &   &   &   &   &\gggray-4 & \gggray6 &\gggray-4 \\
      &   &   &   &   &   &   & \gggray1 &\gggray-4 & \gggray6 \\
\end{array}\right]_{\textstyle,}
\end{equation*}
which is block-diagonal with blocks given by the covariance matrices with $\tau=1$.

\subsection{Weight Computation}

Block-diagonal matrices can be inverted in a block-by-block fashion \cite[Ch. 0.9.2]{horn2012matrix}.
% \begin{align}
%     n = \tau\underbrace{\big\lfloor \tfrac{n}{\tau} \big\rfloor}_{\mathclap{\text{quotient}}} + \underbrace{n-\tau\big\lfloor \tfrac{n}{\tau} \big\rfloor}_{\text{remainder}} \in [N-k],
% \end{align}
As $n\in[N-\tau k]$ is the $\big\lfloor \frac{n}{\tau} \big\rfloor$th element in $\overline{n}\cap[N-\tau k]$, recalling \eqref{sumOfColumns} we obtain
\begin{align}
     \big[\bC_{k,\tau}^{-1}\one\big]_{n}
     &= \frac{{\lfloor \frac{n}{\tau} \rfloor + k\choose k} {|\overline{n}\cap[N-\tau k]|+k-\lfloor \frac{n}{\tau} \rfloor-1 \choose k}}{{2k \choose k}}\\
     &= \frac{{\lfloor \frac{n}{\tau} \rfloor + k\choose k} {\lceil \frac{N-n}{\tau} \rceil-1 \choose k}}{{2k \choose k}},
\end{align}
where
\begin{align}
    |\overline{n}\cap[N-\tau k]| &= \big\lceil \tfrac{N-\tau k-n}{\tau} \big\rceil + \big\lfloor \tfrac{n}{\tau} \big\rfloor\\
    &= \big\lceil \tfrac{N-n}{\tau} \big\rceil + \big\lfloor \tfrac{n}{\tau} \big\rfloor - k
\end{align}
is applied in the last step \cite[Ch. 3.4]{graham1994concrete}.
Therefore,
\begin{align}
    u_{k,\tau}(n) \propto [n\in[N-\tau k]]{\lfloor \frac{n}{\tau} \rfloor + k\choose k} {\lceil \frac{N-n}{\tau} \rceil-1 \choose k}
\end{align}
with the normalizing factor determined from $\sum_n u_{k,\tau}(n) = 1$. 

\section{}
\label{app:repeatedApplicationProof}

The proposed estimator in Alg. \ref{algo:multiLag} can be described in a recursive manner:
\begin{align}
    \hat{b}_\bk(y(\bn),\cT) \equiv \sum_{\btau\in\cT} \delta_{\bk,\btau}(y(\bn),\cT),
\end{align}
where $\delta_{\bk,\btau}(s(\bn),\cT)$ is given by \eqref{delta}. 
\begin{figure*}
\begin{align}
    \frac{1}{2\pi \btau^\bk} \arg\!\bigg(\!\mu_{\bk}\bigg(\cD_{\btau}^{\bk}\bigg(\! s(\bn) \exp \! \bigg(\!-j2\pi \!\!\sum_{\bmm \succ \bk}\! \hat{b}_{\bmm}(s(\bn),\cT){\bn \choose \bmm}\!\bigg)\exp \! \bigg(\!-j2\pi\!\! \sum_{\btau' \prec \btau}\! \delta_{\bk,\btau'}(s(\bn),\cT){\bn \choose \bk}\!\bigg)\!\bigg)\!\bigg)\!\bigg) \label{delta}
\end{align}
\hrulefill
\end{figure*}
Using induction on $\bk$ and $\btau$,
it shall next be proved that
\begin{align}
    \delta_{\bk,\btau}(y(\bn),\cT) - [\btau=\one]b_\bk - \delta_{\bk,\btau}(y(\bn)e^{-j2\pi x(\bn)},\cT) \in \bbZ \label{deltaInduction} .
\end{align}
 The above trivially holds for the maximum element $\bk\in\cM$ and $\btau=\one$ thanks to \eqref{circularProperty}. Let us now assume that the above also holds for $(\bmm,\btau')\in\cM\times\cT$ for the cases i) $\bmm \succ \bk$, and ii) $\bmm = \bk$ and $\btau'\prec \btau$. 
\begin{figure*}
\begin{align}
    &\delta_{\bk,\btau}(y(\bn),\cT) \nonumber\\
    &= \frac{1}{2\pi \btau^\bk}\arg\!\bigg(\!\mu_{\bk}\bigg(\cD_{\btau}^{\bk}\bigg[ y(\bn) \exp \! \bigg(\!-j2\pi \!\!\sum_{\bmm \succ \bk}\! \hat{b}_{\bmm}(y(\bn),\cT){\bn \choose \bmm}\!\bigg)\exp \! \bigg(\!-j2\pi\!\! \sum_{\btau' \prec \btau}\! \delta_{\bk,\btau'}(y(\bn),\cT){\bn \choose \bk}\!\bigg)\bigg]\bigg)\!\bigg)\label{deltaStart}\\
    &= \frac{1}{2\pi \btau^\bk}\arg\!\bigg(\!\mu_{\bk}\bigg(\cD_{\btau}^{\bk}\bigg[ y(\bn) \exp \! \bigg(\!-j2\pi \!\!\sum_{\bmm \succ \bk}\! \big(b_\bmm + \hat{b}_{\bmm}(y(\bn)e^{-j2\pi x(\bn)},\cT)\big){\bn \choose \bmm}\!\bigg)\label{deltaInductionHypothesis}\\
    &\quad \cdot \exp \! \bigg(\!-j2\pi\!\! \sum_{\btau' \prec \btau}\! \big(\delta_{\bk,\btau'}(y(\bn)e^{-j2\pi x(\bn)},\cT)+[\btau'=\one]b_\bk \big){\bn \choose \bk}\!\bigg)\bigg]\bigg)\!\bigg) \nonumber\\
    &= \frac{1}{2\pi \btau^\bk}\arg\!\bigg(\!\mu_{\bk}\bigg(\cD_{\btau}^{\bk}\bigg[ \exp \! \bigg(\!-j2\pi \sum_{\bmm \succ \bk}b_\bmm {\bn \choose \bmm} \bigg)\! \exp \! \bigg(\!-j2\pi\sum_{\btau' \prec \btau}\!  [\btau'=\one] b_\bk{\bn \choose \bk} \bigg) \bigg]\\
    &\quad \cdot \cD_{\btau}^{\bk}\bigg[y(\bn)  \exp \! \bigg(\!-j2\pi \!\!\sum_{\bmm \succ \bk}\! \hat{b}_{\bmm}(y(\bn)e^{-j2\pi x(\bn)},\cT){\bn \choose \bmm}\!\bigg) \exp \! \bigg(\!-j2\pi\!\! \sum_{\btau' \prec \btau}\! \delta_{\bk,\btau'}(y(\bn)e^{-j2\pi x(\bn)},\cT){\bn \choose \bk}\!\bigg)\bigg]\bigg)\!\bigg) \nonumber\\
    &= \frac{1}{2\pi \btau^\bk}\arg\!\bigg(\!\mu_{\bk}\bigg(\cD_{\btau}^{\bk}\bigg[e^{j2\pi x(\bn)} \exp \! \bigg(\!-j2\pi \sum_{\bmm \succ \bk}b_\bmm {\bn \choose \bmm} \bigg)\! \exp \! \bigg(\!-j2\pi\sum_{\btau' \prec \btau}\!  [\btau'=\one] b_\bk{\bn \choose \bk} \bigg) \bigg]\\
    &\quad \cdot \cD_{\btau}^{\bk}\bigg[y(\bn)e^{-j2\pi x(\bn)}  \exp \! \bigg(\!-j2\pi \!\!\sum_{\bmm \succ \bk}\! \hat{b}_{\bmm}(y(\bn)e^{-j2\pi x(\bn)},\cT){\bn \choose \bmm}\!\bigg) \exp \! \bigg(\!-j2\pi\!\! \sum_{\btau' \prec \btau}\! \delta_{\bk,\btau'}(y(\bn)e^{-j2\pi x(\bn)},\cT){\bn \choose \bk}\!\bigg)\bigg]\bigg)\!\bigg) \nonumber\\
    &= \frac{1}{2\pi \btau^\bk}\arg\!\bigg(\!\mu_{\bk}\bigg(\cD_{\btau}^{\bk}\bigg[ \exp \! \bigg(\!j2\pi \sum_{\bmm \prec \bk}b_\bmm {\bn \choose \bmm} \bigg)\bigg] \cD_{\btau}^{\bk}\bigg[\exp \! \bigg(\!j2\pi  [\btau=\one] b_\bk{\bn \choose \bk} \bigg)\bigg]\label{deltaSimplification}\\
    &\quad \cdot \cD_{\btau}^{\bk}\bigg[y(\bn)e^{-j2\pi x(\bn)} \exp \! \bigg(\!-j2\pi \!\!\sum_{\bmm \succ \bk}\! \hat{b}_{\bmm}(y(\bn)e^{-j2\pi x(\bn)},\cT){\bn \choose \bmm}\!\bigg) \exp \! \bigg(\!-j2\pi\!\! \sum_{\btau' \prec \btau}\! \delta_{\bk,\btau'}(y(\bn)e^{-j2\pi x(\bn)},\cT){\bn \choose \bk}\!\bigg)\bigg]\bigg)\!\bigg) \nonumber\\
    &= \frac{1}{2\pi \btau^\bk}\arg\!\bigg(\!e^{j2\pi[\btau=\one]\btau^\bk b_\bk}\mu_{\bk}\bigg(\cD_{\btau}^{\bk}\bigg[y(\bn)e^{-j2\pi x(\bn)}\label{deltaEnd}\\
    &\quad\cdot    \exp \! \bigg(\!-j2\pi \!\!\sum_{\bmm \succ \bk}\! \hat{b}_{\bmm}(y(\bn)e^{-j2\pi x(\bn)},\cT){\bn \choose \bmm}\!\bigg) \exp \! \bigg(\!-j2\pi\!\! \sum_{\btau' \prec \btau}\! \delta_{\bk,\btau'}(y(\bn)e^{-j2\pi x(\bn)},\cT){\bn \choose \bk}\!\bigg)\bigg]\bigg)\!\bigg) \nonumber
\end{align}
\hrulefill
\end{figure*}
We then have \eqref{deltaStart}--\eqref{deltaEnd}, which concludes the proof; \eqref{deltaInductionHypothesis} is a direct consequence of the induction hypothesis, \eqref{deltaSimplification} follows from
\begin{align}
    &e^{j2\pi x(\bn)}\exp \! \bigg(\!-j2\pi \!\sum_{\bmm \succ \bk}\!b_\bmm {\bn \choose \bmm} \!\bigg) \nonumber \\
    &\qquad\qquad\qquad\qquad \cdot\exp \! \bigg(\!-j2\pi\!\sum_{\btau' \prec \btau}\!  [\btau'=\one] b_\bk{\bn \choose \bk} \!\bigg) \nonumber\\
    &=\exp \! \bigg(\!j2\pi \!\sum_{\bmm \prec \bk}b_\bmm {\bn \choose \bmm} \!\bigg)\exp \! \bigg(\!j2\pi b_\bk {\bn \choose \bk} \!\bigg) \nonumber \\
    &\quad \cdot  \exp \! \bigg(\!-j2\pi  [\btau\neq\one] b_\bk{\bn \choose \bk} \!\bigg)  \\
    &=\exp \! \bigg(\!j2\pi \!\sum_{\bmm \prec \bk}\!b_\bmm {\bn \choose \bmm} \!\bigg)\exp \! \bigg(\!j2\pi  [\btau=\one] b_\bk{\bn \choose \bk} \!\bigg)
\end{align}
and \eqref{deltaEnd} holds because
\begin{align}
    &\cD_{\btau}^{\bk}\bigg[ \exp \! \bigg(\!j2\pi\! \sum_{\bmm \prec \bk}b_\bmm {\bn \choose \bmm} \!\bigg)\bigg] = 1 \\
    &\cD_{\btau}^{\bk}\bigg[\exp \! \bigg(\!j2\pi  [\btau=\one] b_\bk{\bn \choose \bk} \!\bigg)\bigg] = e^{j2\pi[\btau=\one]\btau^\bk b_\bk}.
\end{align}
% \section{}
% \label{app:modifiedEstimatorMonomialProof}

% In this appendix, we present a proof for \eqref{modificationCircularMonomial}. 

% On the other hand, for $\bk\in\cM$, vectorization gives 
% \begin{align}
%     &\bE\big(\hat{\ba}\big(y(\bn),\cM\big)-\ba-\hat{\ba}\big(y(\bn)e^{-j2\pi x(\bn)},\cM\big)\big) \nonumber\\
%     &=\bE(\bE^\top(\bT(\bJ')^{-1}\bT^\top)^{-1}\bE)^{-1} \\
%     &\qquad\qquad \cdot\bE^\top(\bT(\bJ')^{-1}\bT^\top)^{-1} \hat{\ba}\big(y(\bn),\cM'\big)\nonumber\\
%     & - \bE(\bE^\top(\bT(\bJ')^{-1}\bT^\top)^{-1}\bE)^{-1} \bE^\top(\bT(\bJ')^{-1}\bT^\top)^{-1}\bE\ba \nonumber\\
%     & -\bE(\bE^\top(\bT(\bJ')^{-1}\bT^\top)^{-1}\bE)^{-1} \nonumber\\
%     &\qquad\qquad \cdot\bE^\top(\bT(\bJ')^{-1}\bT^\top)^{-1} \hat{\ba}\big(y(\bn)e^{-j2\pi x(\bn)},\cM'\big) \nonumber\\
%     &=\bE(\bE^\top\!(\bT(\bJ')^{-1}\bT^\top)^{-1}\bE)^{-1} \bE^\top\!(\bT(\bJ')^{-1}\bT^\top)^{-1} \\
%     &\qquad \cdot\big(\hat{\ba}\big(y(\bn),\cM'\big) - \bE\ba - \hat{\ba}\big(y(\bn)e^{-j2\pi x(\bn)},\cM'\big)\big). \nonumber
% \end{align}

% \begin{align}
%     &\!\bE\big(\hat{\ba}_\bk\big(y(\bn),\cM\big)\!-\!\ba_\bk\!-\!\hat{\ba}_\bk\big(y(\bn)e^{-j2\pi x(\bn)},\cM\big)\!\big)\\
%     &\!-\!\big(\hat{\ba}_\bk\big(y(\bn),\cM'\big)\!-\!\ba_\bk\!-\!\hat{\ba}_\bk\big(y(\bn)e^{-j2\pi x(\bn)},\cM'\big)\!\big)\!\in\!\bT\bbZ^{|\cM'|}\!, \nonumber
% \end{align}

\bibliographystyle{IEEEtran}
\bibliography{ref}

\end{document}